\documentstyle[preprint,aps,epsf]{revtex}
\tightenlines
\begin{document}
\title{Scalar Quarkonium Masses and Mixing with the Lightest Scalar Glueball}
\author{W.\ Lee \cite{LANL}
and D.\ Weingarten}
\address{ IBM Research, P.O.~Box 218,
Yorktown Heights, NY 10598}
\maketitle

\begin{abstract}

We evaluate the continuum limit of the valence (quenched) approximation
to the mass of the lightest scalar quarkonium state for a range of
different quark masses and to the mixing energy between these states
and the lightest scalar glueball. Our results support the interpretation
of $f_0(1710)$ as composed mainly of the lightest scalar glueball.

\end{abstract}
\pacs{12.38.Gc, 13.25.-k, 14.40.-n}
\narrowtext

\section{Introduction}\label{sect:intro}

Evidence that $f_0(1710)$ is composed mainly of the lightest scalar
glueball is now given by two different sets of numerical determinations
of QCD predictions using the theory's lattice formulation in the valence
(quenched) approximation.  A calculation on GF11~\cite{Sexton95} of the
width for the lightest scalar glueball to decay to all possible
pseudoscalar pairs, on a $16^3 \times 24$ lattice with $\beta$ of 5.7,
corresponding to a lattice spacing $a$ of 0.140(4) fm, gives 108(29)
MeV.  This number combined with any reasonable guesses for the effect of
finite lattice spacing, finite lattice volume, and the remaining width
to multibody states yields a total width small enough for the lightest
scalar glueball to be seen easily in experiment.  For the infinite
volume continuum limit of the lightest scalar glueball mass, a
reanalysis~\cite{latestglue} of a calculation on GF11~\cite{Vaccarino},
using 25000 to 30000 gauge configurations, gives 1648(58) MeV.	An
independent calculation by the UKQCD-Wuppertal~\cite{Livertal}
collaboration, using 1000 to 3000 gauge configurations, when
extrapolated to the continuum limit according to
Refs.~\cite{latestglue,Weingarten94} yields 1567(88) MeV.  A more recent
calculation using an improved action~\cite{Morningstar} gives 1730(94)
MeV.  The three results combined become 1656(47) MeV. A phenomenological
model of the glueball spectrum which supports this prediction is
discussed in Ref.~\cite{Brisudova}.

Among established resonances with the quantum numbers to be a scalar
glueball, all are clearly inconsistent with the mass calculations except
$f_0(1710)$ and $f_0(1500)$.  Between these two, $f_0(1710)$ is favored
by the mass result with largest statistics, by the combined result, and
by the expectation~\cite{Weingarten97} that the valence approximation
will lead to an underestimate of the scalar glueball's mass.
Refs.~\cite{Sexton95,Weingarten97} interpret $f_0(1500)$ as dominantly
composed of strange-antistrange, $s\overline{s}$, scalar quarkonium.  A
possible objection to this interpretation, however, is that $f_0(1500)$
apparently does not decay mainly to states containing an $s$ and an
$\overline{s}$ quark~\cite{Amsler1}. In part for this reason,
Ref.~\cite{Amsler2} interprets $f_0(1500)$ as composed mainly of the
lightest scalar glueball and $f_0(1710)$ as largely $s\overline{s}$
scalar quarkonium.  A second objection is that while the Hamiltonian of
full QCD couples quarkonium and glueballs, so that physical states
should be linear combinations of both, mixing is not treated
quantitatively in Ref.~\cite{Sexton95}. In the extreme, mixing could
lead to $f_0(1710)$ and $f_0(1500)$ each half glueball and half
quarkonium.

Using the valence approximation for a fixed lattice period $L$ of about
1.6 fm and a range of different values of quark mass, we have now
calculated the continuum limit of the mass of the lightest scalar
$q\overline{q}$ states and the continuum limit of the mixing energy
between these states and the lightest scalar glueball. Our calculations
have been done with four different choices of lattice spacing.
Continuum predictions are found by extrapolation of results obtained
from the three smallest values of lattice spacing. For the two choices
of lattice spacing we have also calculated scalar $q\overline{q}$ masses
on lattices with $L$ of about 2.3 fm, and for one choice of lattice
spacing we have found scalar quarkonium-glueball mixing energies on a
lattice with $L$ of about 2.3 fm.  Preliminary versions of this work are
reported in Refs.~\cite{Weingarten97,Weingarten98,Lee97}.

Our results provide answers to the objections to the interpretation of
$f_0(1710)$ as largely the lightest scalar glueball.  For the valence
approximation to the infinite volume continuum limit of the
$s\overline{s}$ scalar mass we find a value significantly below the
valence approximation scalar glueball mass. This prediction makes
improbable, in our opinion, the identification~\cite{Amsler2} of
$f_0(1500)$ as primarily a glueball and $f_0(1710)$ as primarily
$s\overline{s}$ quarkonium. Our calculation of glueball-quarkonium
mixing energy, combined with the simplification of considering mixing
only among the lightest discrete isosinglet scalar states, then yields a
mixed $f_0(1710)$ which is 73.8(9.5)\% glueball and a mixed $f_0(1500)$
which is 98.4(1.4)\% quarkonium, mainly $s\overline{s}$. The glueball
amplitude which leaks from $f_0(1710)$ goes almost entirely to the state
$f_0(1390)$, which remains mainly $n\overline{n}$, normal-antinormal,
the abbreviation we adopt for $(u\overline{u} +
d\overline{d})/\sqrt{2}$. We find also that $f_0(1500)$ acquires an
$n\overline{n}$ amplitude with sign opposite to its $s\overline{s}$
component suppressing, by interference, the state's decay to
$K\overline{K}$ final states.  Assuming SU(3) flavor symmetry before
mixing for the decay couplings of scalar quarkonium to pairs of
pseudoscalars, the $K\overline{K}$ decay rate of $f_0(1500)$ is
suppressed by a factor of 0.39(16) in comparison to the rate of an
unmixed $s\overline{s}$ scalar.	 This suppression is consistent, within
uncertainties, with the experimentally observed suppression.

It perhaps is useful to discuss briefly at this point a proposed
calculation of mixing between valence approximation quarkonium and
glueball states through common decay channels~\cite{Boglione} which
forms the basis for additional objections to the identification of
$f_0(1710)$ as primarily a glueball.  A detailed examination of problems
with the calculation of Ref.~\cite{Boglione} appears in
Ref.~\cite{Lee99}.  One defect of the work in Ref.~\cite{Boglione} is
the omission of quarkonium to glueball transitions by direct
annihilation of the quarkonium's quark and antiquark into chromoelectric
field. Direct annihilation is the leading valence approximation
contribution to mixing and is evaluated in the present paper.  On the
other hand, the transitions through two-pseudoscalar intermediate states
which Ref.~\cite{Boglione} considers include an extra closed quark loop
in addition to the quark paths of the direct quark-antiquark
annihilation process. Thus according to a systematic scheme for
evaluating all quark loop corrections to the valence
approximation~\cite{Lee99}, the decay channel mixing calculation is part
of the one-quark-loop correction to the direct annihilation mixing
amplitude.  As shown in detail in Ref.~\cite{Lee99}, however, the
corrections to the direct annihilation amplitude must include also a
counterterm proportional to the pure gauge action.  This counterterm is
required to compensate for the shift between the screened effective
gauge coupling used in the valence approximation and the unscreened bare
coupling of full QCD. The counterterm is entirely absent from the
calculation of Ref.~\cite{Boglione}. As a consequence of this omission
and of the omission of the direct quark-antiquark annihilation term, we
believe the mixing calculation of Ref.~\cite{Boglione} is not correct.

The remainder of this paper is organized as follows. In
Section~\ref{sect:ensembles} we describe the Monte Carlo ensembles of
gauge field configurations we use. In Section~\ref{sect:mass} we present
the calculation of scalar quarkonium masses. In Section~\ref{sect:glb}
we describe a glueball mass calculation.  In Section~\ref{sect:mix} we
present a calculation of quarkonium-glueball mixing energy. In
Section~\ref{sect:states} we consider the physical mixed glueball and
quarkonium states. Finally, Section~\ref{sect:decay} briefly examines
consequences of quarkonium-glueball mixing for glueball decay.

\section{Monte Carlo Ensembles}\label{sect:ensembles}

Our calculations, using Wilson fermions and the plaquette action, were
done for four choices of $\beta$ with two different lattice sizes at
each of the two smallest $\beta$ giving a total of six combinations of
$\beta$ and lattice structure. These are listed in
Table~\ref{tab:lattices}.  For each combination of $\beta$ and lattice
structure, calculations were typically done with five different choices
of $\kappa$. These are listed in Table~\ref{tab:kappas} along with the
corresponding Monte Carlo ensemble sizes.  In all cases, a sufficient
number of updating sweeps was skipped between successive Monte Carlo
configuration to leave no statistically significant correlations between
successive pairs.  The ensemble of 599 configurations used for two
hopping constant choices at $\beta$ of 6.4 is a subset of the 1003
configuration ensemble used for the three other $\kappa$ at $\beta$ of
6.4. At $\beta$ of 5.70 on a lattice $16^3 \times 24$, the 3870 member
ensemble at $\kappa$ of 0.1600 is a subset of the 12186 member ensemble
at $\kappa$ of 0.1650, which has no overlap with the 1972 member
ensemble used at $\kappa$ of 0.1625 and 0.1650. For all other entries in
Table~\ref{tab:kappas}, all $\kappa$ values share a single ensemble of
gauge field configurations.

From the smallest to largest $\beta$, the lattice
spacing varies by nearly a factor of 2.7. The smaller lattices with
$\beta$ of 5.70 and 5.93, and the lattices with $\beta$ of 6.17 and 6.40
have nearly the same periods in the two (or three) equal space
directions and thereby permit extrapolations to zero lattice spacing
with nearly constant physical volume.

The values of lattice spacing and lattice period in
Table~\ref{tab:lattices} and conversions from lattice to physical units
in the remainder of this article, are
determined~\cite{latestglue,Vaccarino} from the exact solution to the
two-loop zero-flavor Callan-Symanzik equation for
$\Lambda^{(0)}_{\overline{MS}} a$ with $\Lambda^{(0)}_{\overline{MS}}$
of 234.9(6.2) MeV determined from the continuum limit of
$(\Lambda^{(0)}_{\overline{MS}} a)/(m_{\rho} a)$ in Ref.~\cite{Butler}.
For $\beta$ from 5.70 to 6.17, the ratio $(\Lambda^{(0)}_{\overline{MS}}
a)/(m_{\rho} a)$ in Ref.~\cite{Butler} was found to be constant within
statistical errors, thus our results are, within errors, almost
certainly the same as those we would have obtained by converting to
physical units using values of $m_{\rho} a$.  We chose to convert using
$\Lambda^{(0)}_{\overline{MS}} a$, however, since Ref.~\cite{Butler} did
not find $m_{\rho} a$ at $\beta$ of 6.40, which would be needed for our
present calculations.

\section{Quarkonium Masses}\label{sect:mass}

For each ensemble of gauge fields, with two exceptions at $\beta$ of
5.70, we evaluated correlation functions using smeared Coulomb gauge
quark and antiquark fields incorporating random sources following
Ref.~\cite{Lee97}. From these fields we constructed pseudoscalar and
scalar quarkonium propagators.  Averaged over the random sources, the
propagators we calculated become
\begin{eqnarray}
\label{defC}
C_{ff}(t) & = & \sum_{\vec{x}} < f_q(\vec{x},t) f_r(0,0)>,
\end{eqnarray}
where $f$ is either $\pi$ or $\sigma$ and $\pi_v(\vec{x},t)$ and
$\sigma_v(\vec{x},t)$ are, respectively, the smeared pseudoscalar and
scalar operators of Ref.~\cite{Lee97} with smearing size $v$.  At
$\beta$ of 5.70 on a lattice $16^3 \times 24$, for the 3870 member gauge
ensemble with $\kappa$ of 0.1600 and for the 12186 member ensemble at
$\kappa$ of 0.1650, the propagators of Eq.~(\ref{defC}) were evaluated
directly without use of random sources.  Evidence in Ref.~\cite{Lee97}
suggests that for equal statistical uncertainties, propagators found
using random sources require about half the computer time needed for a
direct calculation. The values used for $r$ for each $\beta$ and lattice
are listed in Table~\ref{tab:smearing}.

For sufficiently large values of $t$ and the lattice time period $T$,
$C_{\pi \pi}(t)$ and $C_{\sigma \sigma}(t)$ are expected to approach the
asymptotic form
\begin{eqnarray}
\label{diagZm}
C_{ff}(t) & \rightarrow & Z_f [ exp(-m_f a t) + exp(m_f a t - m_f a T)] ,
\end{eqnarray}
where $f$ can be either $\pi$ or $\sigma$.  Fitting the large $t$
behavior of $C_{\pi \pi}(t)$ and $C_{\sigma \sigma}(t)$ to
Eq.~(\ref{diagZm}) we obtained the masses, in lattice units, $m_{\pi} a$
and $m_{\sigma} a$, and the field strength renormalization constants
$Z_{\pi}$ and $Z_{\sigma}$.

\begin{figure}
\epsfxsize=\textwidth
\epsfbox{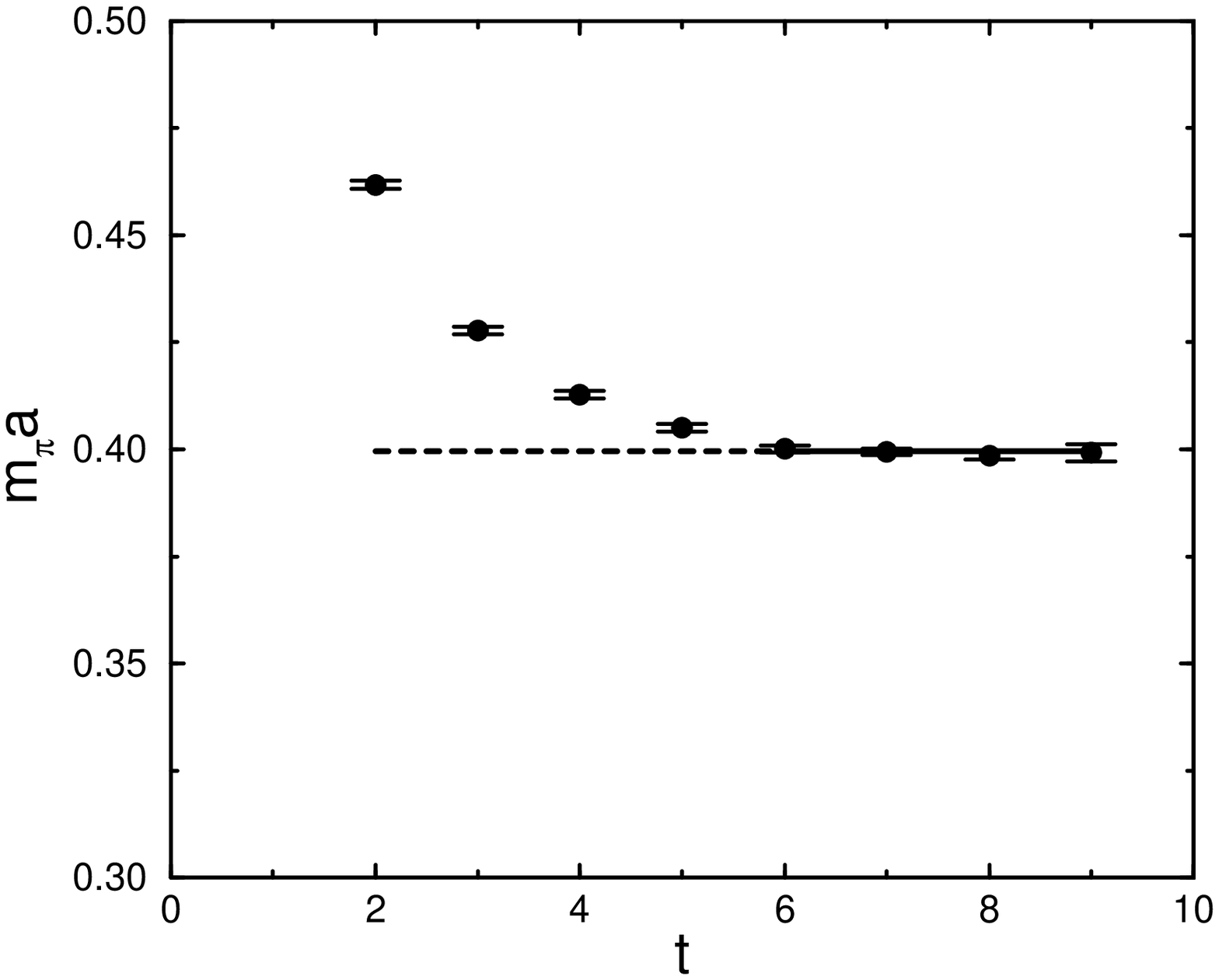}
\caption{Pseudoscalar effective masses and fitted mass for $\beta$ of
5.93 and $\kappa$ of 0.1554 on a lattice $16^2 \times 14 \times 20$.}
\label{fig:psb593x16k1554}
\end{figure}

\begin{figure}
\epsfxsize=\textwidth
\epsfbox{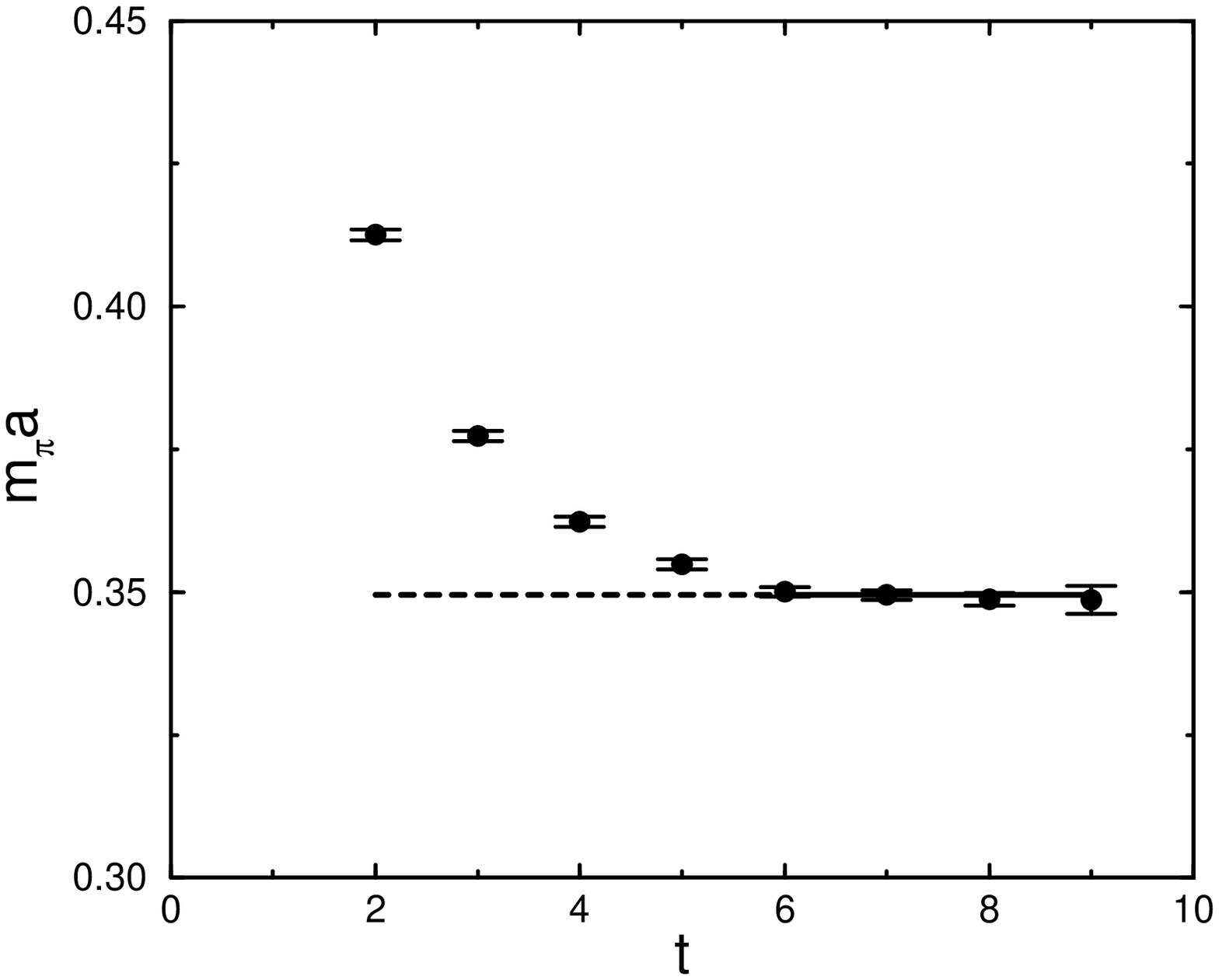}
\caption{Pseudoscalar effective masses and fitted mass for $\beta$ of
5.93 and $\kappa$ of 0.1562 on a lattice $16^2 \times 14 \times 20$.}
\label{fig:psb593x16k1562}
\end{figure}

\begin{figure}
\epsfxsize=\textwidth
\epsfbox{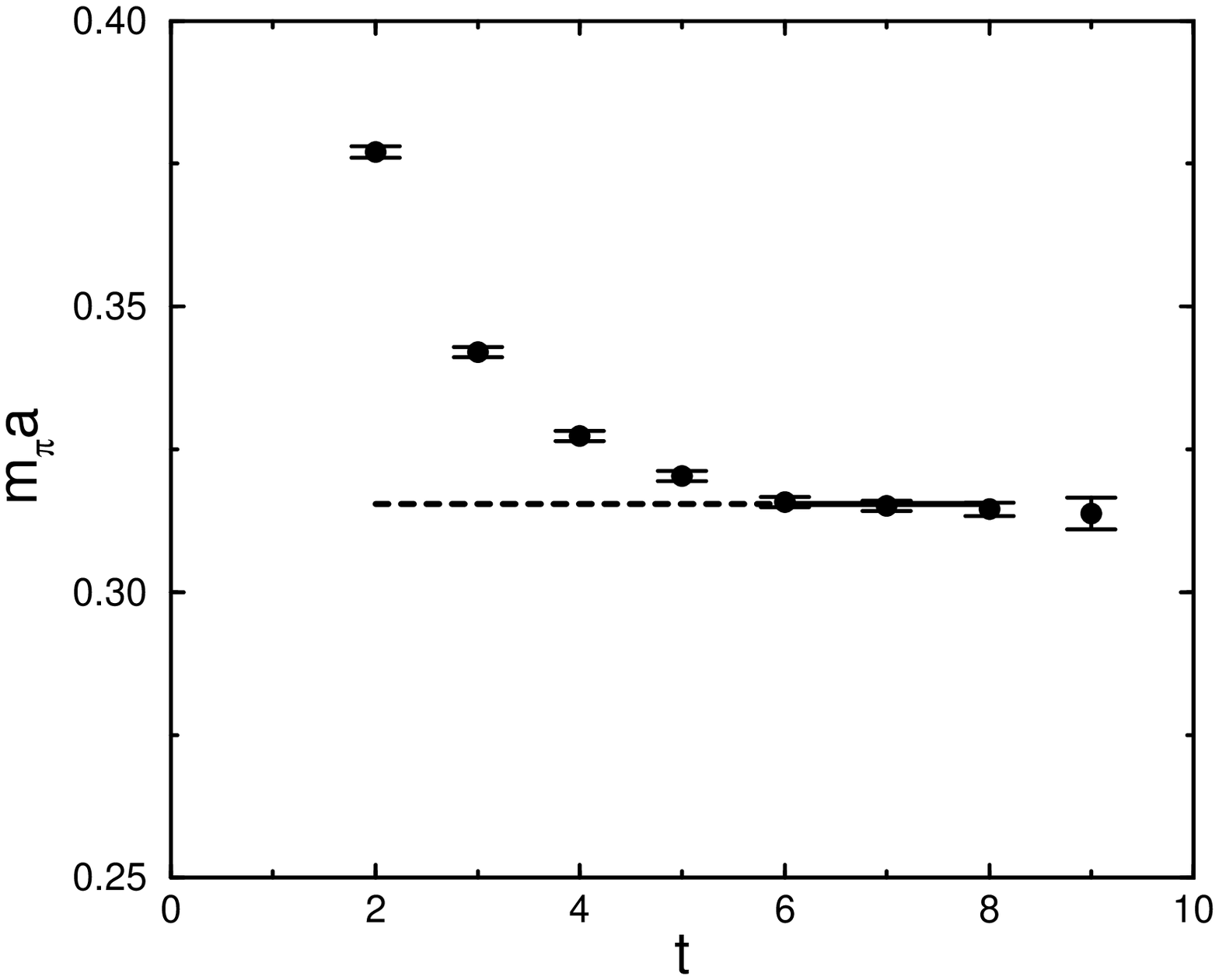}
\caption{Pseudoscalar effective masses and fitted mass for $\beta$ of
5.93 and $\kappa$ of 0.1567 on a lattice $16^2 \times 14 \times 20$.}
\label{fig:psb593x16k1567}
\end{figure}

\begin{figure}
\epsfxsize=\textwidth
\epsfbox{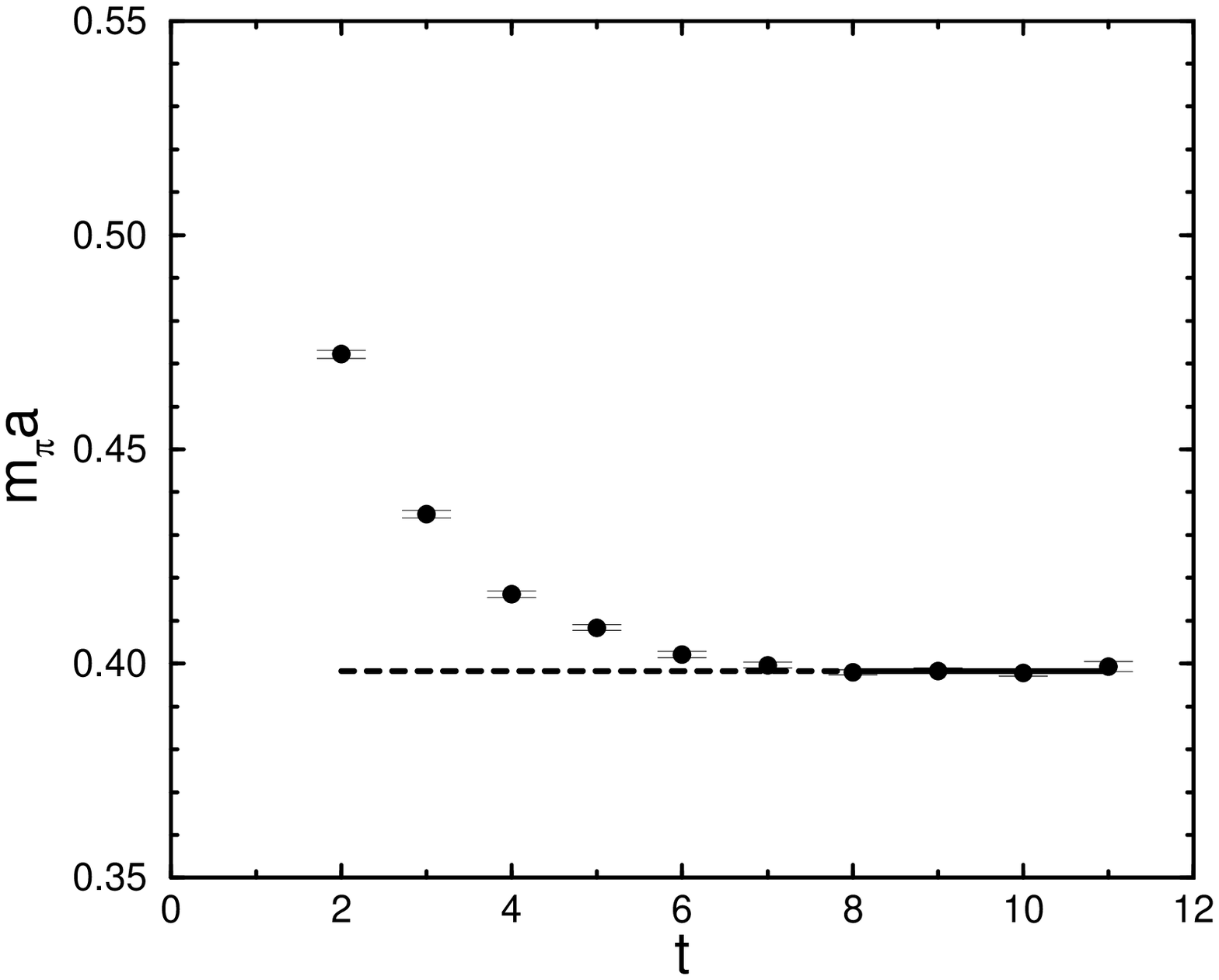}
\caption{Pseudoscalar effective masses and fitted mass for $\beta$ of
5.93 and $\kappa$ of 0.1554 on a lattice $24^4$.}
\label{fig:psb593x24k1554}
\end{figure}

\begin{figure}
\epsfxsize=\textwidth
\epsfbox{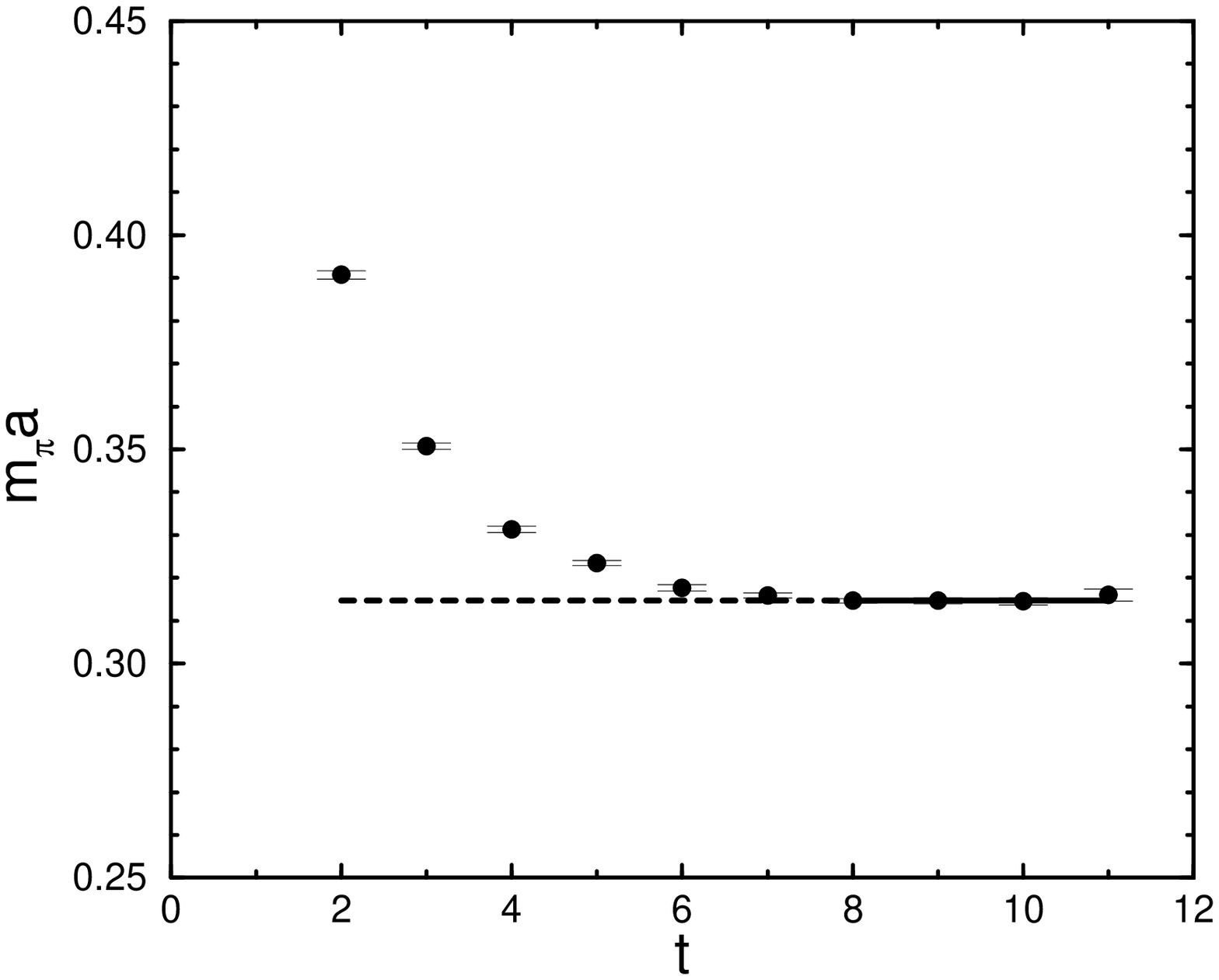}
\caption{Pseudoscalar effective masses and fitted mass for $\beta$ of
5.93 and $\kappa$ of 0.1567 on a lattice $24^4$.}
\label{fig:psb593x24k1567}
\end{figure}

\begin{figure}
\epsfxsize=\textwidth
\epsfbox{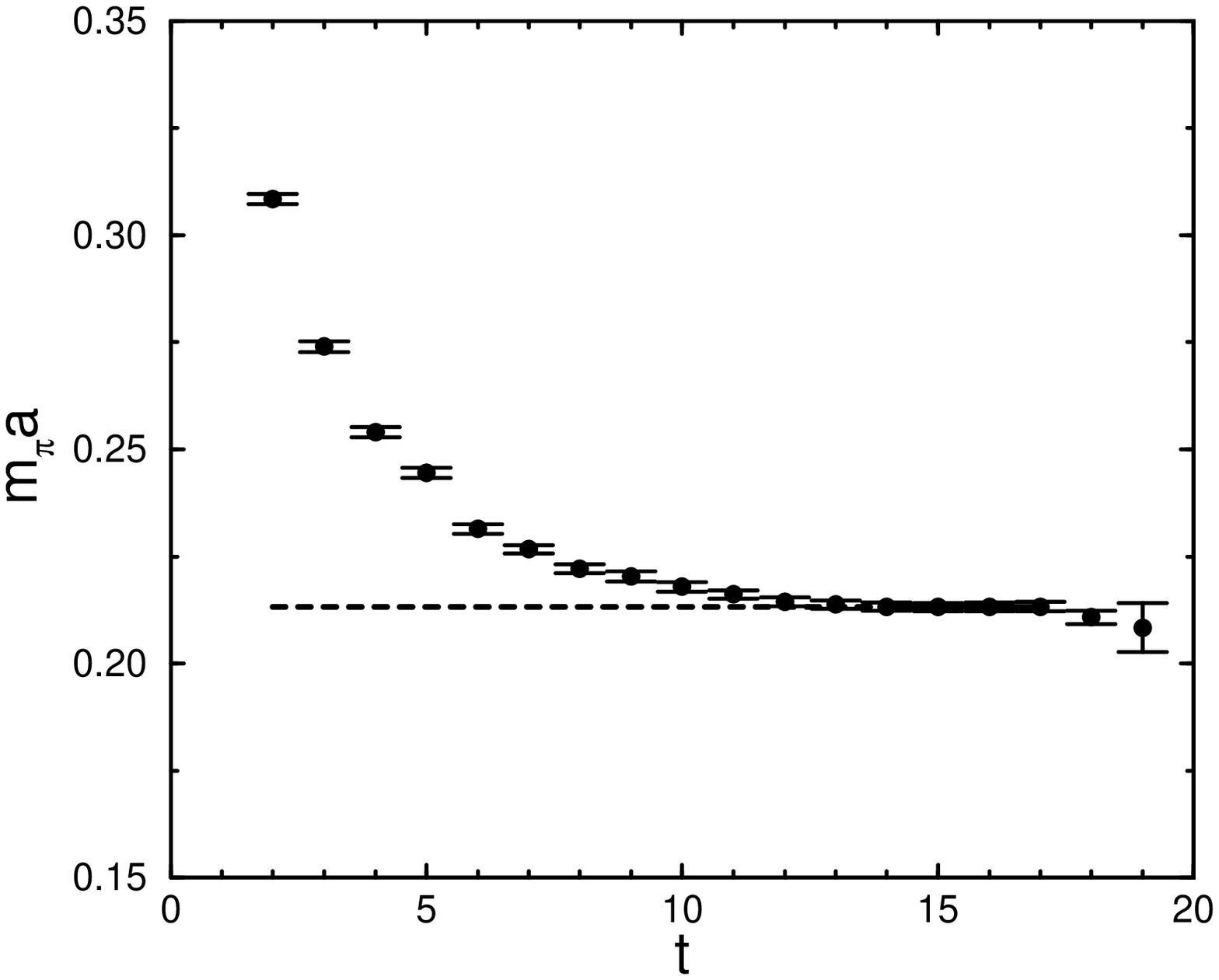}
\caption{Pseudoscalar effective masses and fitted mass for $\beta$ of
6.40 and $\kappa$ of 0.1491 on a lattice $32^2 \times 28 \times 40$.}
\label{fig:psb64x32k1491}
\end{figure}

\begin{figure}
\epsfxsize=\textwidth
\epsfbox{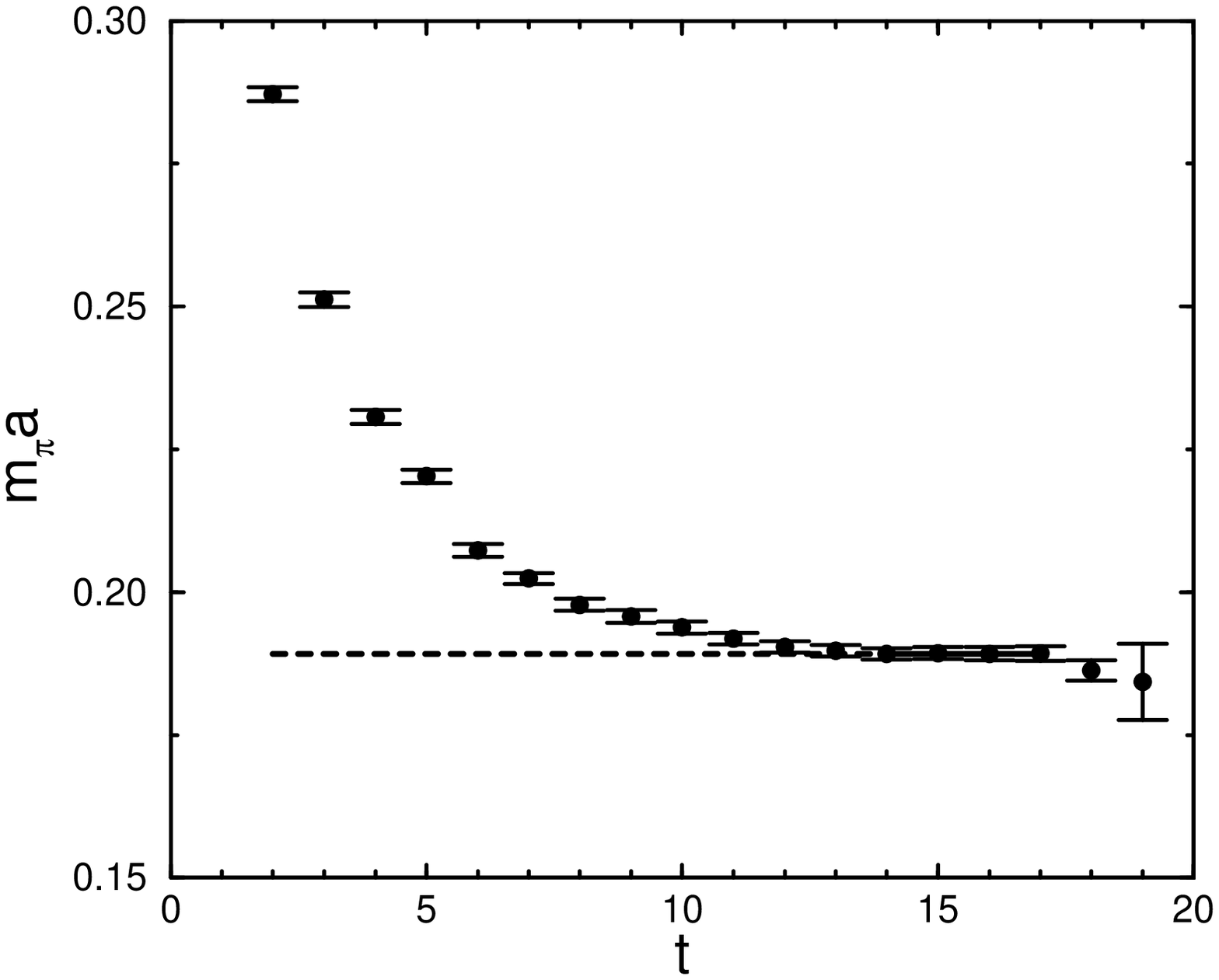}
\caption{Pseudoscalar effective masses and fitted mass for $\beta$ of
6.40 and $\kappa$ of 0.1494 on a lattice $32^2 \times 28 \times 40$.}
\label{fig:psb64x32k1494}
\end{figure}

\begin{figure}
\epsfxsize=\textwidth
\epsfbox{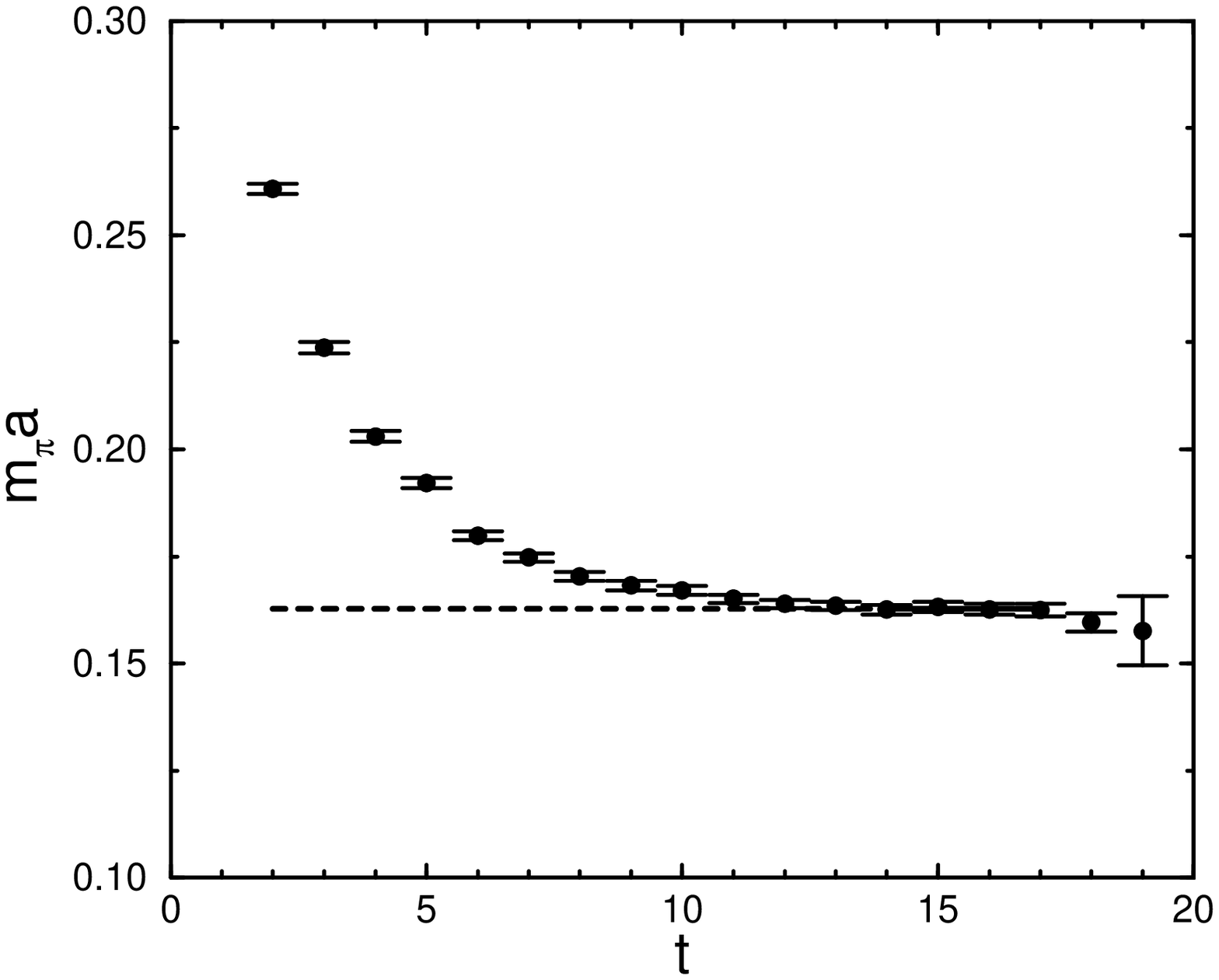}
\caption{Pseudoscalar effective masses and fitted mass for $\beta$ of
6.40 and $\kappa$ of 0.1497 on a lattice $32^2 \times 28 \times 40$.}
\label{fig:psb64x32k1497}
\end{figure}

\begin{figure}
\epsfxsize=\textwidth
\epsfbox{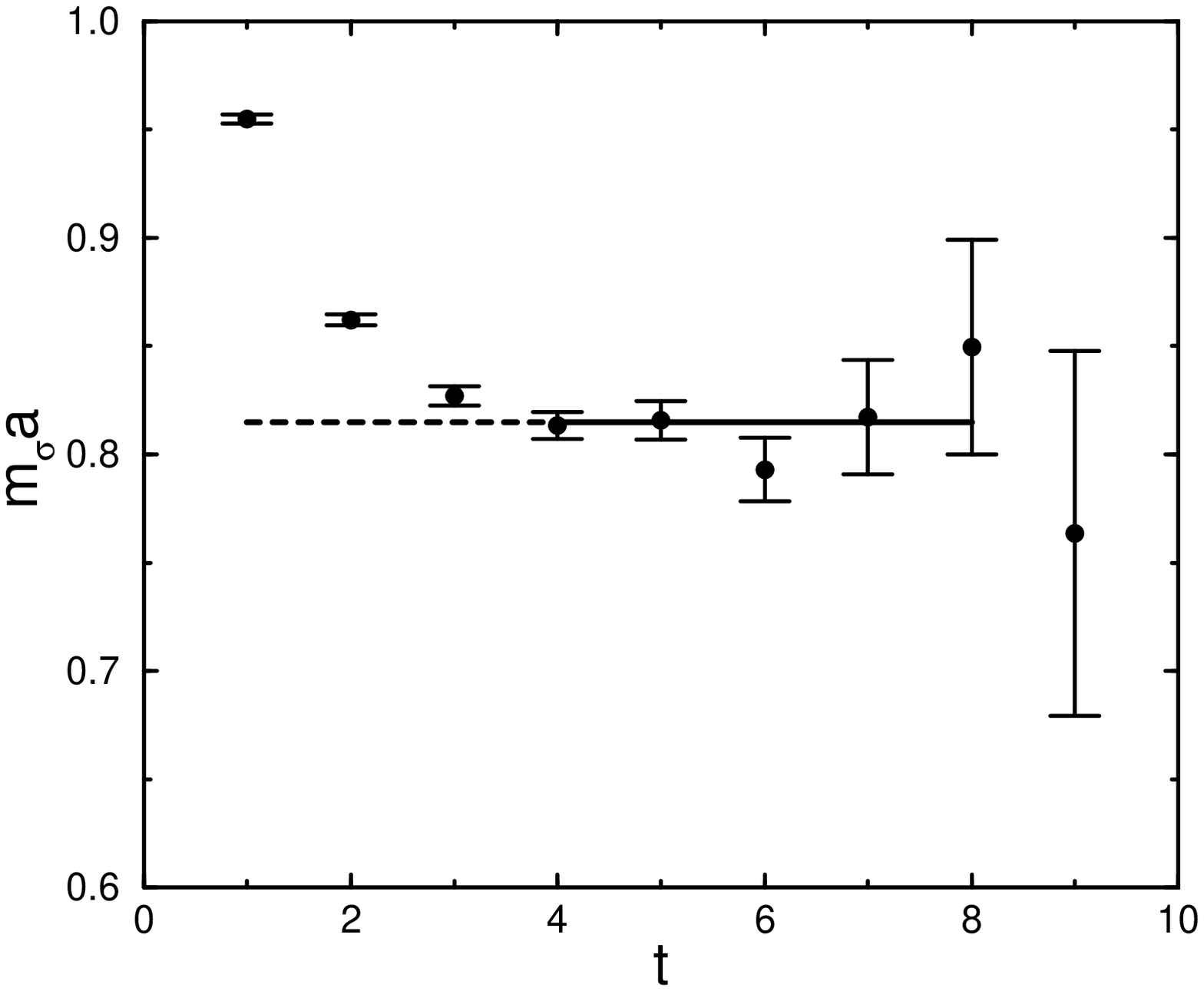}
\caption{Scalar effective masses and fitted mass for $\beta$ of
5.93 and $\kappa$ of 0.1554 on a lattice $16^2 \times 14 \times 20$.}
\label{fig:scb593x16k1554}
\end{figure}

\begin{figure}
\epsfxsize=\textwidth
\epsfbox{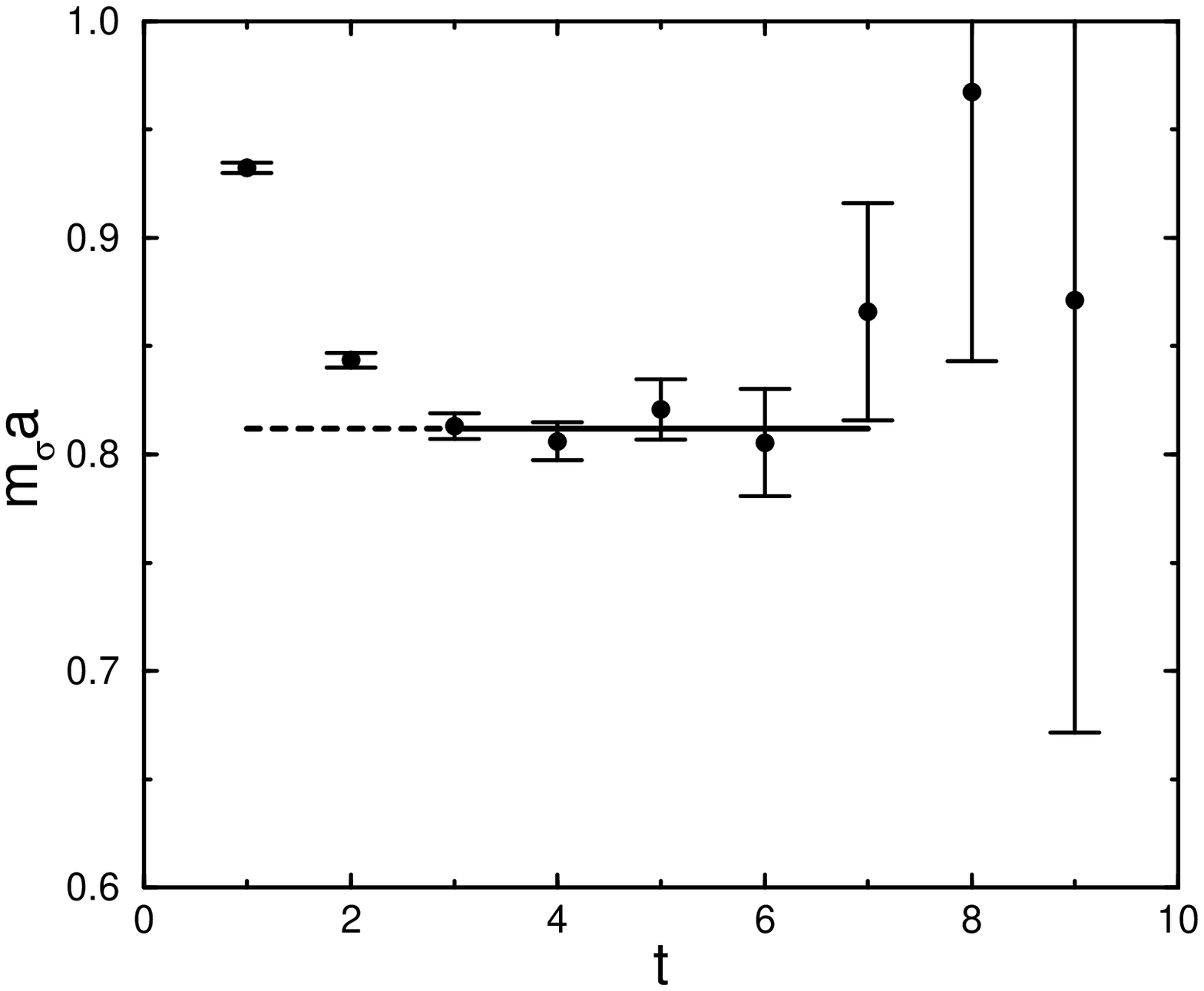}
\caption{Scalar effective masses and fitted mass for $\beta$ of
5.93 and $\kappa$ of 0.1562 on a lattice $16^2 \times 14 \times 20$.}
\label{fig:scb593x16k1562}
\end{figure}

\begin{figure}
\epsfxsize=\textwidth
\epsfbox{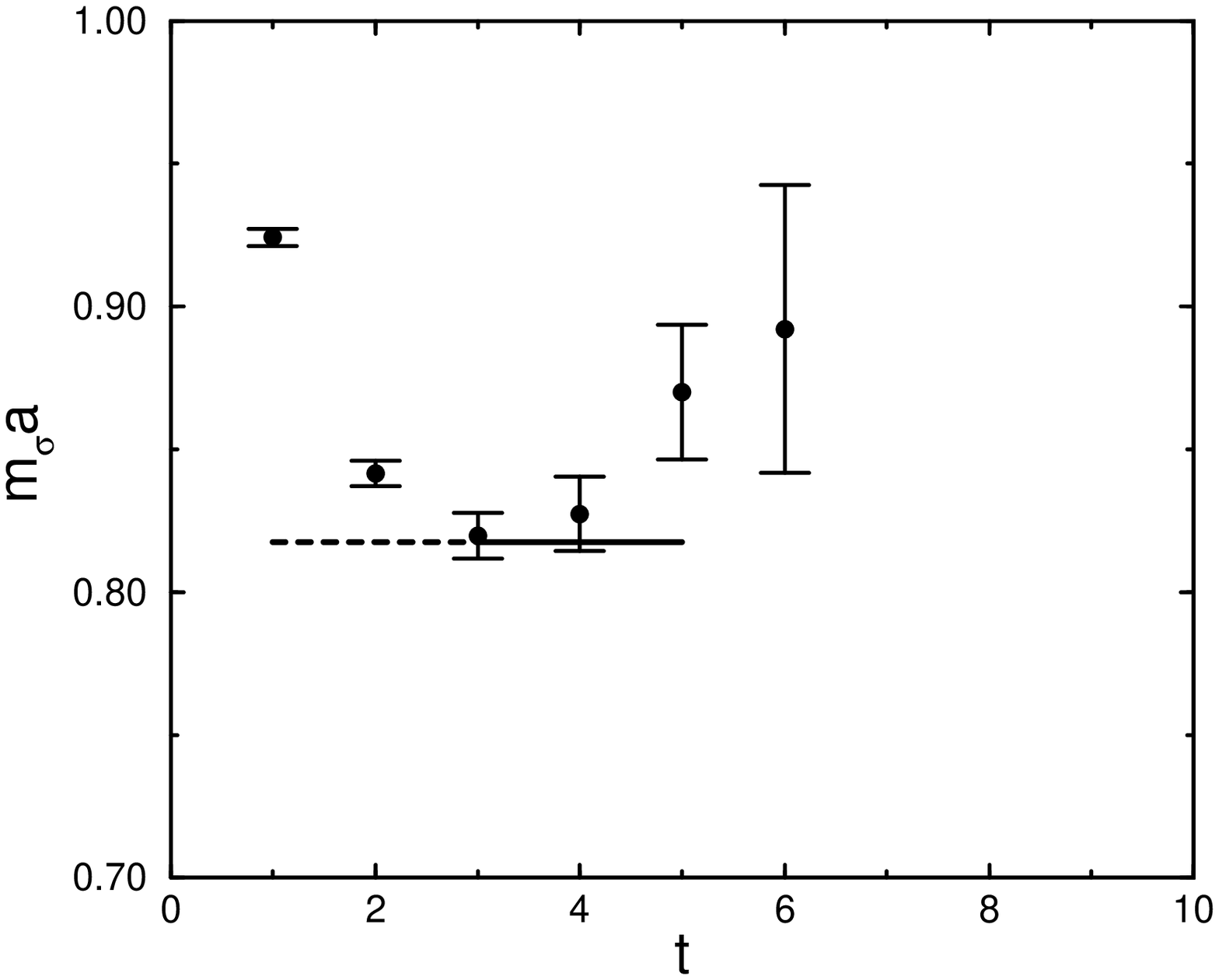}
\caption{Scalar effective masses and fitted mass for $\beta$ of
5.93 and $\kappa$ of 0.1567 on a lattice $16^2 \times 14 \times 20$.}
\label{fig:scb593x16k1567}
\end{figure}

\begin{figure}
\epsfxsize=\textwidth
\epsfbox{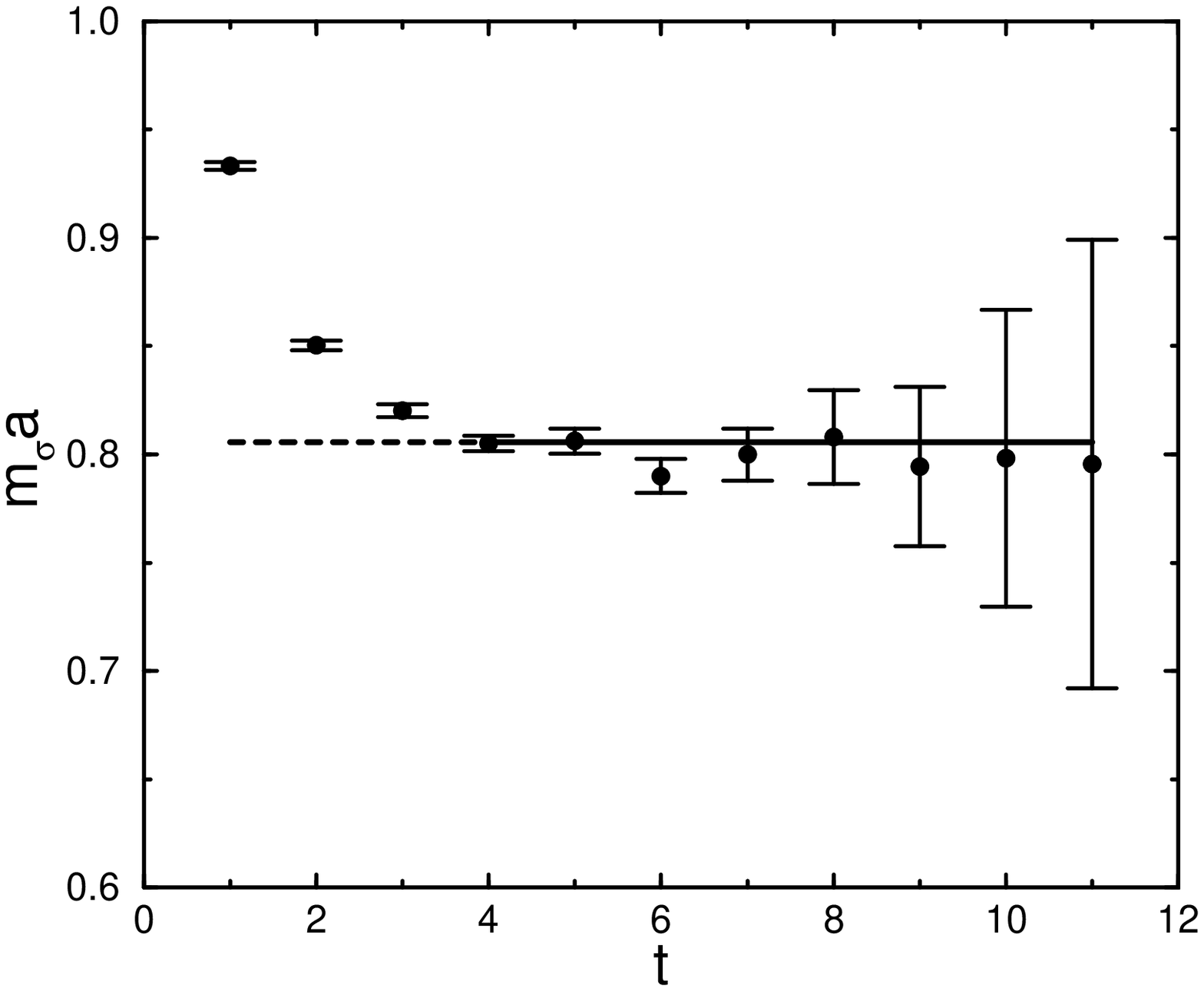}
\caption{Scalar effective masses and fitted mass for $\beta$ of
5.93 and $\kappa$ of 0.1554 on a lattice $24^4$.}
\label{fig:scb593x24k1554}
\end{figure}

\begin{figure}
\epsfxsize=\textwidth
\epsfbox{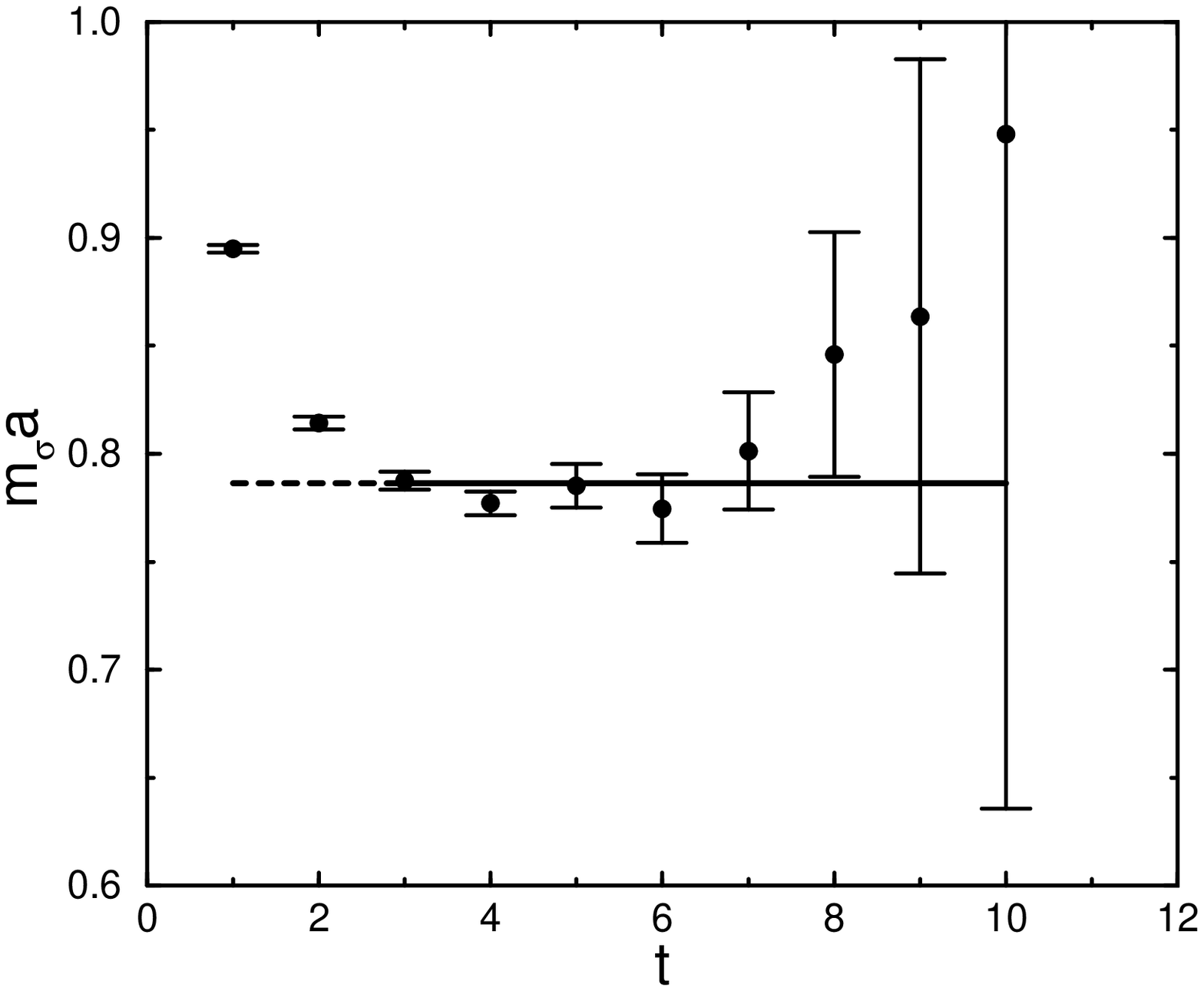}
\caption{Scalar effective masses and fitted mass for $\beta$ of
5.93 and $\kappa$ of 0.1567 on a lattice $24^4$.}
\label{fig:scb593x24k1567}
\end{figure}

\begin{figure}
\epsfxsize=\textwidth
\epsfbox{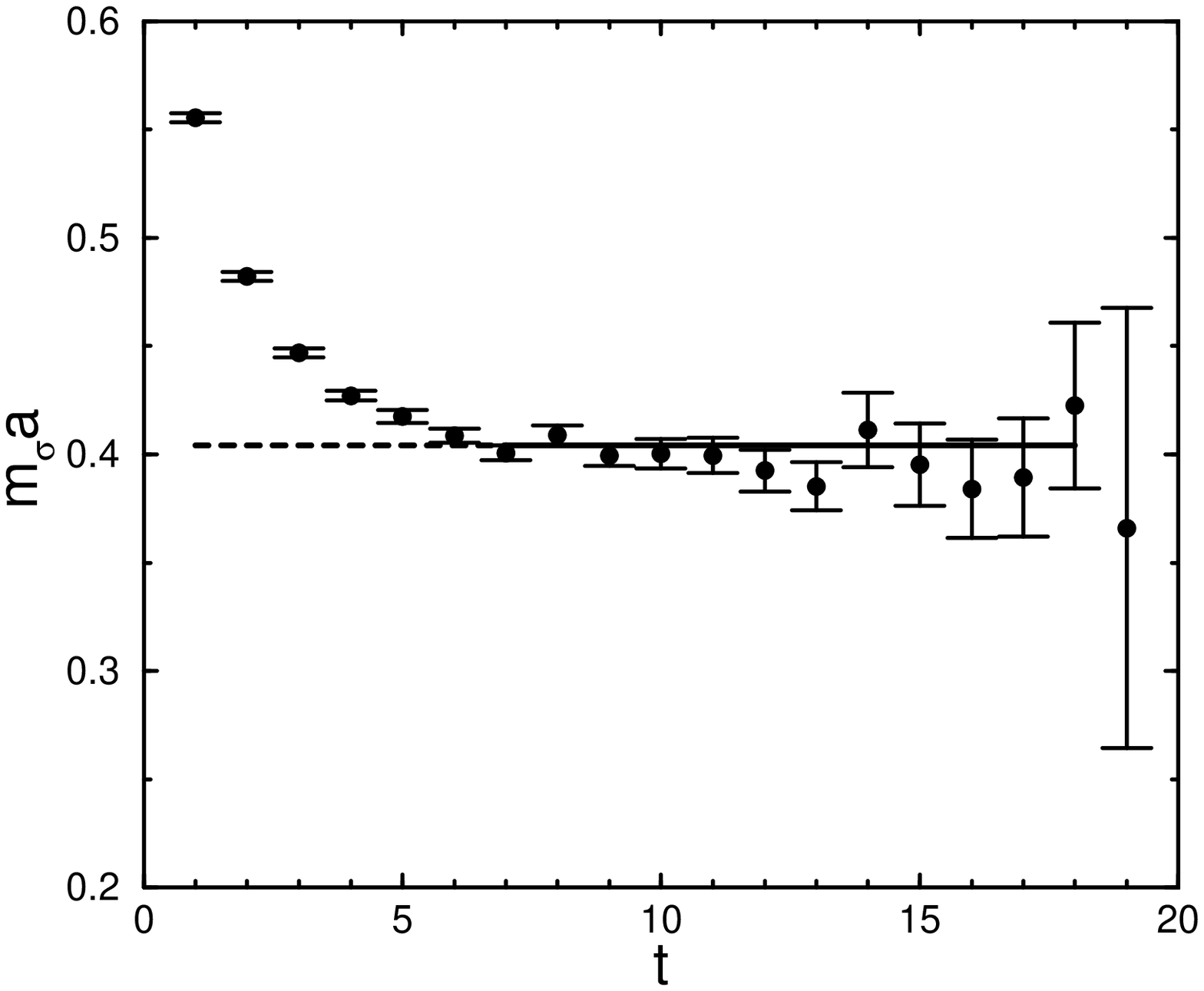}
\caption{Scalar effective masses and fitted mass for $\beta$ of
6.40 and $\kappa$ of 0.1491 on a lattice $32^2 \times 28 \times 40$.}
\label{fig:scb64x32k1491}
\end{figure}

\begin{figure}
\epsfxsize=\textwidth
\epsfbox{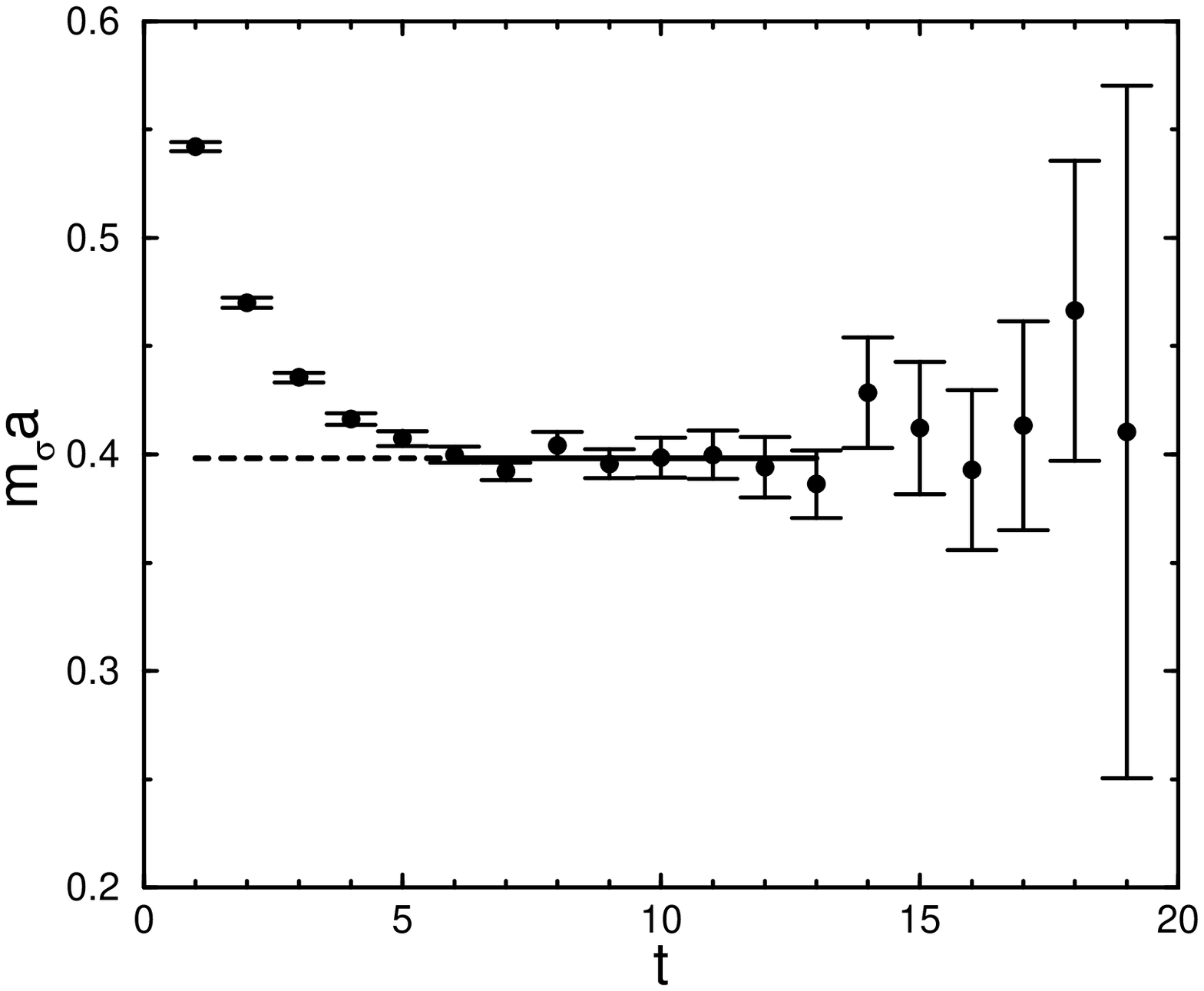}
\caption{Scalar effective masses and fitted mass for $\beta$ of
6.40 and $\kappa$ of 0.1494 on a lattice $32^2 \times 28 \times 40$.}
\label{fig:scb64x32k1494}
\end{figure}

\begin{figure}
\epsfxsize=\textwidth
\epsfbox{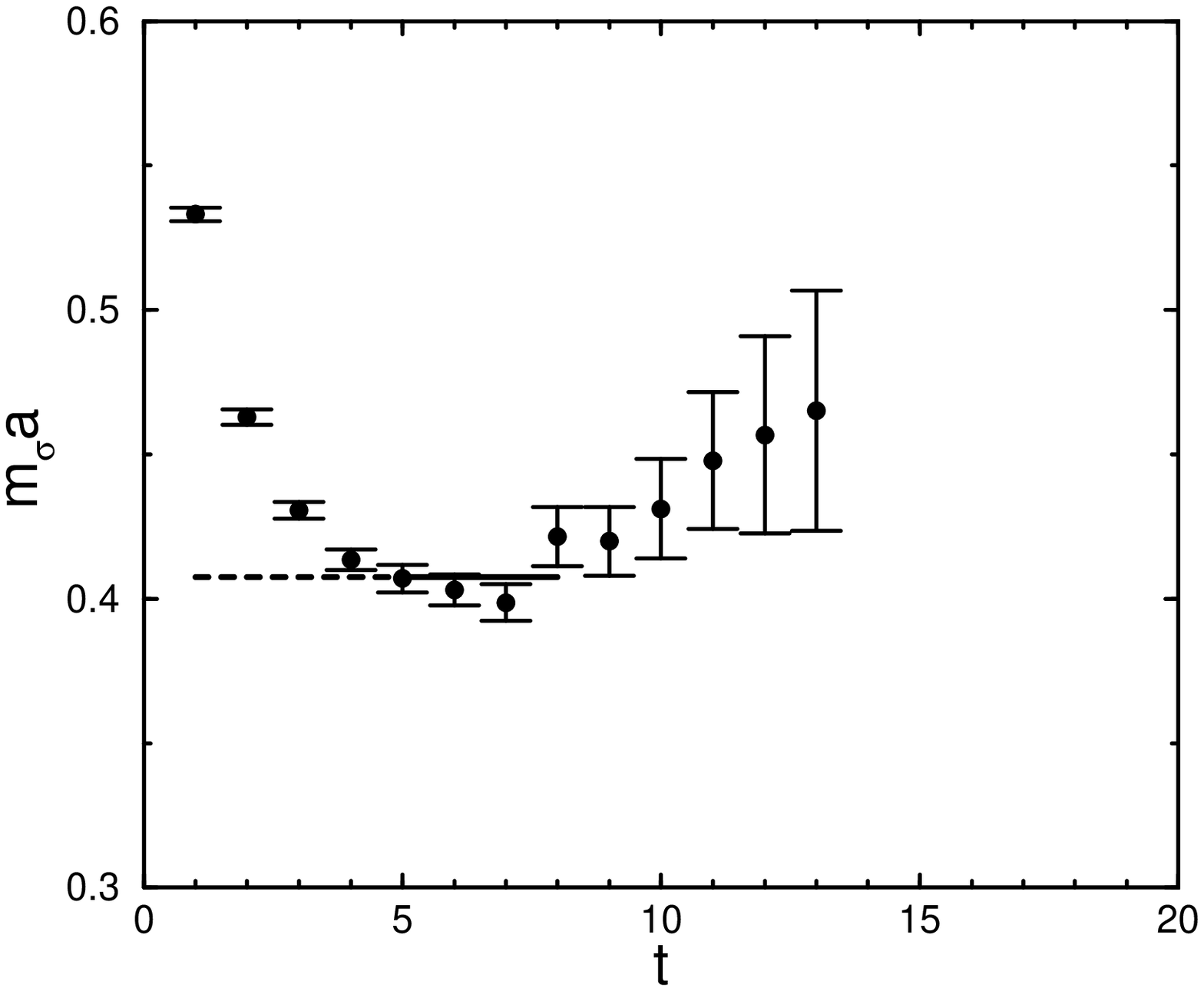}
\caption{Scalar effective masses and fitted mass for $\beta$ of
6.40 and $\kappa$ of 0.1497 on a lattice $32^2 \times 28 \times 40$.}
\label{fig:scb64x32k1497}
\end{figure}

Fitting $C_{\pi \pi}(t)$ and $C_{\sigma \sigma}(t)$ to
Eq.~(\ref{diagZm}) at pairs of neighboring time slices $t$ and $t+1$
gives the effective masses $m_{\pi}(t)$ and $m_{\sigma}(t)$, which at
large $t$ approach $m_{\pi}$ and $m_{\sigma}$, respectively.  To
determine $m_{\pi}$, $Z_{\pi}$, $m_{\sigma}$, and $Z_{\sigma}$, we began
by examining effective mass graphs for a range of $q$, in
Eq.~(\ref{defC}), to find smearing sizes for which $m_{\pi}(t)$ and
$m_{\sigma}(t)$ show clear evidence of approaching constants at large
$t$.  In all but one case we found satisfactory effective mass plateaus
with $q$ the same as $r$ of Table~\ref{tab:smearing}.  For $C_{\pi \pi}$
at $\beta$ of 6.17 a smearing size of 9.0 was used for $q$.  Typical
effective mass graphs are shown in Figures~\ref{fig:psb593x16k1554} -
\ref{fig:scb64x32k1497}.

Trial time intervals on which to fit $C_{\pi \pi}(t)$ and $C_{\sigma
\sigma}(t)$ to Eq.~(\ref{diagZm}) were chosen from effective mass graphs
by eliminating large values of $t$ with large statistical uncertainties
in effective masses and eliminating small $t$ at which effective masses
have clearly not yet reached the large $t$ plateau.  Fits were then made
to Eq.~(\ref{diagZm}) on all subintervals of 3 or more consecutive $t$
within the trial range. The fit for each interval was chosen to minimize
the $\chi^2$ taking into account all correlations among the fitted
data. Correlations were determined by the bootstrap method. The final
fitting interval for each propagator was chosen to be the interval with
the smallest $\chi^2$ per degree of freedom.

Final fitting intervals and fitted masses are shown by solid lines in
Figures~\ref{fig:psb593x16k1554} -
\ref{fig:scb64x32k1497}. Dashed lines extend the solid lines toward
smaller times to display the approach of effective masses to the final
fitted masses.

Tables~\ref{tab:psb57x12} - \ref{tab:scb64x32} list the final
pseudoscalar and scalar masses obtained. The statistical uncertainties
for the masses in these tables, and in all other Monte Carlo results in
this article, are determined by the bootstrap method.

\begin{figure}
\epsfxsize=\textwidth
\epsfbox{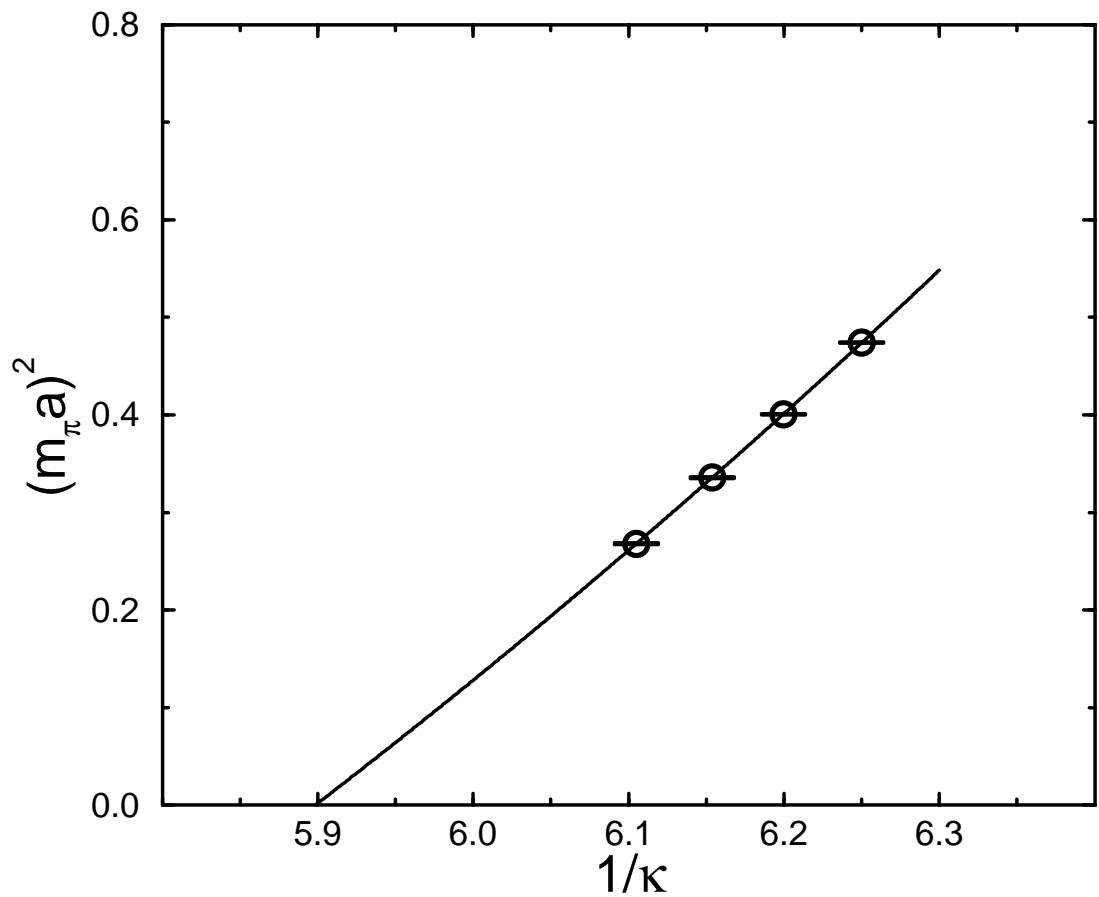}
\caption{
Pseudoscalar quarkonium mass squared as a function of $1/\kappa$ for $\beta$ of
5.70 on a lattice $12^2 \times 10 \times 24$. Results for $16^3 \times
24$ are nearly identical.}
\label{fig:ps57}
\end{figure}

\begin{figure}
\epsfxsize=\textwidth
\epsfbox{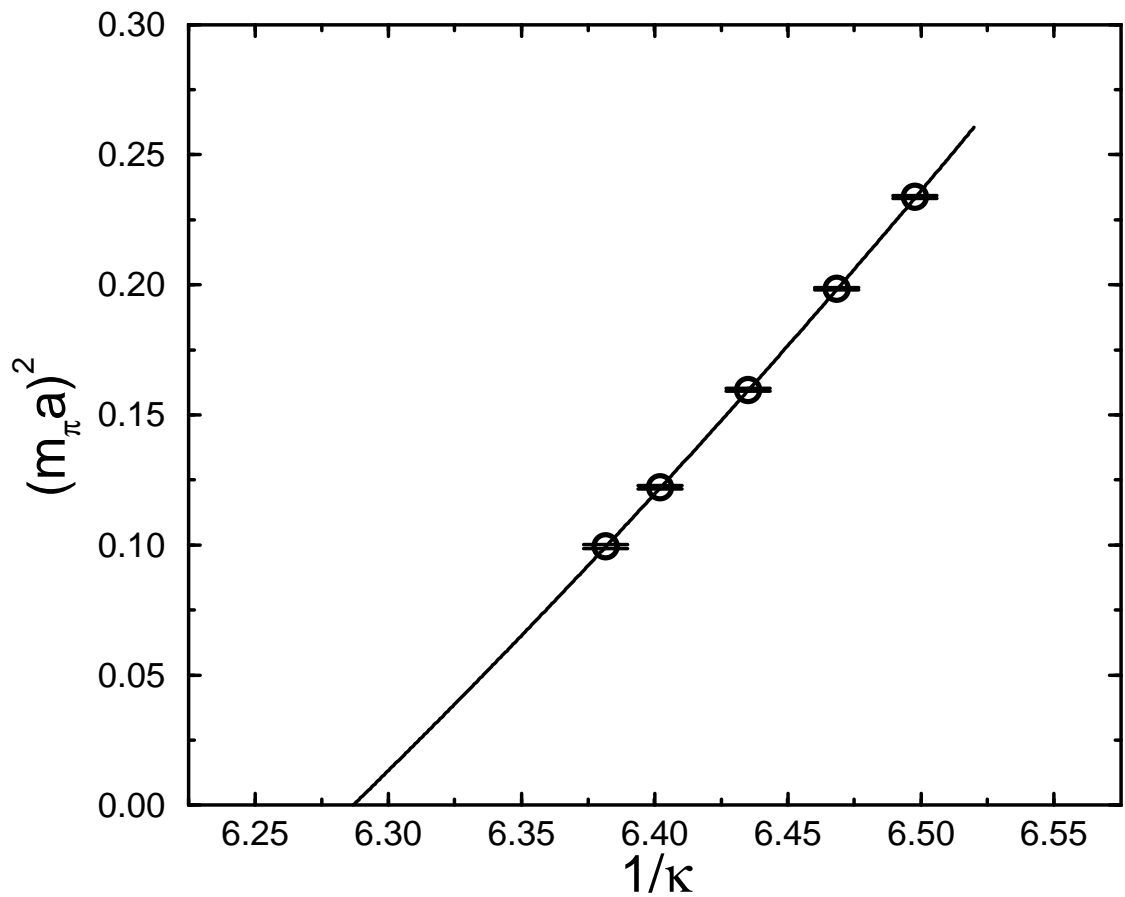}
\caption{
Pseudoscalar quarkonium mass squared as a function of $1/\kappa$ for $\beta$ of
5.93 on a lattice $16^2 \times 14 \times 20$. Results for $24^4$ are
nearly identical.}
\label{fig:ps593}
\end{figure}

\begin{figure}
\epsfxsize=\textwidth
\epsfbox{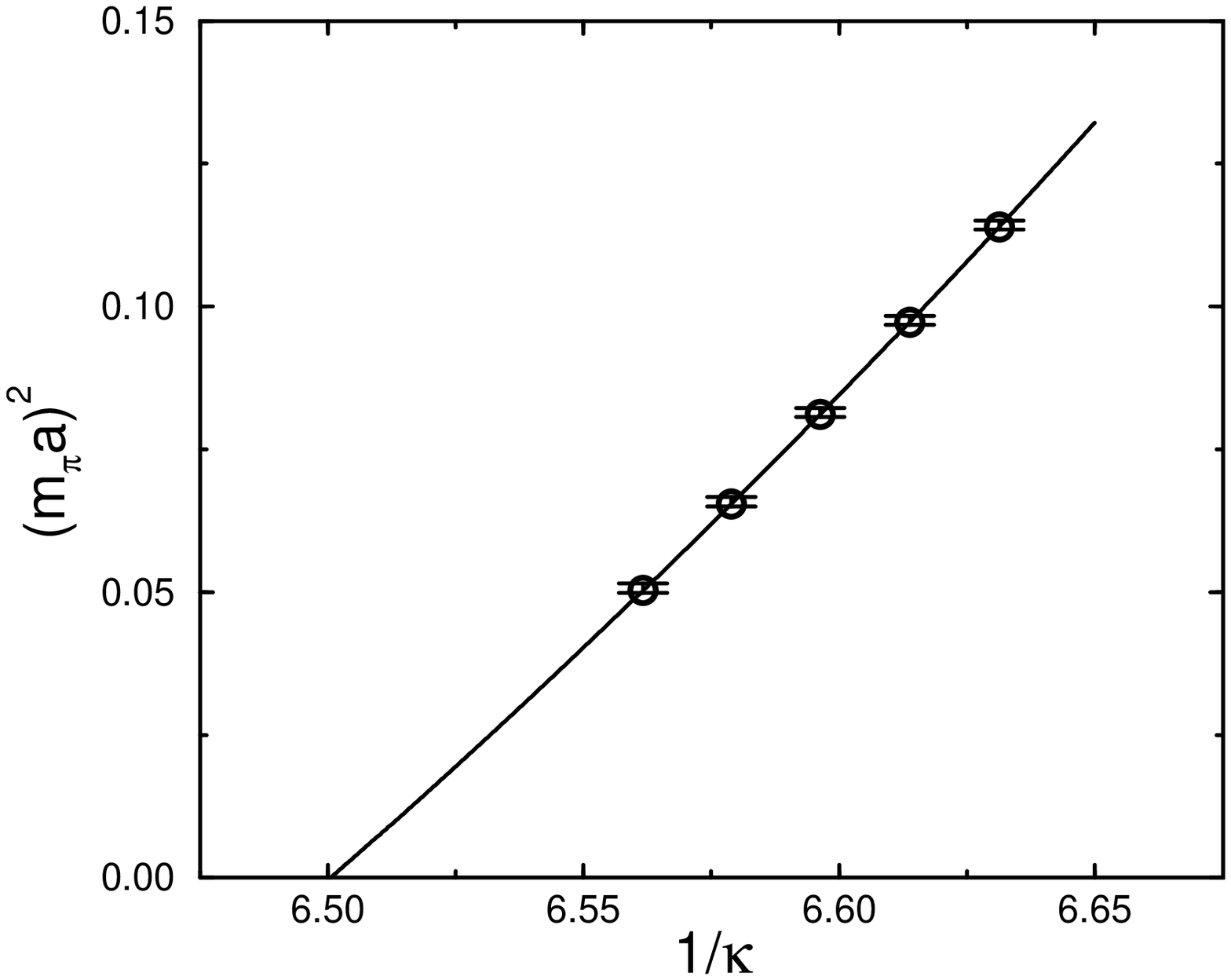}
\caption{Pseudoscalar quarkonium mass squared as a function of $1/\kappa$ for
$\beta$ of 6.17 on the lattice $24^2 \times 20 \times 32$.}
\label{fig:ps617}
\end{figure}

\begin{figure}
\epsfxsize=\textwidth
\epsfbox{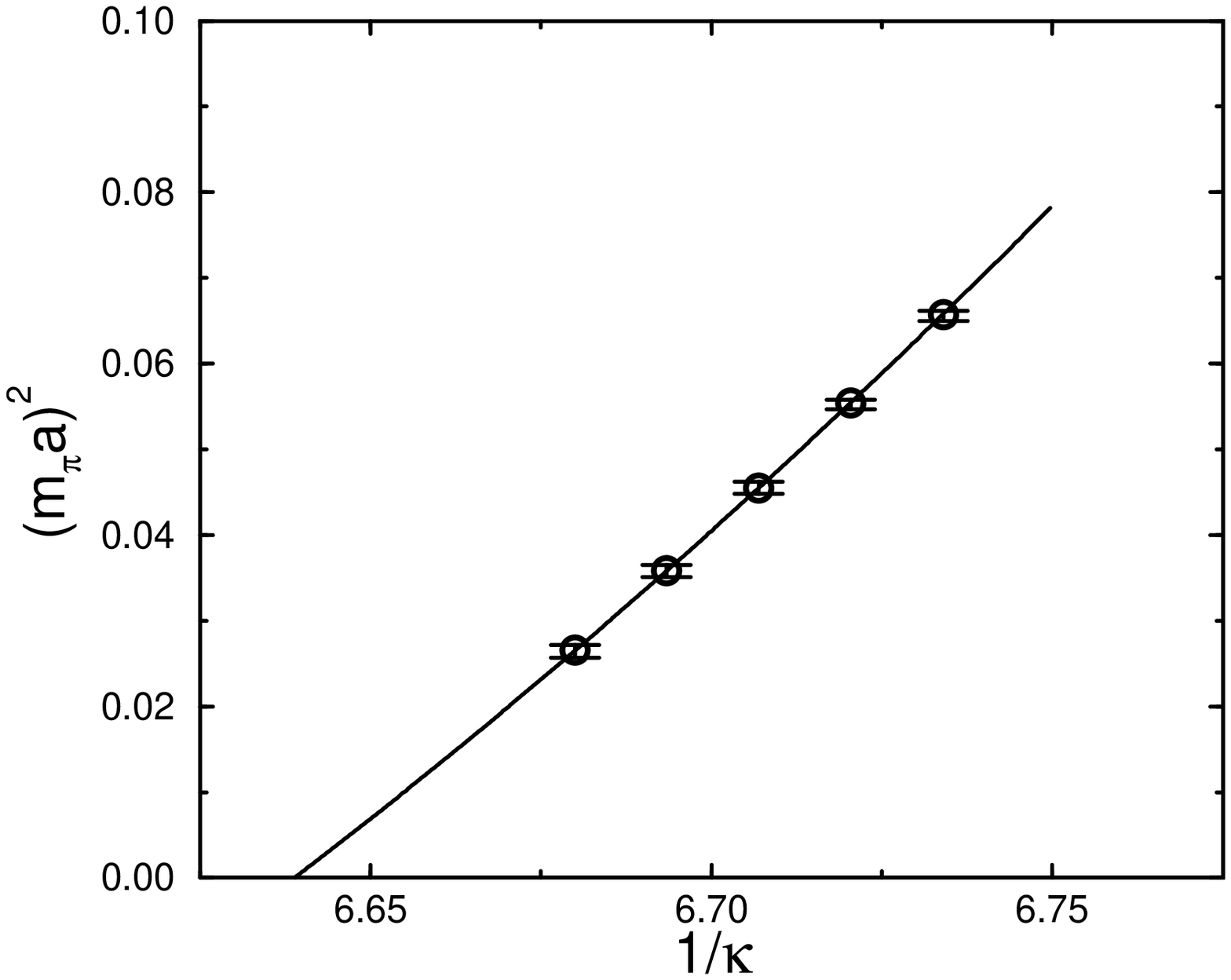}
\caption{Pseudoscalar quarkonium mass squared as a function of $1/\kappa$ for
$\beta$ of 6.4 on the lattice $32^2 \times 28 \times 40$.}
\label{fig:ps64}
\end{figure}

For $\beta$ of 5.70, 5.93, 6.17 and 6.4, Figures~\ref{fig:ps57},
\ref{fig:ps593}, \ref{fig:ps617}, and \ref{fig:ps64}, respectively,
show the pseudoscalar mass squared $m_{\pi}^2$ as a function of $1/\kappa$.
The solid line in each figure shows a fit of $m_{\pi}^2$ to a quadratic 
function of $1/\kappa$ used to determine the strange quark hopping
constant $\kappa_s$ at which
\begin{eqnarray}
\label{defms}
m_{\pi}^2 = 2 M_K^2 - M_{\pi}^2,
\end{eqnarray}
where $M_K$ and $M_{\pi}$ are the observed neutral kaon and pion masses,
respectively. The quadratic fits in $1/\kappa$ were used also to
determine the critical hopping constant $\kappa_{crit}$ at which $m_{\pi}$
is zero. Although the determination of $\kappa_{crit}$ depends on
extrapolation of each fit beyond the $kappa$ interval in which we have
data, the determination of $\kappa_s$ does not and uses the fits only to
interpolate between measurements.  From $\kappa_{crit}$ we define the quark
mass for each $\kappa$ to be
\begin{eqnarray}
\label{defmq}
\mu a = \frac{1}{2 \kappa} - \frac{1}{2 \kappa_{crit}}.
\end{eqnarray}
Values of $\kappa_s$ and $\kappa_{crit}$ are given in
Table~\ref{tab:kskc}.

\begin{figure}
\epsfxsize=\textwidth
\epsfbox{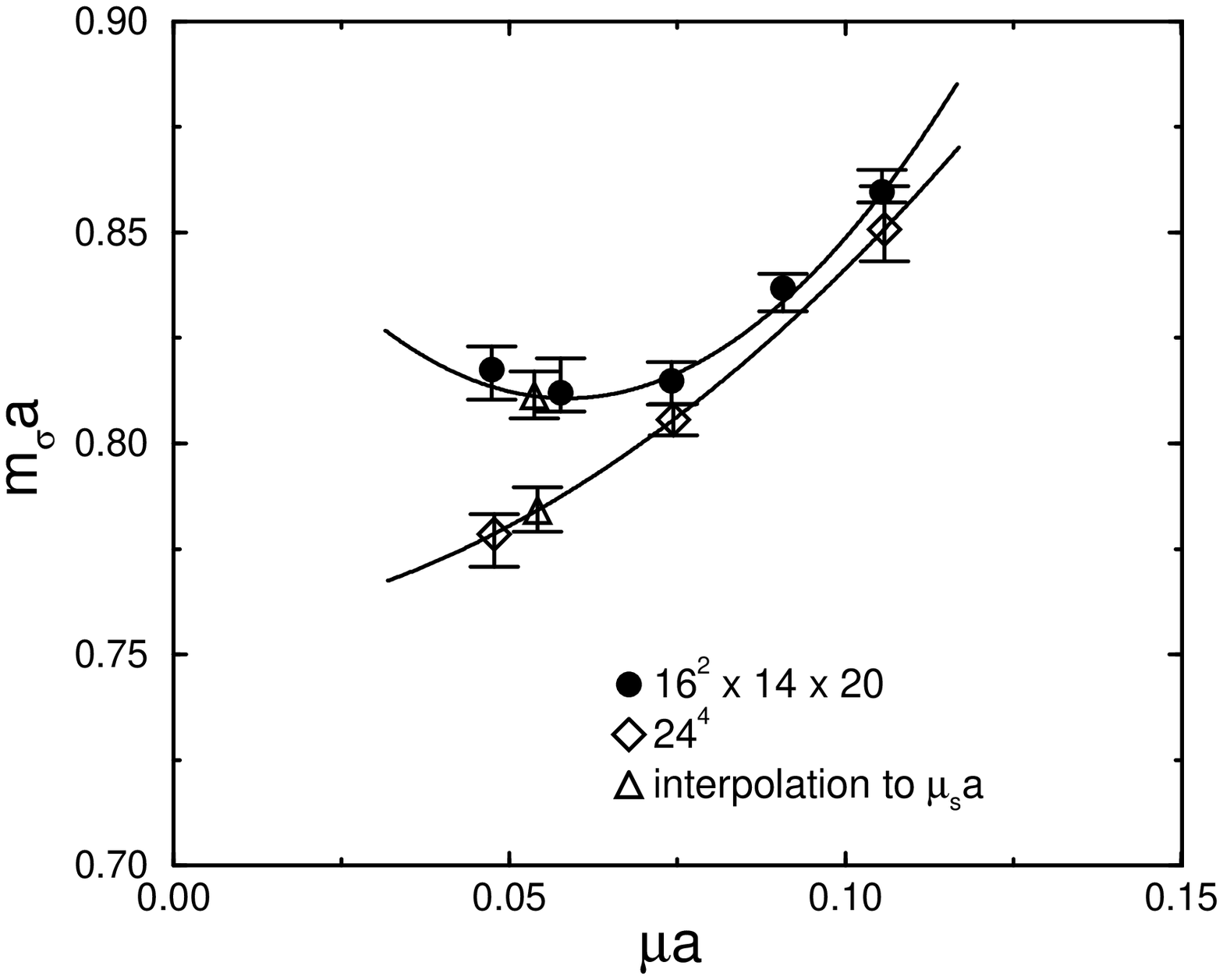}
\caption{Scalar quarkonium mass as a function of quark mass for
$\beta$ of 5.93.}
\label{fig:sc593}
\end{figure}

\begin{figure}
\epsfxsize=\textwidth
\epsfbox{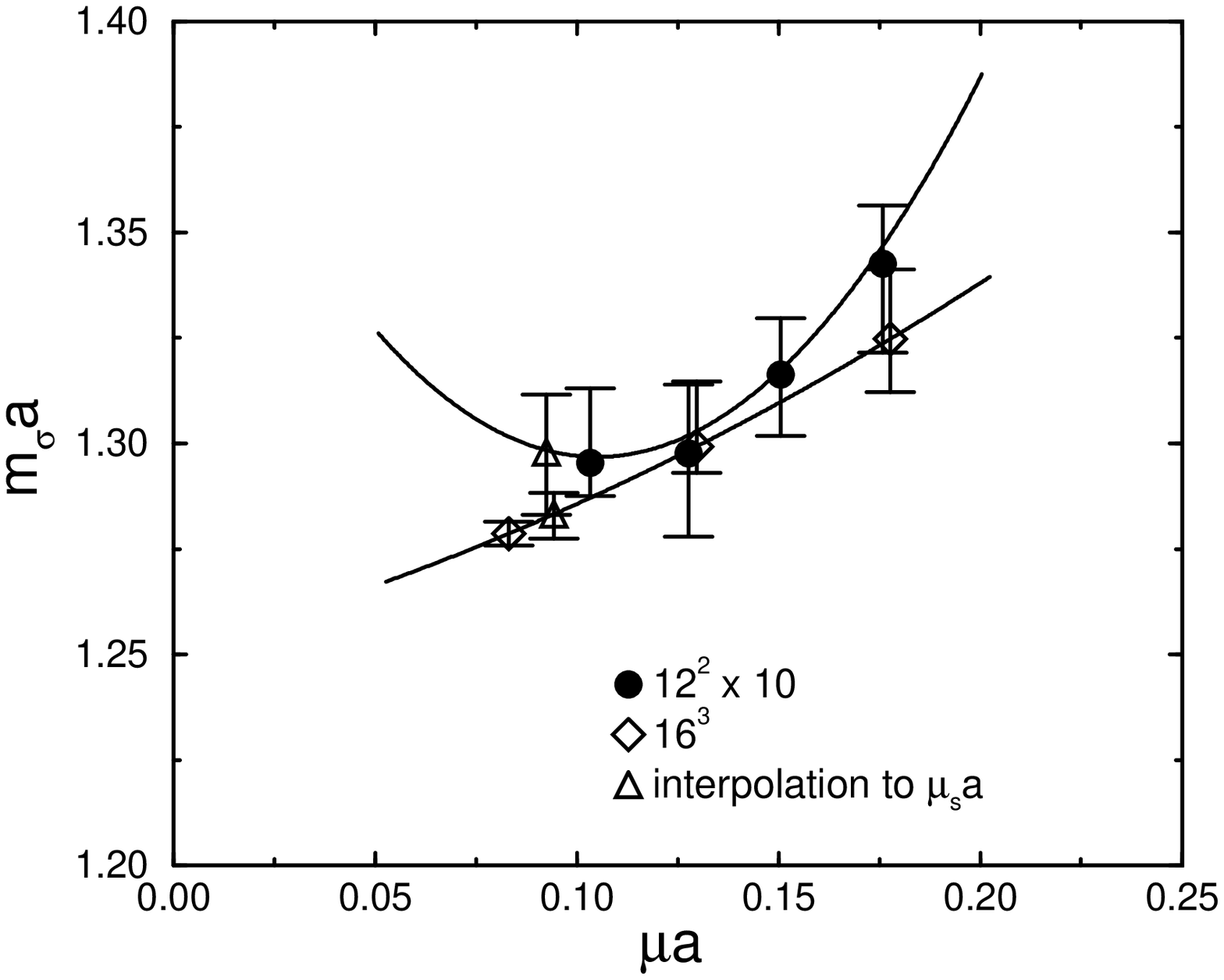}
\caption{Scalar quarkonium mass as a function of quark mass for
$\beta$ of 5.7.}
\label{fig:sc57}
\end{figure}

\begin{figure}
\epsfxsize=\textwidth
\epsfbox{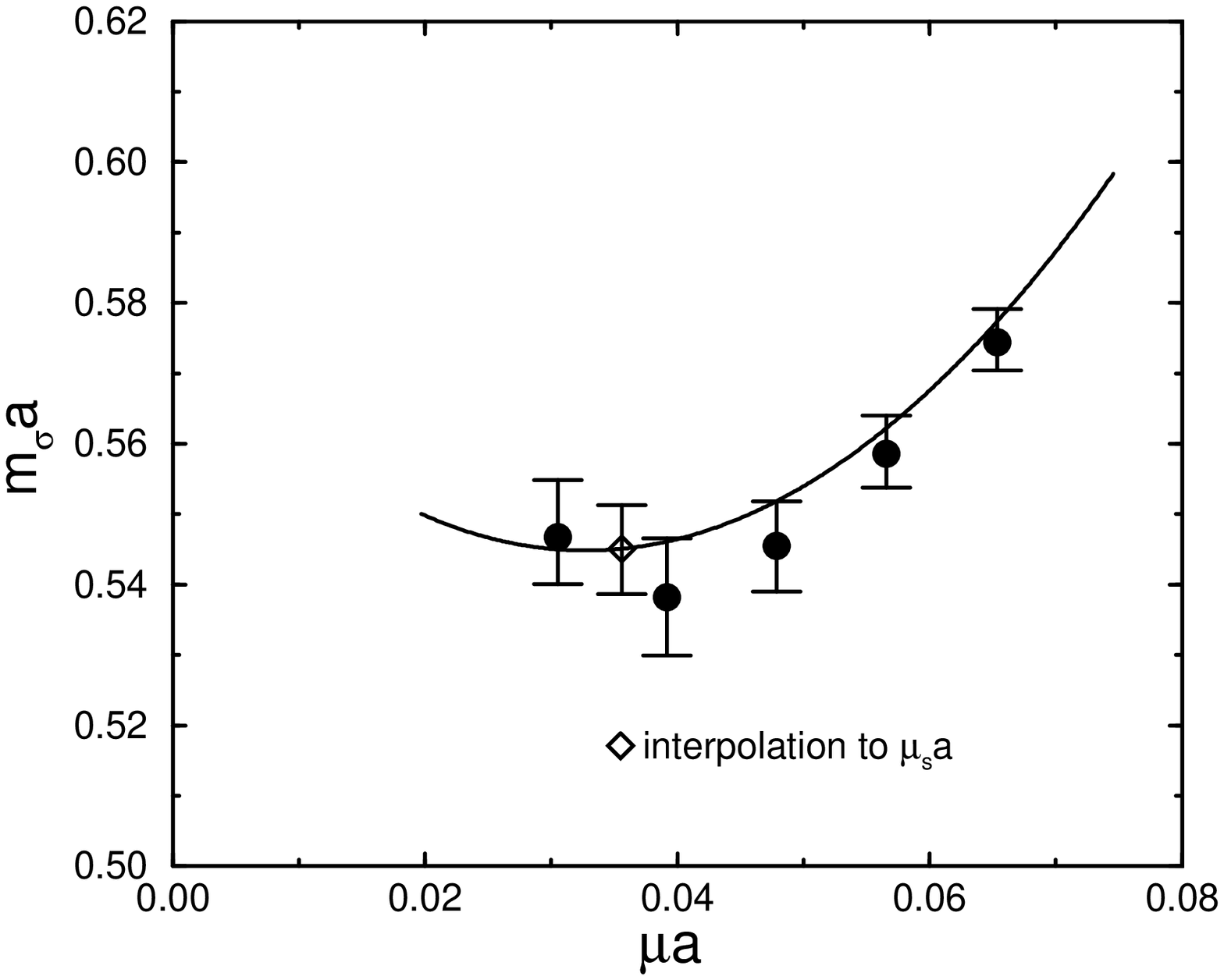}
\caption{Scalar quarkonium mass as a function of quark mass for
$\beta$ of 6.17.}
\label{fig:sc617}
\end{figure}

\begin{figure}
\epsfxsize=\textwidth
\epsfbox{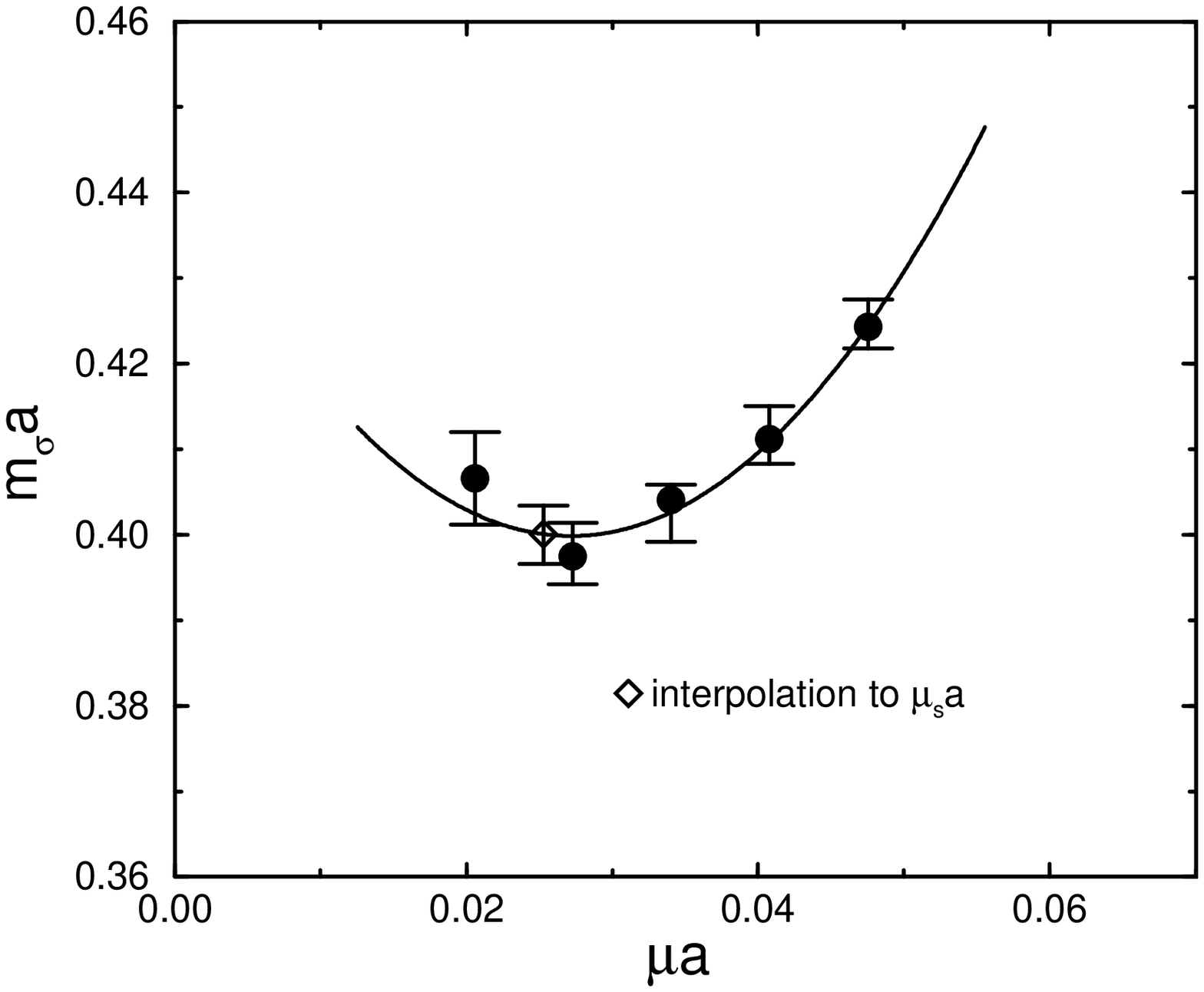}
\caption{Scalar quarkonium mass as a function of quark mass for
$\beta$ of 6.4.}
\label{fig:sc64}
\end{figure}

For the two lattices with $\beta$ of 5.93, Figure~\ref{fig:sc593} shows
the scalar quarkonium mass as a function of quark mass $\mu a$. The
solid lines in Figure~\ref{fig:sc593} are fits of the scalar
mass to quadratic functions of quark mass. The scalar masses found by
interpolation to the strange quark mass are also indicated.  As shown by
the figure, for the lattice $16^2 \times 14 \times 20$ with $L$ of
1.54(4) fm the scalar mass as a function of quark mass flattens out as
quark mass is lowered toward the strange quark mass and then appears to
begin to rise as the quark mass is decreased still further.  This
feature is absent from the data at $\beta$ of 5.93 for the lattice
$24^4$ with $L$ of 2.31(6) fm and is thus a finite-volume artifact.  It
is present in the data at $\beta$ of 5.70 with $L$ of 1.68(5) fm, at
$\beta$ of 6.17 with $L$ of 1.74(5) fm, and at $\beta$ of 6.40 with $L$
of 1.66(5) fm shown in Figures~\ref{fig:sc57}, \ref{fig:sc617} and
\ref{fig:sc64}, respectively.  It is absent, however, in the data at
$\beta$ of 5.70 with $L$ of 2.24(7) fm shown in Figure~\ref{fig:sc57}.
Values of the scalar quarkonium mass interpolated to the strange
quark mass are given in Table~\ref{tab:scs}

The pseudoscalar mass squared $m_{\pi}^2$ shown in
Figures~\ref{fig:ps57} -
\ref{fig:ps64} is nearly a linear function of $1/\kappa$ for all $\beta$
and lattice periods. The difference in $m_{\pi} a$ between the two
lattice at $\beta$ of 5.70 and between the two lattice at $\beta$ of
5.93 is in all cases less than 0.5\%. The anomaly in quark mass dependence
of the scalar mass for $L$ of 1.6 fm, shown in Figure~\ref{fig:sc593},
is absent from the quark mass dependence of the pseudoscalar
mass for this value of $L$. 

\begin{figure}
\epsfxsize=\textwidth
\epsfbox{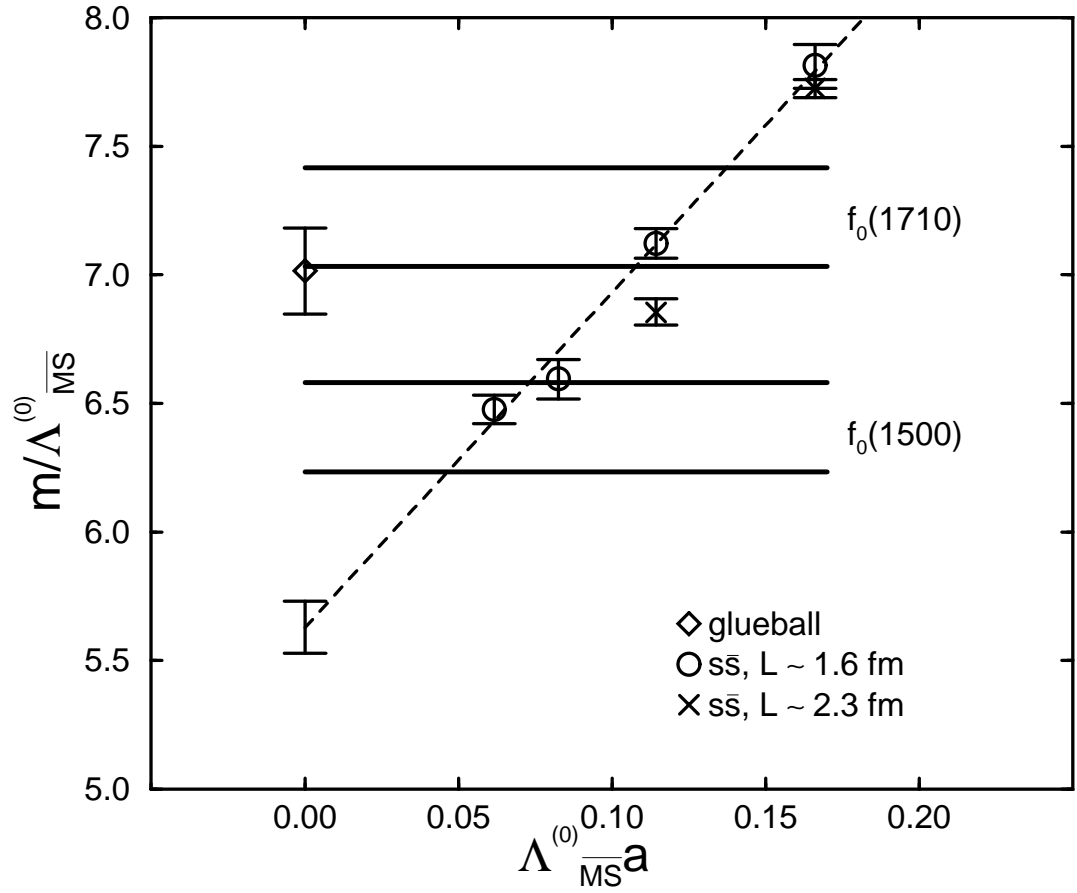}
\caption{
Lattice spacing dependence and continuum limit of the scalar
$s\overline{s}$ mass, continuum limit of the scalar glueball masses, and one
sigma upper and lower bounds on observed masses.}
\label{fig:masscont}
\end{figure}

For $L$ near 1.6 fm, Figure~\ref{fig:masscont} shows the $s\overline{s}$
scalar mass in units of $\Lambda^{(0)}_{\overline{MS}}$ as a function of
lattice spacing in units of $1/\Lambda^{(0)}_{\overline{MS}}$.	A linear
extrapolation of the mass to zero lattice spacing gives 1322(42) MeV,
far below our valence approximation infinite volume continuum glueball
mass of 1648(58) MeV.  For the ratio of the $s\overline{s}$ mass to the
infinite volume continuum limit of the scalar glueball mass we obtain
0.802(24).  Figure~\ref{fig:masscont} shows also values of the
$s\overline{s}$ scalar mass at $\beta$ of 5.70 and 5.93 with $L$ of
2.24(7) and 2.31(6) fm, respectively.  The $s\overline{s}$ mass with $L$
near 2.3 fm lies below the 1.6 fm result for both values of lattice
spacing. Thus the infinite volume continuum $s\overline{s}$ mass should
lie below 1322(42) MeV.	 We believe our data make improbable the
interpretation of $f_0(1500)$ as mainly composed of the lightest scalar
glueball with $f_0(1710)$ consisting mainly of $s\overline{s}$ scalar
quarkonium. For comparison with our data, Figure~\ref{fig:masscont} shows
the valence approximation value for the infinite volume continuum limit
of the scalar glueball mass and the observed value of the mass of
$f_0(1500)$ and of the mass of $f_0(1710)$ The uncertainties shown in
the observed masses in units of $\Lambda^{(0)}_{\overline{MS}}$ arise
mainly from the uncertainty in $\Lambda^{(0)}_{\overline{MS}}$.

\section{Glueball Mass}\label{sect:glb} 

In preparation for a calculation of quarkonium-glueball mixing energy,
from each gauge ensemble we also constructed scalar glueball operators.
On the gauge ensembles at $\beta$ of 5.70, we evaluated smeared Coulomb
gauge scalar glueball operators and at all larger $\beta$ smeared gauge
invariant scalar glueball operators. The operators we used are discussed
in Ref.~\cite{Vaccarino}.  The correlation
function constructed from these is
\begin{eqnarray}
\label{defCgg}
C_{gg}(t) & = & 
\frac{1}{V}\sum_{\vec{x}\vec{y}} [< g(\vec{x},t) g(\vec{y},0)> - 
< g(\vec{x},t)>< g(\vec{y},0)>], \nonumber \\
\end{eqnarray}
where $g(\vec{x},t)$ is the smeared scalar glueball operator and $V$ is
the space direction lattice volume.

\begin{figure}
\epsfxsize=\textwidth
\epsfbox{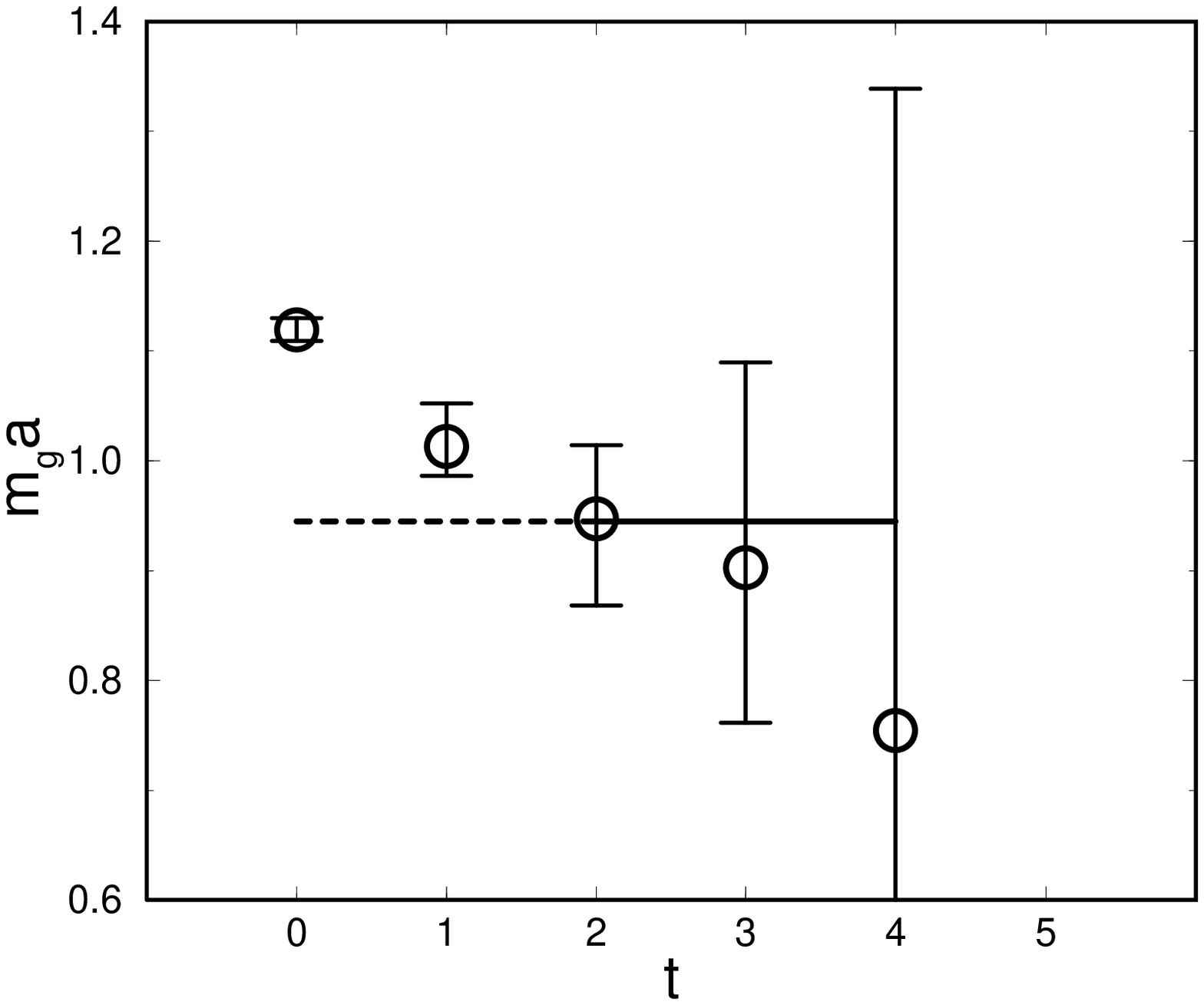}
\caption{Scalar glueball effective masses and fitted mass for $\beta$ of
5.70 on a lattice $12^2 \times 10 \times 24$.}
\label{fig:glb57x12}
\end{figure}

\begin{figure}
\epsfxsize=\textwidth
\epsfbox{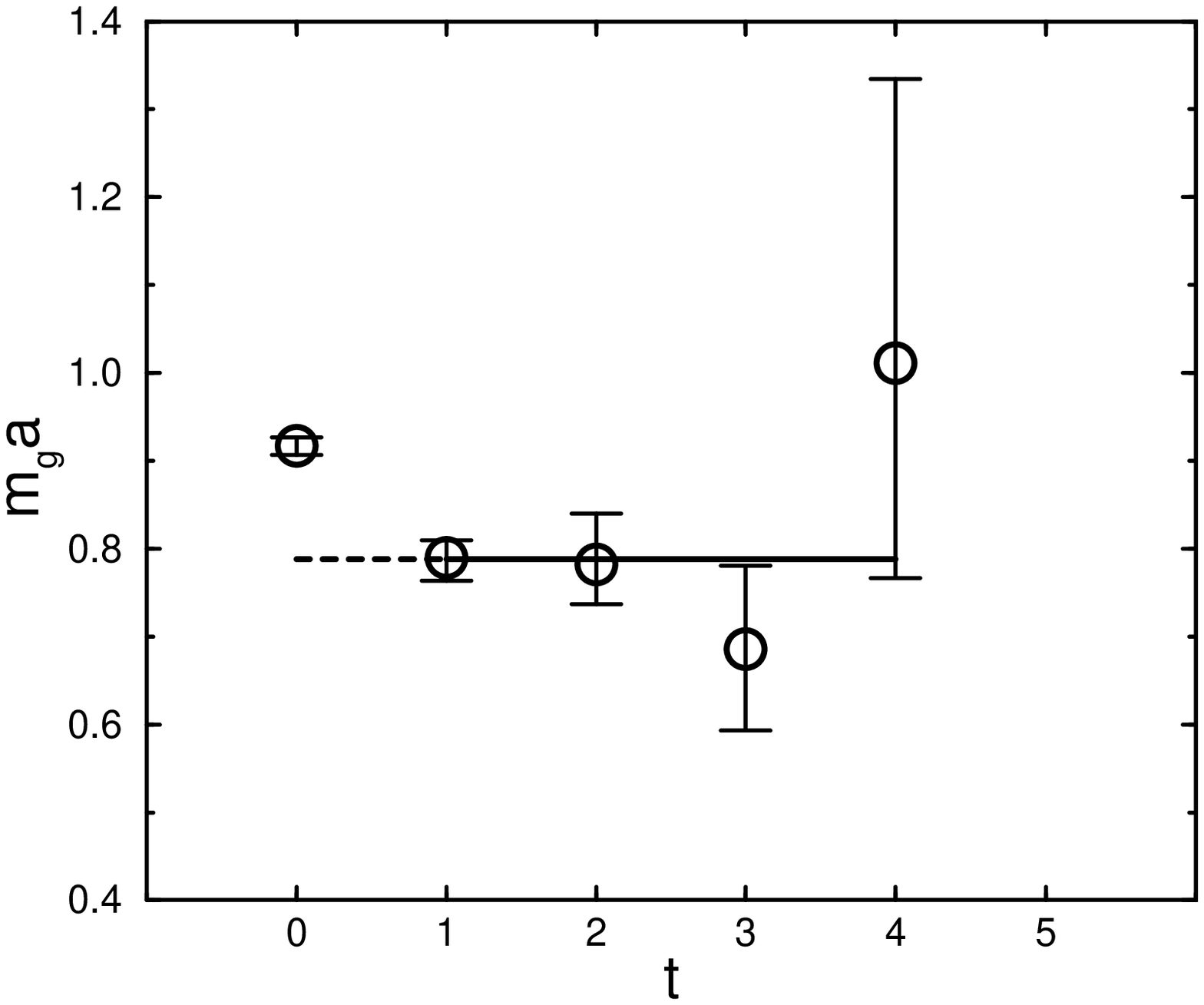}
\caption{Scalar glueball effective masses and fitted mass for $\beta$ of
5.93 on a lattice $16^2 \times 14 \times 20$.}
\label{fig:glb593x16}
\end{figure}

\begin{figure}
\epsfxsize=\textwidth
\epsfbox{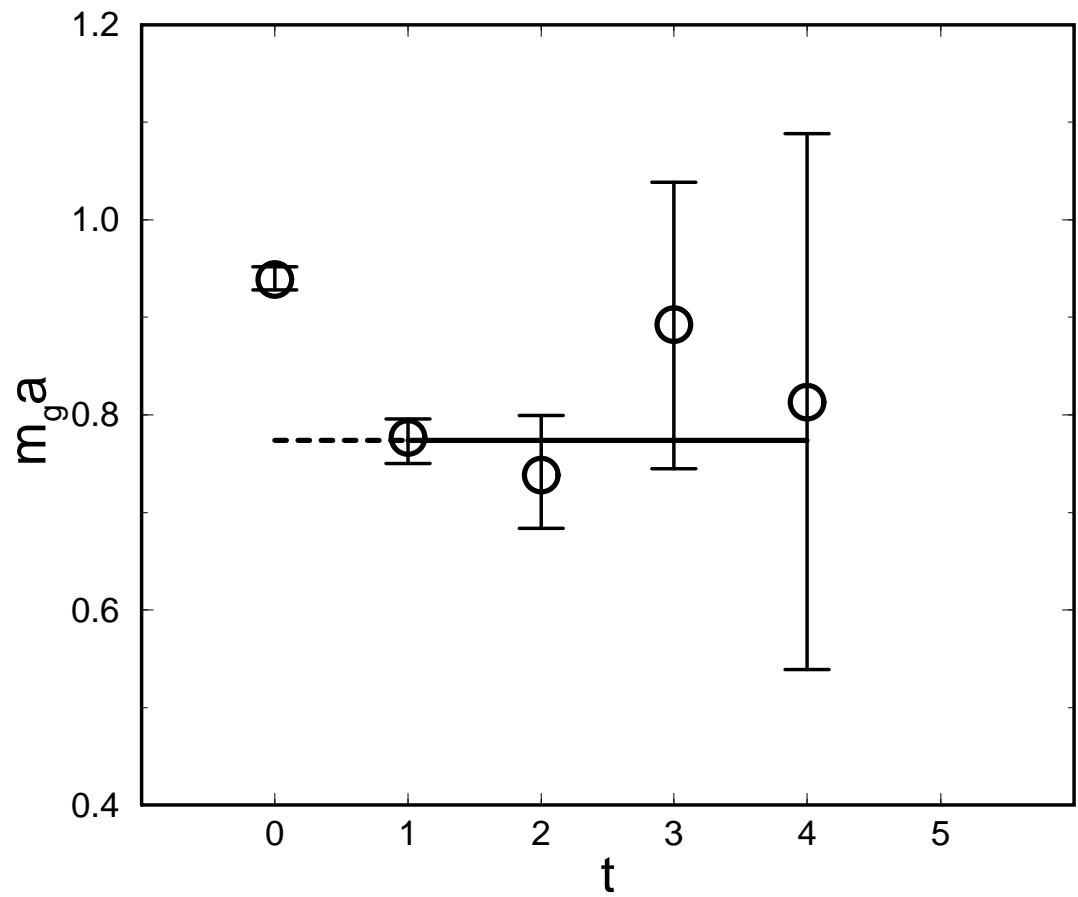}
\caption{Scalar glueball effective masses and fitted mass for $\beta$ of
5.93 on a lattice $24^4$.}
\label{fig:glb593x24}
\end{figure}

\begin{figure}
\epsfxsize=\textwidth
\epsfbox{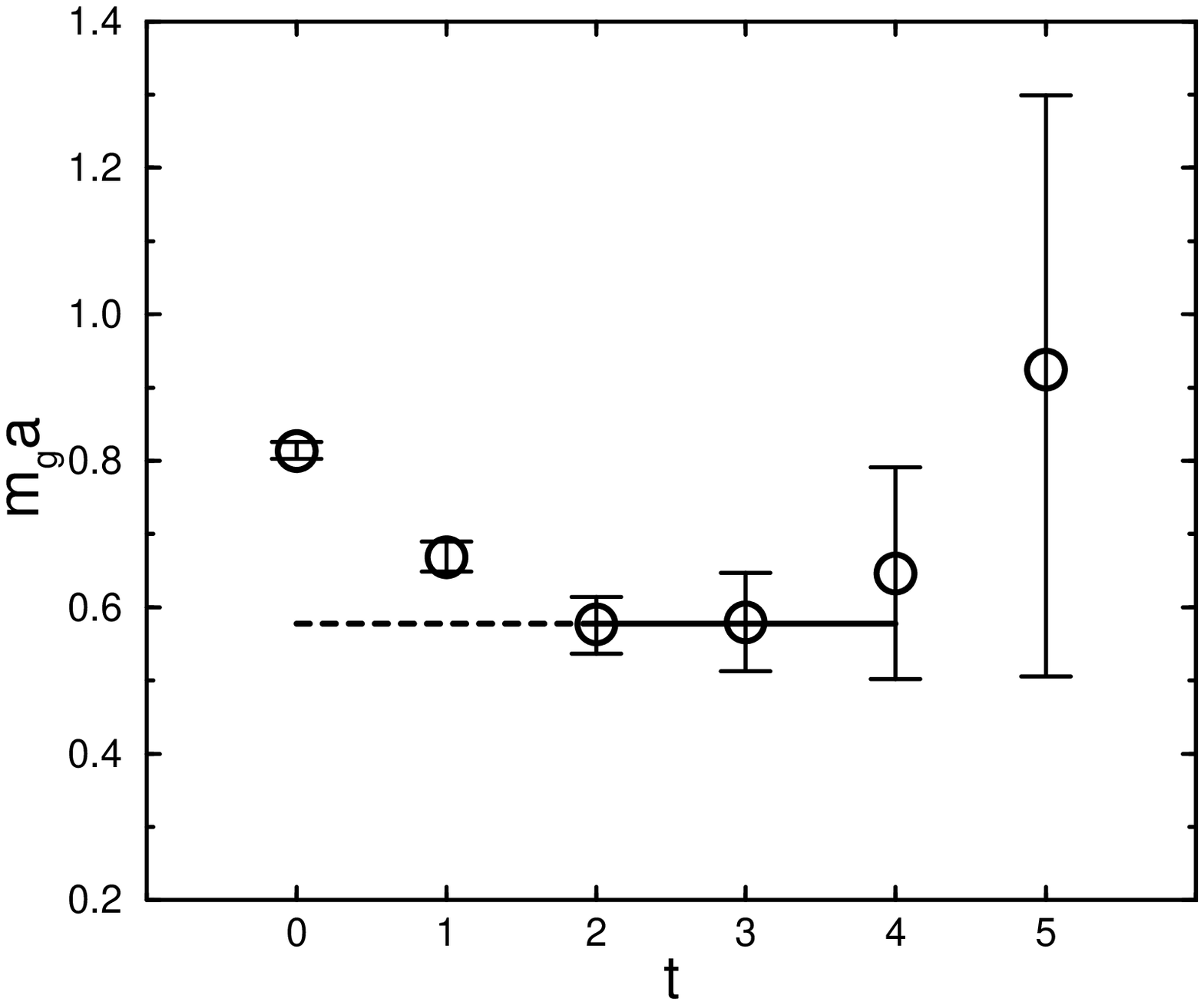}
\caption{Scalar glueball effective masses and fitted mass for $\beta$ of
6.17 on a lattice $24^2 \times 20 \times 32$.}
\label{fig:glb617x24}
\end{figure}

\begin{figure}
\epsfxsize=\textwidth
\epsfbox{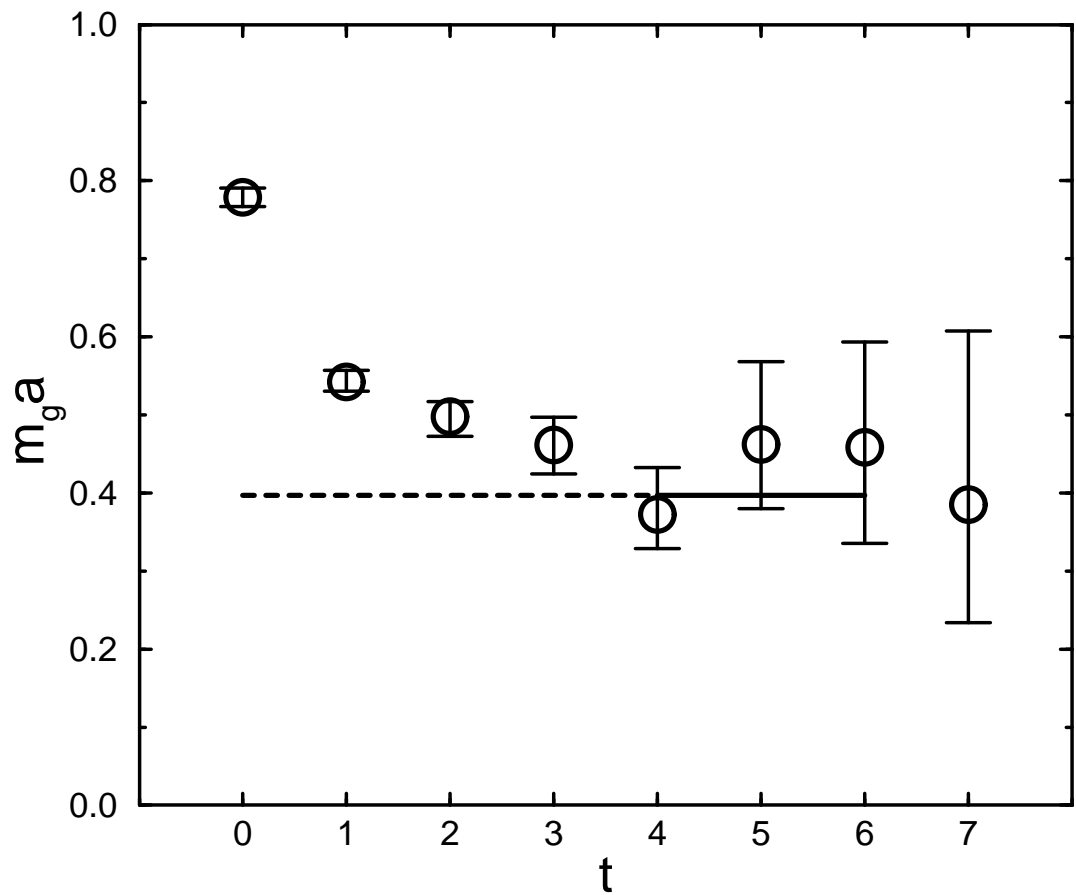}
\caption{Scalar glueball effective masses and fitted mass for $\beta$ of
6.40 on a lattice $32^2 \times 28 \times 40$.}
\label{fig:glb64x32}
\end{figure}

Fitting the the large $t$ behavior of $C_{gg}(t)$ to Eq.~(\ref{diagZm})
for $f$ chosen to be $g$, we obtain the glueball mass $m_g a$ and field
strength renormalization constant $Z_g$. A detailed discussion of
calculations of $m_g a$ and $Z_g$ for the same $\beta$ and nearly the
same lattice sizes considered here, but with much larger Monte Carlo
ensemble sizes, is presented in Ref.~\cite{latestglue}. Using the
calculation of Ref.~\cite{latestglue} to guide the choice of smearing
parameters and time intervals to be fit, we applied the fitting
procedure of Section~\ref{sect:mass}. Smearing parameters we found to
be satisfactory are given in Table~\ref{tab:smearing}. Effective mass
graphs are shown in Figures~\ref{fig:glb57x12} - \ref{fig:glb64x32}.
Fitted masses, fitted time intervals, $\chi^2$ per degree of freedom of
each fit, and corresponding lattice sizes and fitted masses from
Ref.~\cite{latestglue} are given in Table~\ref{tab:glb}.  For the Monte
Carlo ensemble with $\beta$ of 5.70 on a lattice $16^3
\times 24$ the fits of $C_{gg}(t)$ to Eq.~(\ref{diagZm}) yielded either
large $\chi^2$ or large statistical errors, therefore 
no results are given in Table~\ref{tab:glb} for this case.

\section{Mixing Energy}\label{sect:mix}

To determine scalar quarkonium-glueball mixing energies, we evaluated
the correlation between the scalar quarkonium operators of
Section~\ref{sect:mass} and the glueball operators of Section~\ref{sect:glb}. 
For scalar quarkonium operators not containing random variables and for
the random operators when averaged over random variables,
the correlation function we calculated becomes
\begin{eqnarray}
\label{defCgs}
C_{g \sigma}(t) & = & \sum_{\vec{x}} [< g(\vec{x},t) \sigma(0,0)> -
< g(\vec{x},t)>< \sigma(0,0)>.
\end{eqnarray}
The smearing parameters for quark and glueball fields, as before,
are listed in Table~\ref{tab:smearing}.
For large $t$ and time period $T$, the asymptotic behavior of
$C_{g \sigma}(t)$ for $m_{\sigma}$ close to $m_g$ is
\begin{eqnarray}
\label{offdiag}
C_{g \sigma}(t) & \rightarrow &
\sqrt{Z_g Z_{s}} E a \sum_{t'} 
[exp( -m_g a |t - t'|) + exp( m_g a |t - t'| - m_g a T)] \times \nonumber \\
& & [exp( -m_{\sigma} a |t'|) + exp( m_{\sigma} a |t'| - m_{\sigma} a
T)].
\end{eqnarray}

\begin{figure}
\epsfxsize=\textwidth
\epsfbox{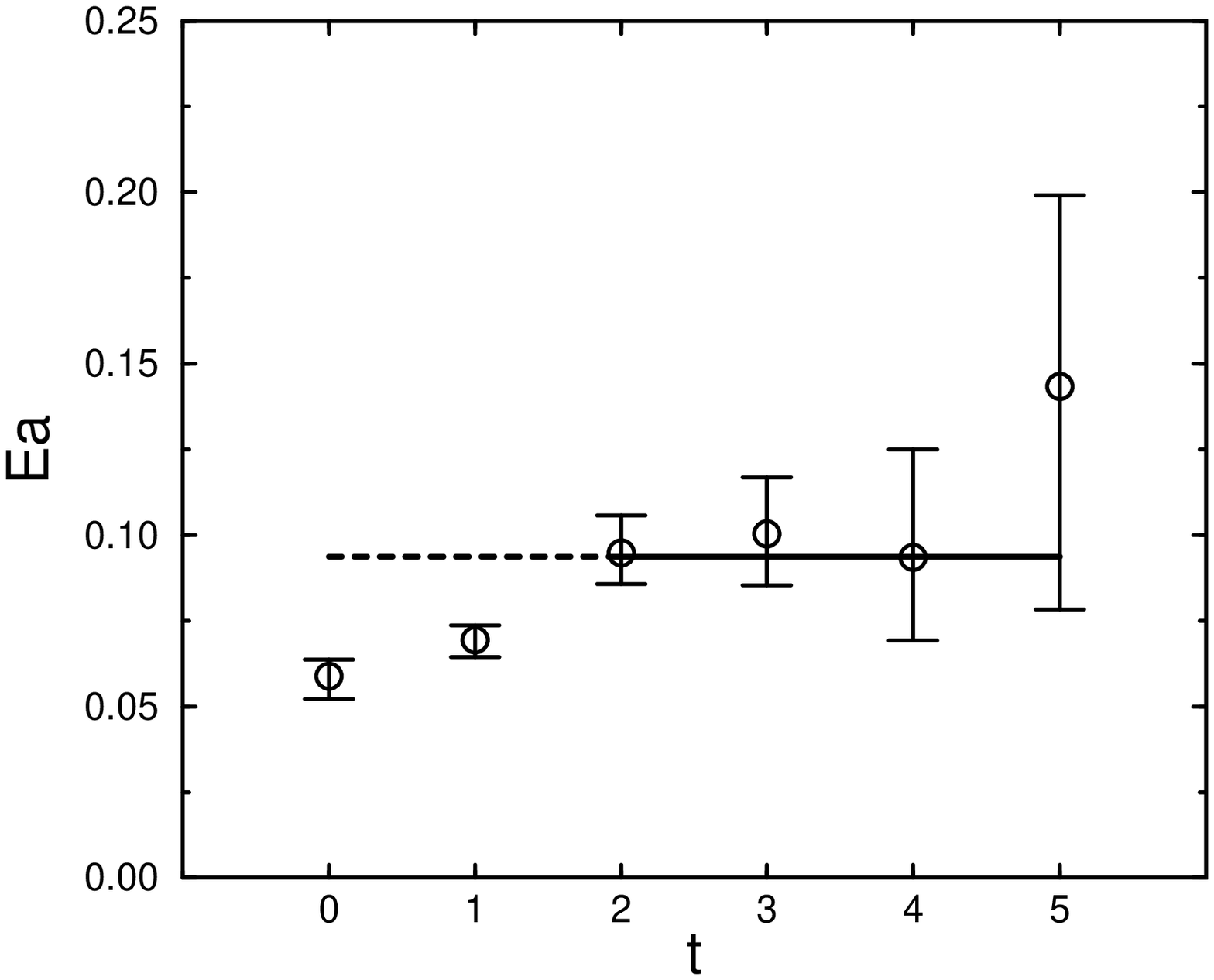}
\caption{Effective mixing  energy and fitted mixing energy for $\beta$ of
5.93 and $\kappa$ of 0.1554 on a lattice $16^2 \times 14 \times 20$.}
\label{fig:Eb593x16k1554}
\end{figure}

\begin{figure}
\epsfxsize=\textwidth
\epsfbox{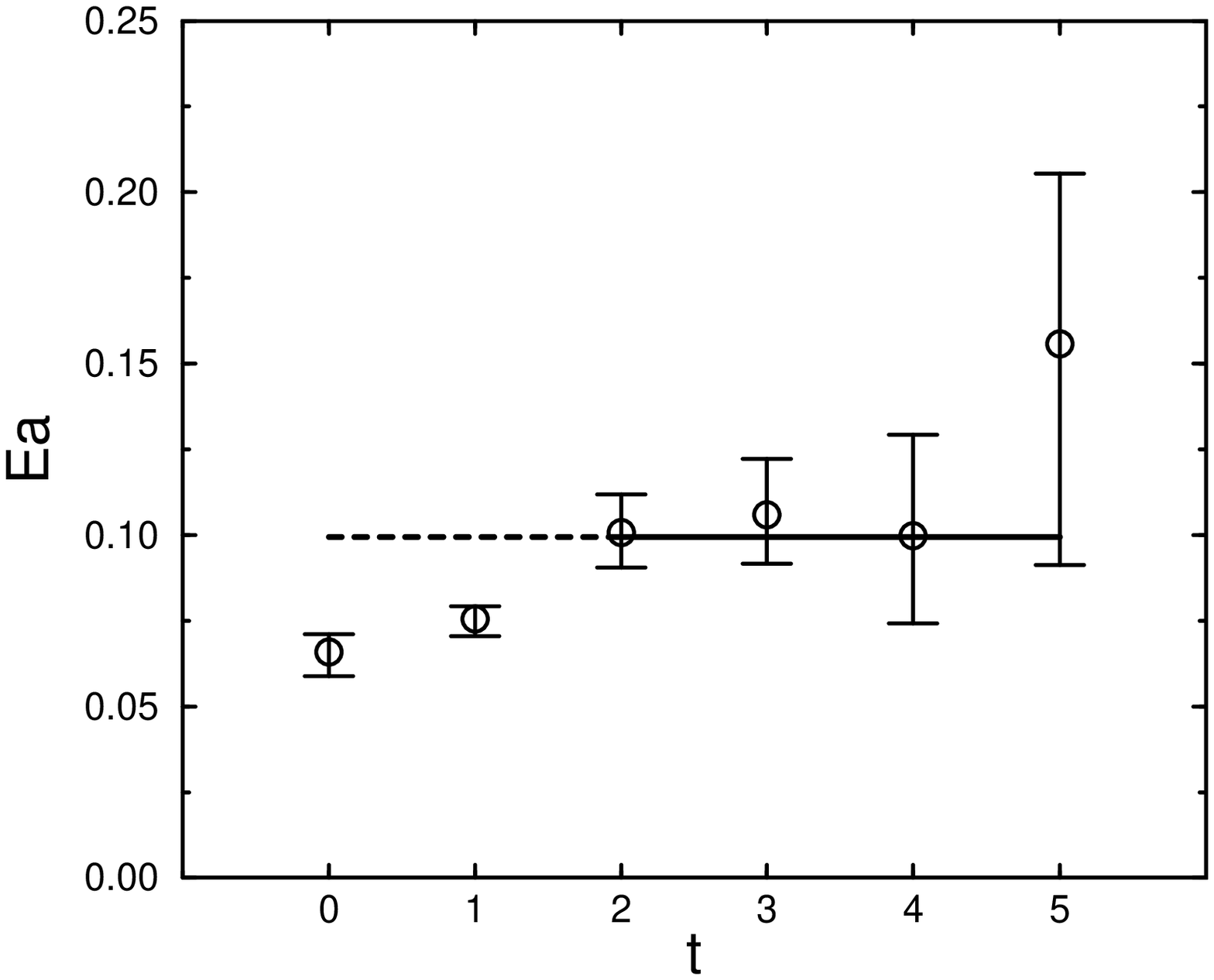}
\caption{Effective mixing  energy and fitted mixing energy for $\beta$ of
5.93 and $\kappa$ of 0.1562 on a lattice $16^2 \times 14 \times 20$.}
\label{fig:Eb593x16k1562}
\end{figure}

\begin{figure}
\epsfxsize=\textwidth
\epsfbox{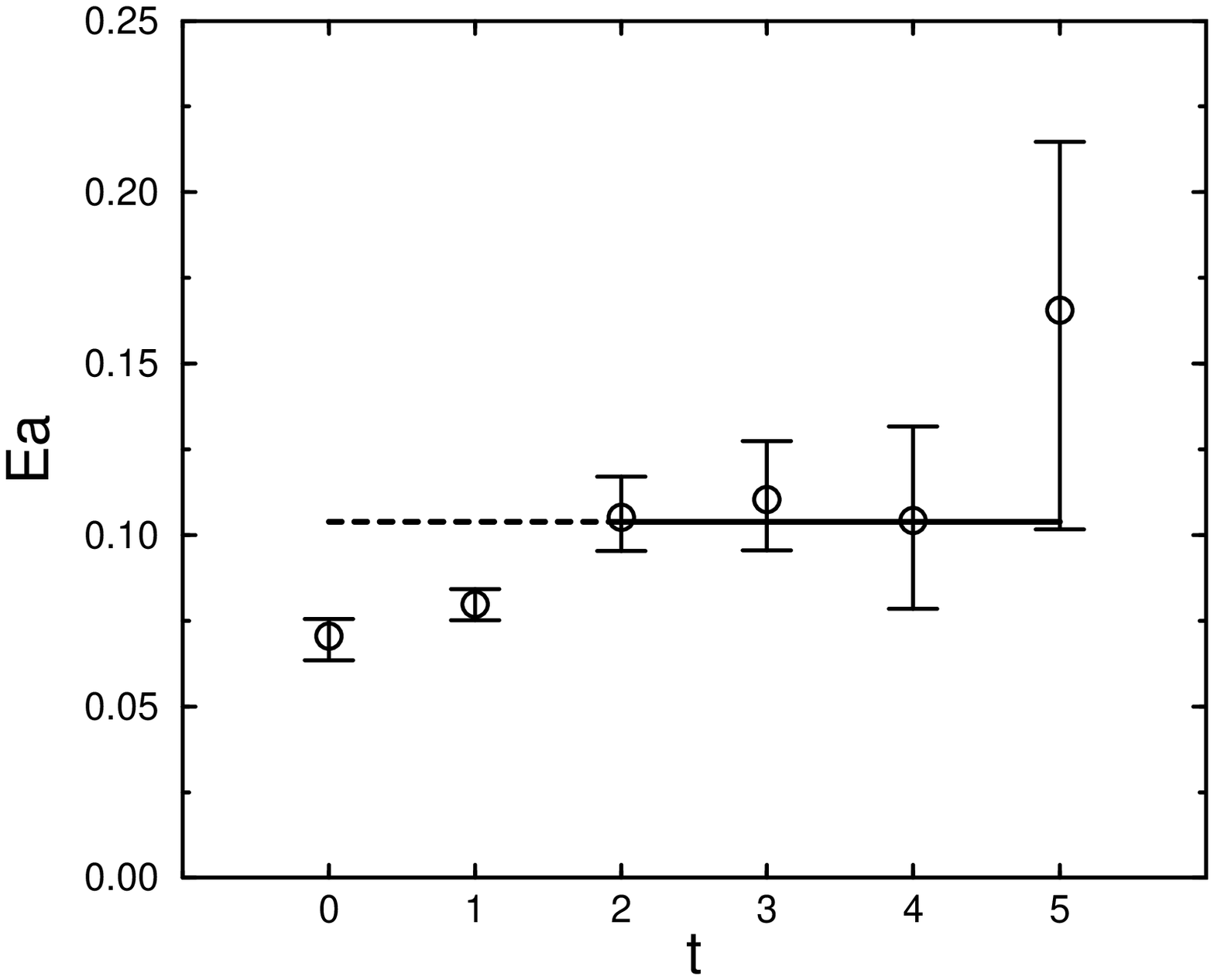}
\caption{Effective mixing  energy and fitted mixing energy for $\beta$ of
5.93 and $\kappa$ of 0.1567 on a lattice $16^2 \times 14 \times 20$.}
\label{fig:Eb593x16k1567}
\end{figure}

\begin{figure}
\epsfxsize=\textwidth
\epsfbox{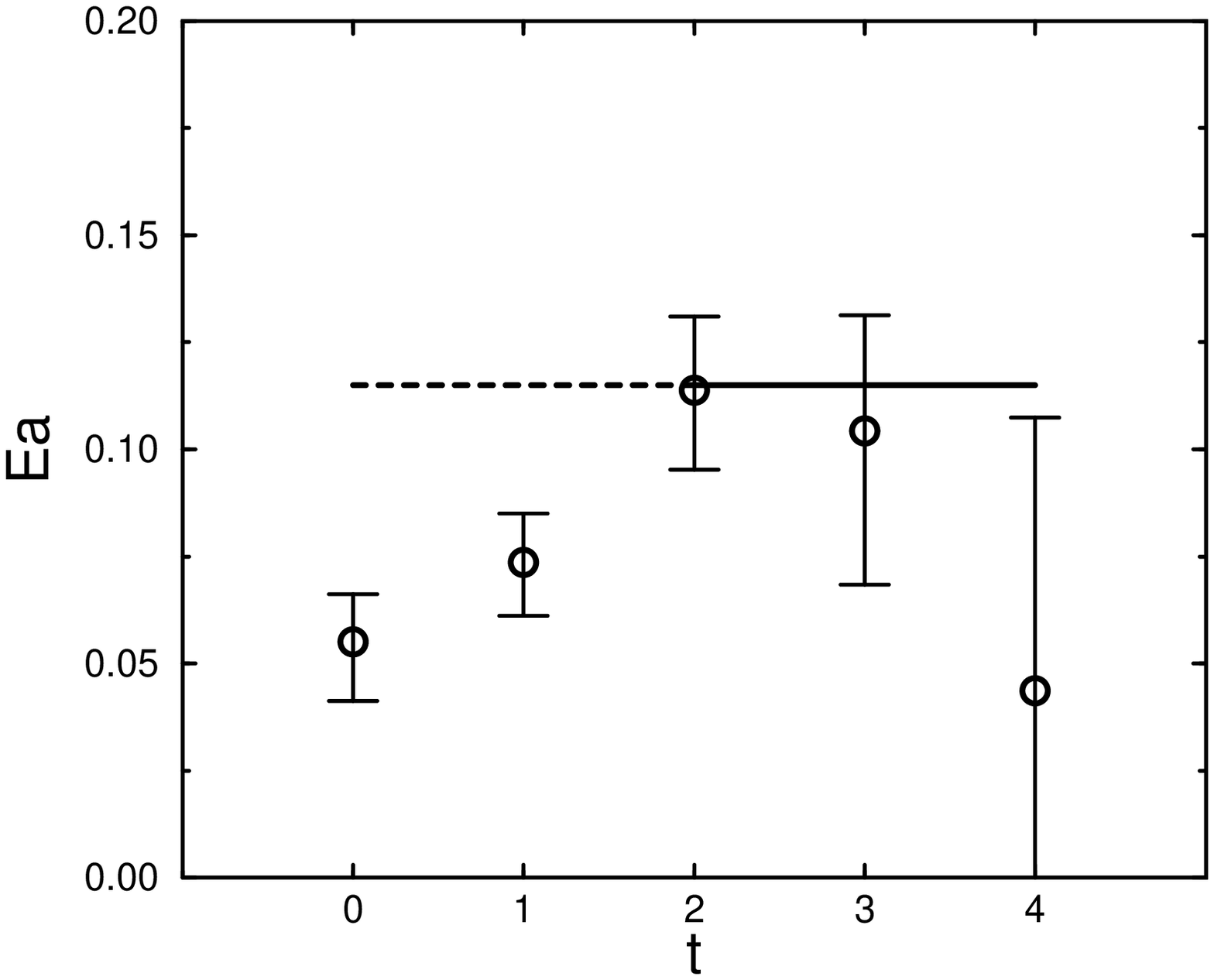}
\caption{Effective mixing  energy and fitted mixing energy for $\beta$ of
5.93 and $\kappa$ of 0.1554 on a lattice $24^4$.}
\label{fig:Eb593x24k1554}
\end{figure}

\begin{figure}
\epsfxsize=\textwidth
\epsfbox{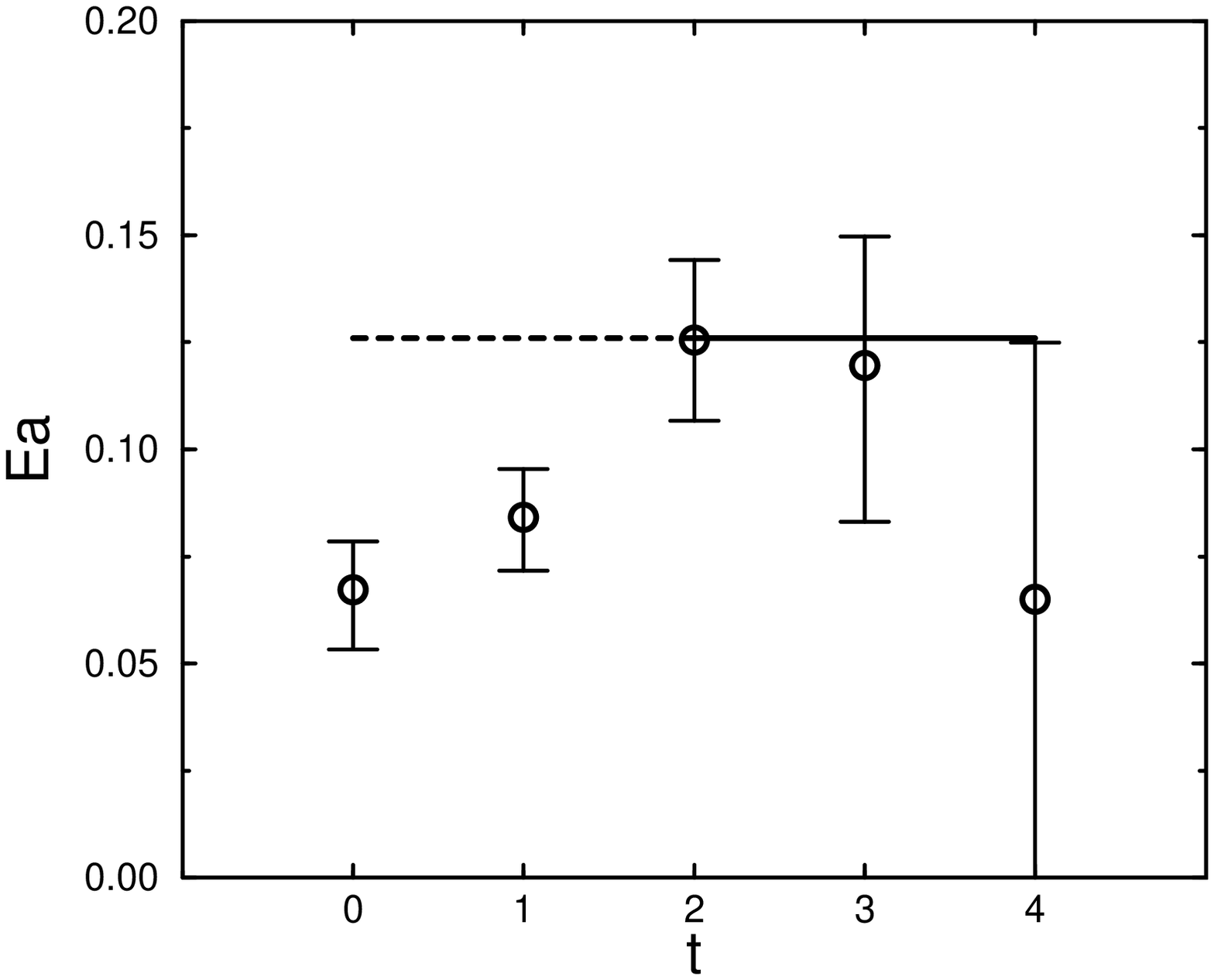}
\caption{Effective mixing  energy and fitted mixing energy for $\beta$ of
5.93 and $\kappa$ of 0.1567 on a lattice $24^4$.}
\label{fig:Eb593x24k1567}
\end{figure}

\begin{figure}
\epsfxsize=\textwidth
\epsfbox{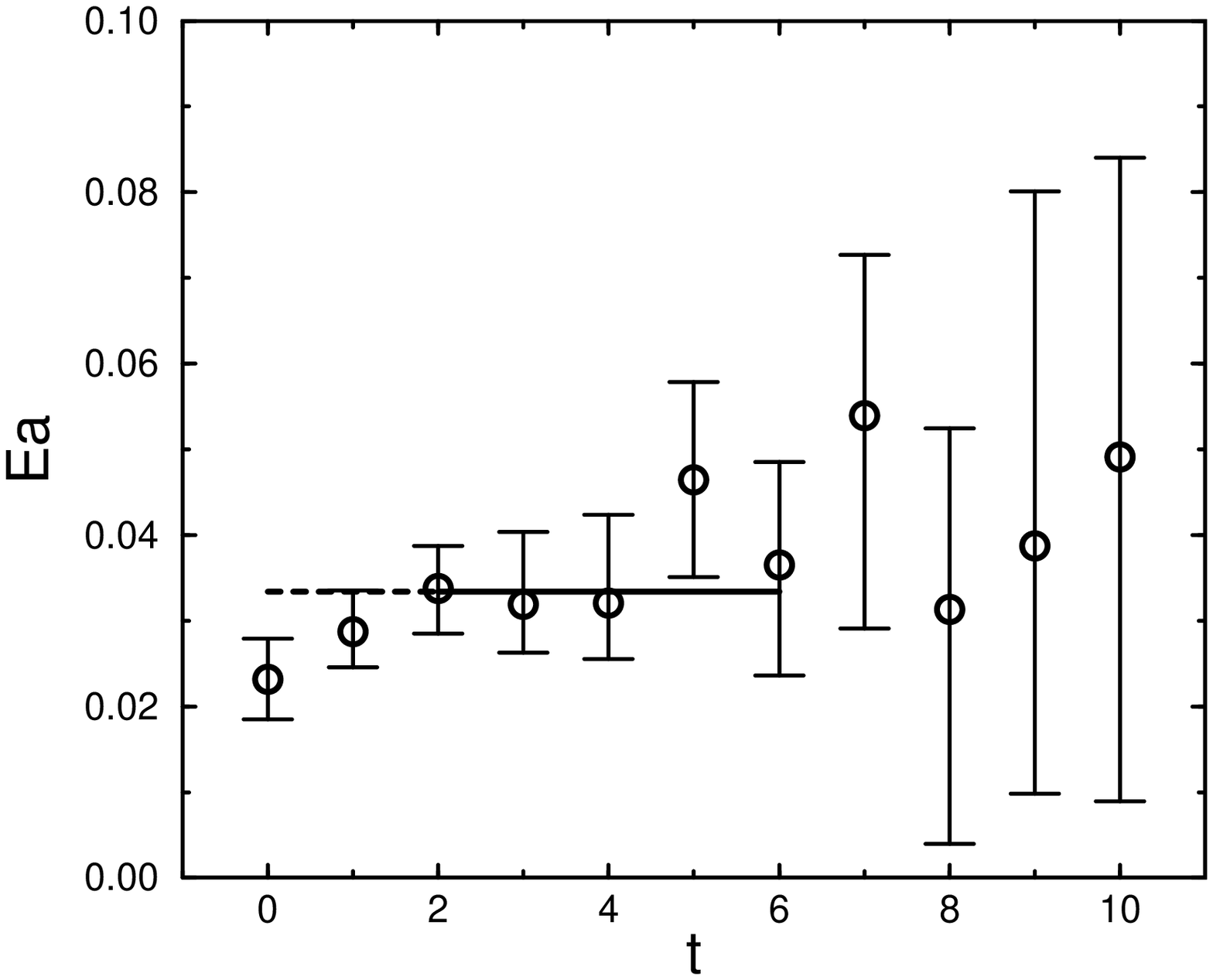}
\caption{Effective mixing  energy and fitted mixing energy for $\beta$ of
6.40 and $\kappa$ of 0.1491 on a lattice $32^2 \times 28 \times 40$.}
\label{fig:Eb64x32k1491}
\end{figure}

\begin{figure}
\epsfxsize=\textwidth
\epsfbox{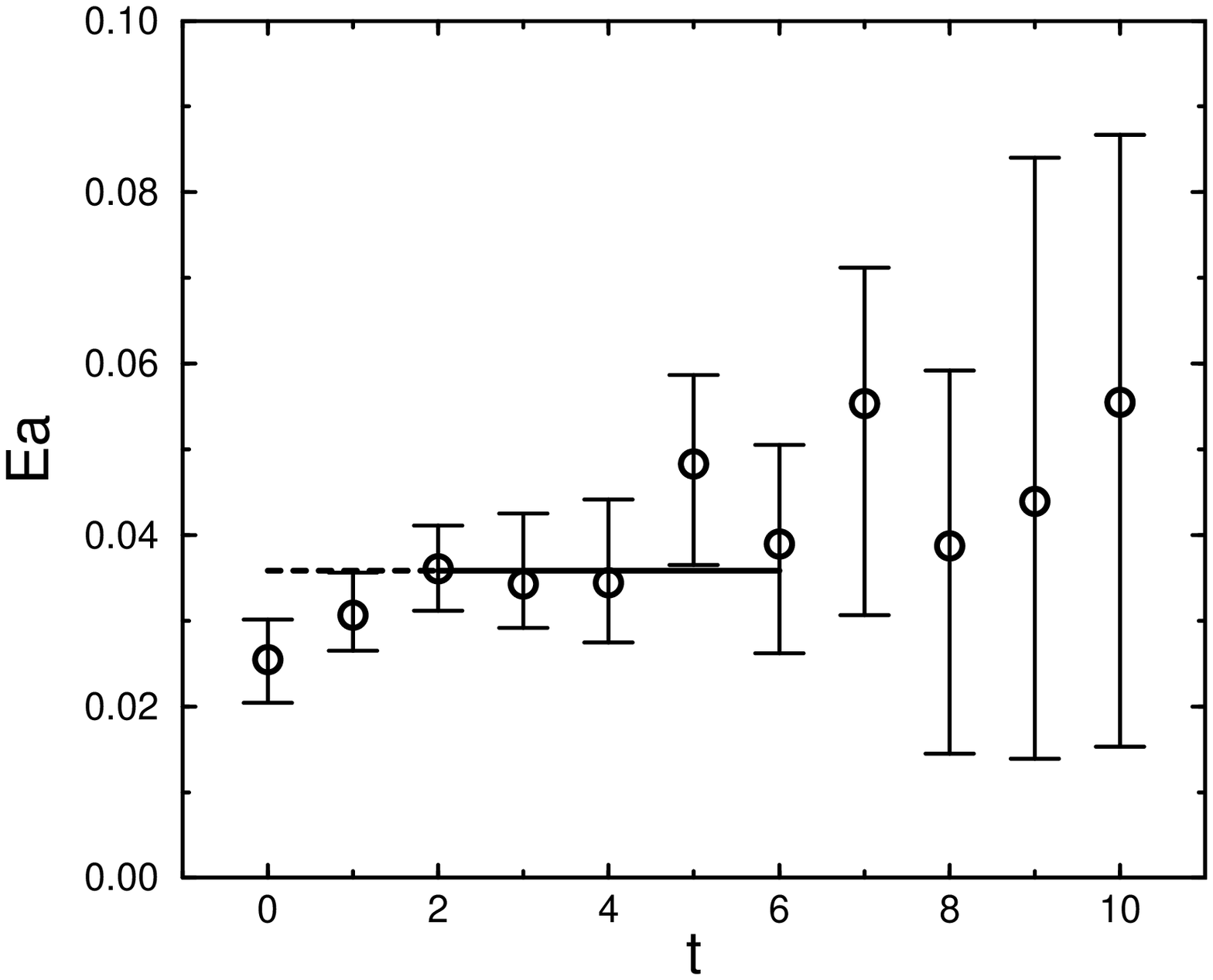}
\caption{Effective mixing  energy and fitted mixing energy for $\beta$ of
6.40 and $\kappa$ of 0.1494 on a lattice $32^2 \times 28 \times 40$.}
\label{fig:Eb64x32k1494}
\end{figure}

\begin{figure}
\epsfxsize=\textwidth
\epsfbox{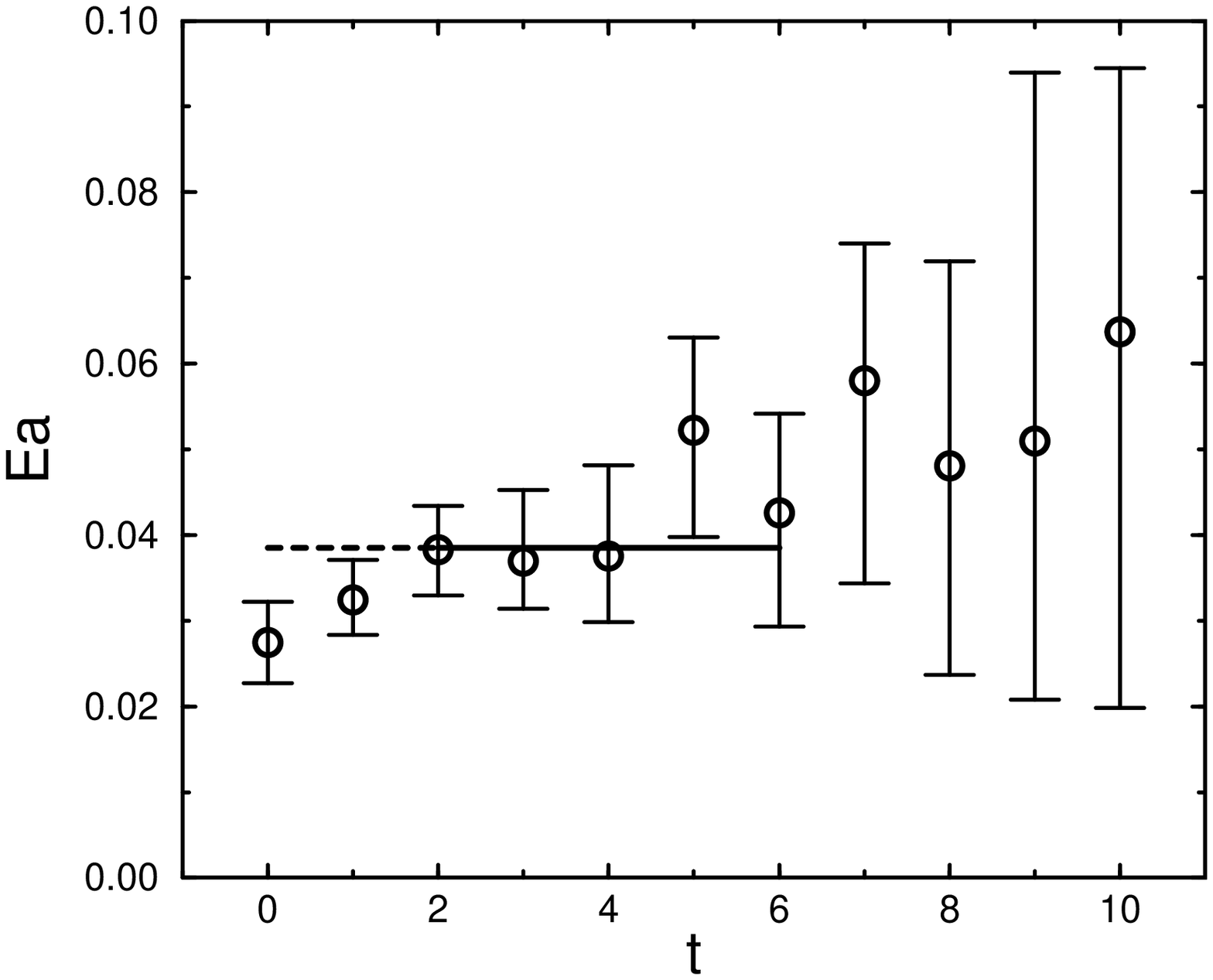}
\caption{Effective mixing  energy and fitted mixing energy for $\beta$ of
6.40 and $\kappa$ of 0.1497 on a lattice $32^2 \times 28 \times 40$.}
\label{fig:Eb64x32k1497}
\end{figure}

Fitting $C_{g\sigma}(t)$ to Eq.~(\ref{offdiag}) using $m_{\sigma}$,
$Z_{\sigma}$, $m_g$ and $Z_g$ from Sections~\ref{sect:mass} and
\ref{sect:glb}, we found the glueball-quarkonium mixing energy in
lattice units $E a$. To choose the $t$ range over which to fit
$C_{g\sigma}(t)$ to Eq.~(\ref{offdiag}), it is convenient to define an
effective mixing energy $E(t)$ by fitting $C_{g\sigma}(t)$ to
Eq.~(\ref{offdiag}) solely at $t$. Typical data for $E(t)$ is shown in
Figures~\ref{fig:Eb593x16k1554} - \ref{fig:Eb64x32k1497}. Trial time
intervals on which to fit $C_{g\sigma}(t)$ to Eq.~(\ref{offdiag}) were
chosen from graphs of $E(t)$, following the fitting procedure of
Section~\ref{sect:mass}, by eliminating large values of $t$ with large
statistical uncertainties in $E(t)$ and eliminating small $t$ at which
$E(t)$ has clearly not yet reached a large $t$ plateau.  Fits with
minimal correlated $\chi^2$ were then made to Eq.~(\ref{offdiag}) on all
subintervals of 2 or more consecutive $t$ within the trial range. The
final fitting interval for each propagator was chosen to give the
smallest $\chi^2$ per degree of freedom.

Final mixing energy values are given in Tables~\ref{tab:Eb57x12} -
\ref{tab:Eb64x32}.  A few of the combinations of $\beta$, $\kappa$ and
lattice size appearing in Table~\ref{tab:kappas} are missing from
Tables~\ref{tab:Eb57x12} - \ref{tab:Eb64x32}.  No results are given for
$\beta$ of 5.7 on the lattice $16^3 \times 24$ since, as mentioned in
Section~\ref{sect:glb}, we were unable to obtain stable values for $m_g$
and $Z_g$ for this data set. We also give no results for $\beta$ of 5.7
and $\kappa$ of 0.1650 on the lattice $12^2 \times 10 \times 24$, for
which the scalar quarkonium fit was poor, and no results for $\beta$ of
6.4 and $\kappa$ of 0.1485 and 0.1488 on $32 \times 28 \times 40$, for
which the Monte Carlo ensembles were too small to give reliable values of
$C_{g\sigma}(t)$.

\begin{figure}
\epsfxsize=\textwidth
\epsfbox{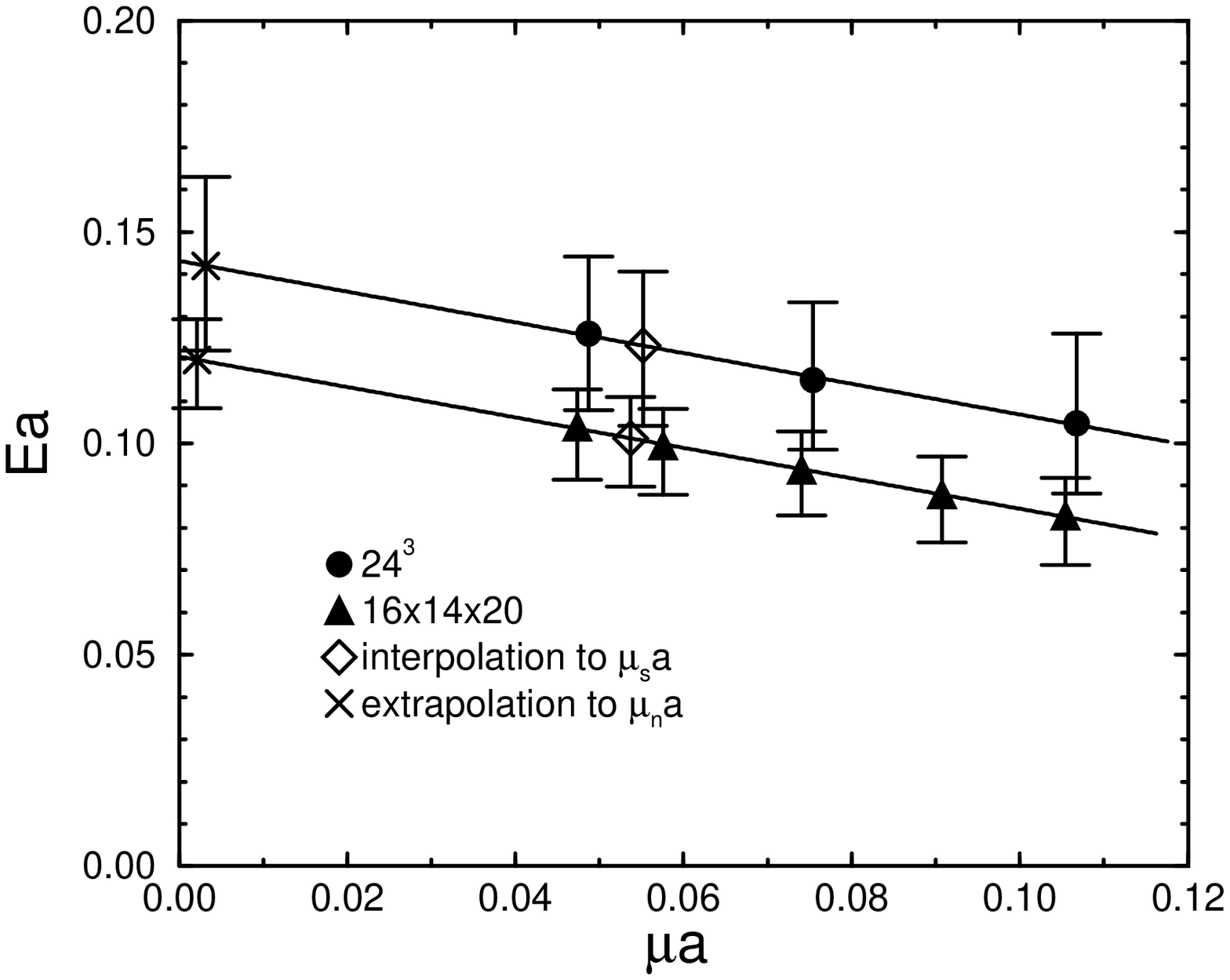}
\caption{Glueball-quarkonium mixing energy as a function of
quark mass for $\beta$ of 5.93.}
\label{fig:E593}
\end{figure}

\begin{figure}
\epsfxsize=\textwidth
\epsfbox{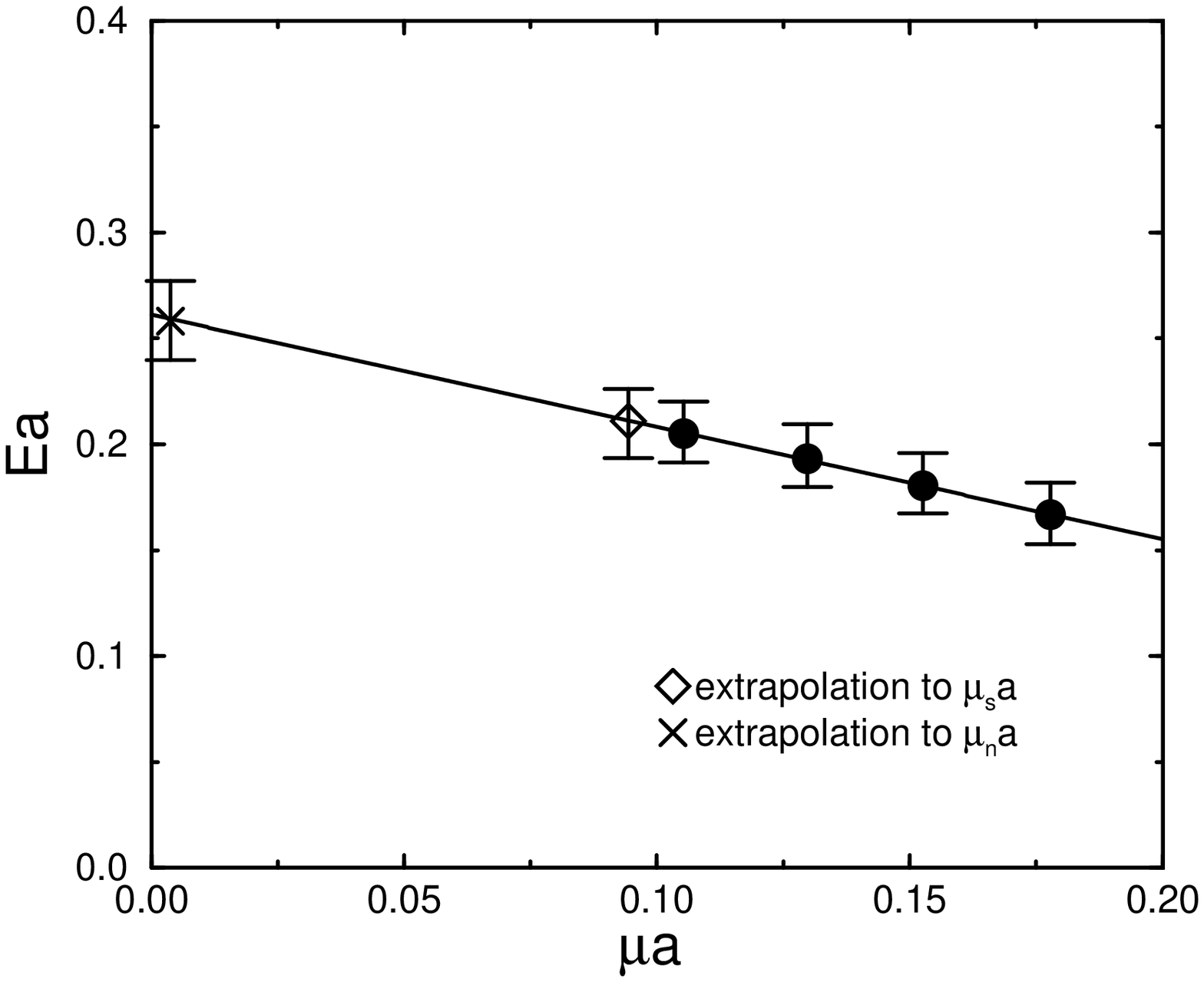}
\caption{Glueball-quarkonium mixing energy as a function of
quark mass for $\beta$ of 5.70.}
\label{fig:E57}
\end{figure}

\begin{figure}
\epsfxsize=\textwidth
\epsfbox{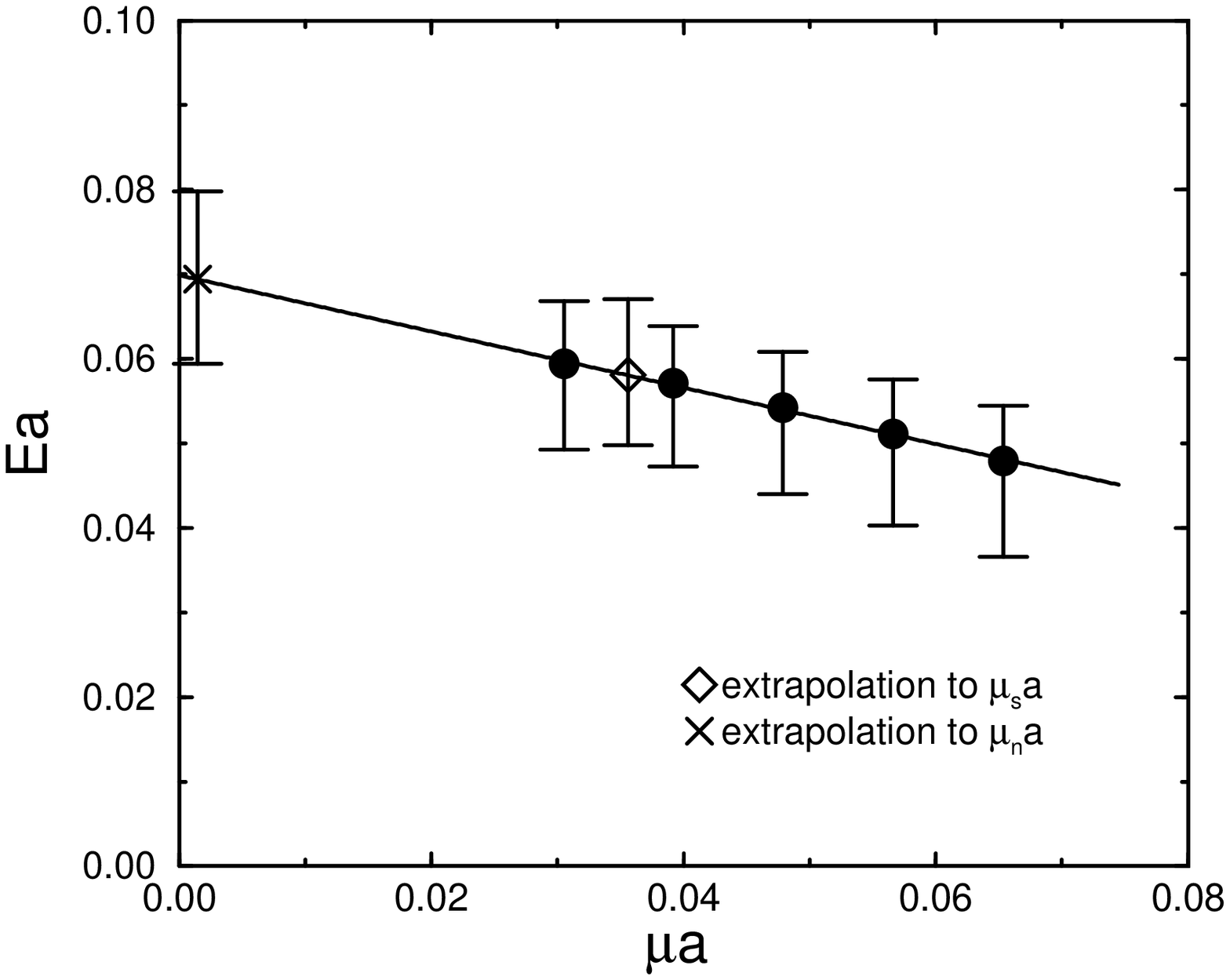}
\caption{Glueball-quarkonium mixing energy as a function of
quark mass for $\beta$ of 6.17.}
\label{fig:E617}
\end{figure}

\begin{figure}
\epsfxsize=\textwidth
\epsfbox{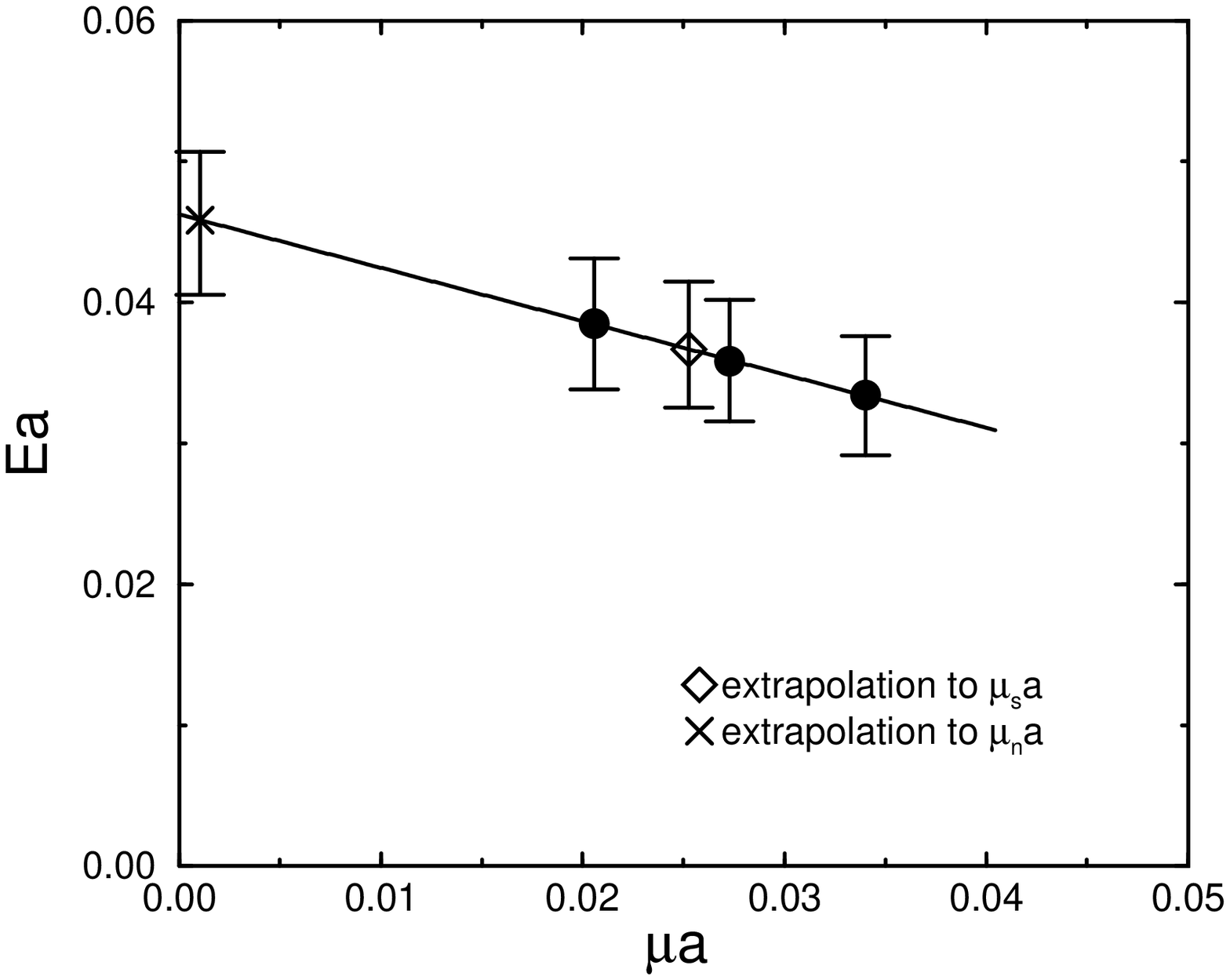}
\caption{Glueball-quarkonium mixing energy as a function of
quark mass for $\beta$ of 6.40.}
\label{fig:E64}
\end{figure}

Figure~\ref{fig:E593} shows the quarkonium-glueball mixing energy as a
function of quark mass for the two different lattices with $\beta$ of
5.93. For neither lattice does there appear to be any sign of the
anomalous quark mass dependence found in Figure~\ref{fig:sc593}. The
mixing energies at different quark masses turn out to be highly
correlated and depend quite linearly on quark mass.
Figures~\ref{fig:E57}, \ref{fig:E617} and \ref{fig:E64} show mixing energy as a function
of quark mass for $\beta$ of 5.70, 6.17 and 6.40, respectively.  For
these values of $\beta$ the mixing energy also shows no sign of the
anomalous quark mass dependence exhibited by the scalar quarkonium mass.
The nearly linear dependence of Figure~\ref{fig:E593} is
repeated.  Thus it appears that the mixing energy can be extrapolated
reliably down to the normal quark mass $\mu_n$, defined to be the quark
mass at which $m_{\pi}$ becomes $M_{\pi}$.  

Table~\ref{tab:Emsmn} gives values of the mixing energy interpolated to
the strange quark mass $\mu_s$, extrapolated down to the normal quark mass
$\mu_n$, and of the ratio of these two energies.  For the data at $\beta$
of 5.93, the ratio changes by less than 3\% from $L$ of 1.54(4) fm to
$L$ of 2.31(6) fm, a difference consistent with the statistical error.
Thus the ratio has at most small volume dependence and seems
already to be near its infinite volume limit with $L$ around 1.6 fm.

\begin{figure}
\epsfxsize=\textwidth
\epsfbox{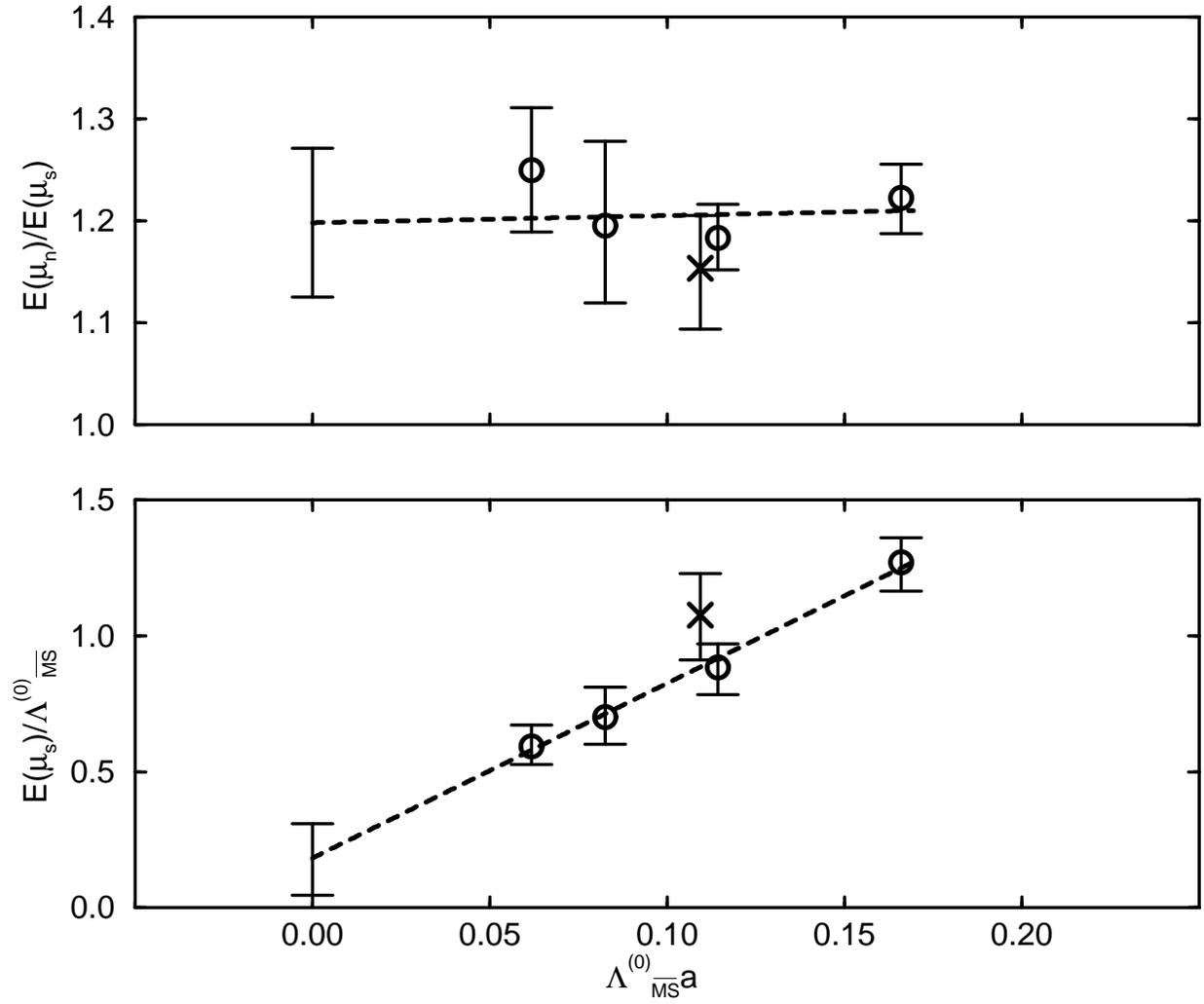}
\caption{Lattice spacing dependence and continuum limit of the
glueball-quarkonium mixing energy $E(\mu_s)$ and of the ratio
$E(\mu_n)/E(\mu_s)$.}
\label{fig:mixcont}
\end{figure}

Figure~\ref{fig:mixcont} shows linear extrapolations to zero lattice
spacing of quarkonium-glueball mixing energy at the strange quark mass
$E(\mu_s)$ and of the ratio $E(\mu_n)/E(\mu_s)$. The zero lattice spacing
prediction $E(\mu_s)$ is 43(31) MeV and of $E(\mu_n)/E(\mu_s)$ is 1.198(72).

\section{Mixed Physical States}\label{sect:states}

We now combine our infinite volume continuum value for $E(\mu_n)/E(\mu_s)$
with a simplified treatment of the mixing among valence approximation
glueball and quarkonium states which arises in full QCD from
quark-antiquark annihilation.  The simplification we introduce is to
permit mixing only between the lightest scalar glueball and the lowest
lying discrete quarkonium states.  We ignore mixing between the lightest
glueball and excited quarkonium states or multiquark continuum states,
and we ignore mixing between the lightest quarkonium states and excited
glueball states or continuum states containing both quarks and
glueballs.

Excited quarkonium and glueball states and states containing both quarks
and glueballs are expected to be high enough in mass that their effect
on the lowest lying states will be much smaller than the effect of
mixing of the lowest lying states with each other. On the other hand, as
mentioned earlier, according to the systematic version of the valence
approximation described in Ref.~\cite{Lee99}, the additional feedback
into mixing among the lowest discrete quarkonium and glueball states
arising as a consequence of the coupling, omitted from our simplified
mixing, of the lowest glueball and scalar quarkonium states to continuum
multi-meson states is a quark loop correction to the direct
glueball-quarkonium mixing amplitude which our simplified mixing
includes.  For low energy QCD properties there is a reasonable amount of
phenomenological evidence that such quark loop corrections are
relatively small.

The structure of the Hamiltonian coupling together the scalar glueball,
the scalar $s\overline{s}$ and the scalar $n\overline{n}$ isosinglet
becomes
\begin{displaymath}
\left|
\begin{array}{ccc}
m_g &  E(\mu_s) & \sqrt{2} r E(\mu_s) \\
E(\mu_s) & m_{\sigma}(\mu_s) & 0 \\
\sqrt{2} r E(\mu_s) & 0 & m_{\sigma}(\mu_n).
\end{array}
\right|
\end{displaymath}
Here $r$ is the ratio $E(\mu_n)/E(\mu_s)$ which we found to be
1.198(72), and $m_g$, $m_{\sigma}(\mu_s)$ and $m_{\sigma}(\mu_n)$
are, respectively, the glueball mass, the $s\overline{s}$ quarkonium
mass and the $n\overline{n}$ quarkonium mass before mixing.

The three unmixed mass parameters we take as unknowns. We will also
treat $E(\mu_s)$ as an unknown since the fractional error bar on our
measured value is large. The four unknown parameters can now be determined from
four observed masses. To leading order in the valence approximation,
with valence quark-antiquark annihilation turned off, corresponding
isotriplet and isosinglet states composed of $u$ and $d$ quarks will be
degenerate. For the scalar meson multiplet, the isotriplet
$(u\overline{u} - d\overline{d})/\sqrt{2}$ state has a mass reported by
the Crystal Barrel collaboration to be 1470(25) MeV~\cite{Amsler1}.
Thus we take $m_{\sigma}(\mu_n)$ to be 1470(25) MeV.  In addition, the
Crystal Barrel collaboration finds an isosinglet mass of 1390(30)
MeV~\cite{Amsler1} from one recent analysis and 1380(40)
MeV~\cite{Abele} from another.	Mark III finds 1430(40)
MeV~\cite{MarkIII}.  We take the mass of the physical mixed state with
largest contribution coming from $n\overline{n}$ to be 1404(24) MeV, the
weighted average of 1390(30) MeV and 1430(40) MeV.  The mass of the
physical mixed states with the largest contributions from
$s\overline{s}$ we take as the mass of $f_0(1500)$, for which the
Particle Data Group's averaged value is 1505(9) MeV.  The mass of the
physical mixed state with the largest contributions from the glueball we
take as the Particle Data group's averaged mass of $f_0(1710)$, 1697(4)
MeV.

Adjusting the parameters in the matrix to give the physical eigenvalues
we just specified, $m_g$ becomes 1622(29) MeV, $m_{\sigma}(\mu_s)$
becomes 1514(11) MeV, and $E(\mu_s)$ becomes 64(13) MeV, with error bars
including the uncertainties in the four input physical masses. The
unmixed $m_g$ is consistent with the world average valence approximation
glueball mass 1656(47) MeV, $E(\mu_s)$ is consistent, within large
errors, with our measured value of 43(31) MeV, and $m_{\sigma}(\mu_s)$
is about 13\% above the valence approximation value 1322(42) MeV for
lattice period 1.6 fm.  This 13\% gap is comparable to the largest
disagreement, about 10\%, found between the valence approximation and
experimental values for the masses of light hadrons. As expected from
the discussion of Ref.~\cite{Weingarten97}, the valence approximation
value lies below the number obtained from experiment.

For the three physical eigenvectors we obtain
\begin{eqnarray}
| f_0(1710) > & = & 0.859(54) | g >  + 0.302(52) | s\overline{s}>
+  0.413(87) |n\overline{n}>, \nonumber	 \\
| f_0(1500) > & = & -0.128(52) | g >  + 0.908(37) | s\overline{s}>
-  0.399(113) | n\overline{n}>,   \\
| f_0(1390) > & = &  -0.495(118) | g >	+ 0.290(91)  | s\overline{s}>
+  0.819(89) | n\overline{n}>. \nonumber
\end{eqnarray}
The mixed $f_0(1710)$ has a glueball content of 73.8(9.5)\%, the mixed
$f_0(1500)$ has a glueball content of 1.6(1.4)\% and the mixed
$f_0(1390)$ has a glueball content of 24.5(10.7)\%.  Since, as well
known, the partial width $\Gamma(J/\Psi \rightarrow \gamma + h)$ is a
measure of the size of the gluon component in the wave function of
hadron $h$, our results imply that $\Gamma(J/\Psi \rightarrow
\gamma + f_0(1710))$ should be significantly larger than $\Gamma(J/\Psi
\rightarrow \gamma + f_0(1390))$ and $\Gamma(J/\Psi \rightarrow \gamma +
f_0(1390))$ should be significantly larger than $\Gamma(J/\Psi
\rightarrow \gamma + f_0(1500))$. These predictions are supported by a
recent reanalysis of Mark III data~\cite{MarkIII}.
In addition, in the state vector for $f_0(1500)$, the relative negative sign
between the $s\overline{s}$ and $n\overline{n}$ components will lead, by
interference, to a suppression of the partial width for this state to
decay to $K\overline{K}$. Assuming SU(3) flavor symmetry for the two
pseudoscalar decay coupling of the scalar quarkonium states, the total
$K\overline{K}$ rate for $f_0(1500)$ is suppressed by a factor of
0.39(16) in comparison to the $K\overline{K}$ rate for an unmixed
$s\overline{s}$ state. This suppression is consistent, within
uncertainties with the experimentally observed suppression.

\section{Glueball Decay Coupling}\label{sect:decay}

We now consider briefly the contribution to scalar glueball decay to
pseudoscalar quarkonium pairs arising from quarkonium-glueball mixing.

In Ref.~\cite{Sexton95} a calculation of scalar decay to pseudoscalar
quarkonium pairs was done on a spatial lattice of $16^3$ at $\beta$ of
5.70 and $\kappa$ of 0.1650 and 0.1675. For these parameters, the
results of Section~\ref{sect:mass} imply the lightest scalar quarkonium
state is significantly heavier than the lightest scalar glueball.  It is
not hard to show that in this circumstance, the valence approximation
decay calculation includes, to first order in the quarkonium-glueball
mixing energy, the contribution arising from mixing of the scalar
glueball with scalar quarkonium.  This first order contribution is
\begin{eqnarray}
\label{deltalambda}
\Delta \lambda( g \rightarrow \pi + \pi), & = & 
  \frac{E}{m_{\sigma} - m_g} \lambda( \sigma \rightarrow \pi + \pi), 
\end{eqnarray}
where, as before, $\sigma$ is the lightest scalar quark-antiquark state
and $\pi$ is the lightest pseudoscalar quark-antiquark state all with a
single common value of $\kappa$.

Although we do not have values for $\lambda( \sigma \rightarrow \pi +
\pi)$ at $\beta$ of 5.70, a rough estimate of the order of magnitude of
$\Delta \lambda( \sigma \rightarrow \pi + \pi)$ can be made by taking
$\lambda( \sigma \rightarrow \pi + \pi)$ from experiment.  Assuming
SU(3) flavor symmetry for scalar quarkonium decay couplings, the
observed decay width of the scalar $K^*(1430)$ yields $\lambda( \sigma
\rightarrow \pi + \pi)$ of about 8 GeV.  Combining this number with $E
a$ of about 0.2, and $m_{\sigma} - m_g$ of about 0.3, we get $\Delta
\lambda (g \rightarrow \pi + \pi)$ of about 5 GeV.  The $\lambda (g
\rightarrow \pi + \pi)$ found in Ref.~\cite{Sexton95} range from about
1.5 to 3 GeV. It thus highly probable that the glueball decay couplings
of Ref.~\cite{Sexton95} include significant contributions from mixing of
the scalar glueball with scalar quarkonium. It appears possible that the
decay couplings may arise entirely from the mixing contribution.  A
lattice calculation of $\lambda( \sigma \rightarrow \pi + \pi)$ would
confirm or refute this possibility. If glueball decay were found at
$\beta$ of 5.70 to occur entirely through mixing, a reasonable guess
would be that this is also the decay mechanism in the real world. 

\begin{table}
\begin{tabular}{cccc}
\hline
$\beta$ & lattice & a (fm) & period (fm) \\
\hline
5.70 & $12^2 \times 10 \times 24$ & 0.140(4) & 1.68(5)\\
\hline
5.70 & $16^3 \times 24$	 & 0.140(4) & 2.24(6) \\
\hline
5.93 & $16^2 \times 14 \times 20$ & 0.0961(25) & 1.54(4) \\
\hline
5.93 & $24^4$ & 0.0961(25) & 2.31(6)\\
\hline
6.17 & $24^2 \times 20 \times 32$ & 0.0694(18) & 1.74(5) \\
\hline
6.40 & $32^2 \times 28 \times 40$ & 0.0519(14) & 1.66(5) \\
\hline
\end{tabular}
\caption{For each $\beta$ and lattice structure, the
corresponding lattice spacing and lattice period in
the two (or three) equal space directions in fermi.}
\label{tab:lattices}
\end{table}

\begin{table}
\begin{tabular}{cccc}
\hline
$\beta$ & lattice & $\kappa$ & ensemble size\\
\hline
5.70 & $12^2 \times 10 \times 24$ & 0.1600, 0.1613,
0.1625, 0.1638 & 2749  \\
 \hline
5.70 & $16^3 \times 24$ & 0.1600 & 3870	 \\
     &			& 0.1625 & 1972  \\ 
     &                  & 0.1650 & 1972, 12186 \\
\hline
5.93 & $16^2 \times 14 \times 20$ & 0.1539, 0.1546,
0.1554, 0.1562, 0.1567 & 2328 \\
\hline
5.93 & $24^4$ & 0.1539, 0.1554, 0.1567 & 1733  \\
\hline
6.17 & $24^2 \times 20 \times 32$ & 0.1508, 0.1523,
0.1516, 0.1520, 0.1524 & 1000 \\
\hline
6.40 & $32^2 \times 28 \times 40$ & 0.1485, 0.1488 & 599 \\
     &				  & 0.1491, 0.1494, 0.1497 & 1003 \\
\hline
\end{tabular}
\caption{Monte Carlo ensemble size for each $\beta$, lattice structure, and 
$\kappa$.}
\label{tab:kappas}
\end{table}

\begin{table}
\begin{tabular}{cccccc}
\hline
$\beta$ & lattice & quark smearing & N & S & C \\
\hline
5.70 & $12^2 \times 10 \times 24$ & 2.0 & &  & 1\\
\hline
5.70 & $16^3 \times 24$	 & 2.0 & &  & 1\\
\hline
5.93 & $16^2 \times 14 \times 20$ & 3.0 & 7 & 6 & \\
\hline
5.93 & $24^4$ & 3.0 & 7 & 6 & \\
\hline
6.17 & $24^2 \times 20 \times 32$ & 4.5 & 7 & 7 & \\
\hline
6.40 & $32^2 \times 28 \times 40$ & 6.0 & 8 & 9 & \\
\hline
\end{tabular}
\caption{Quark smearing parameters, glueball gauge invariant
smearing parameters N and S, and glueball Coulomb gauge smearing parameter C.}
\label{tab:smearing}
\end{table}

\begin{table}
\begin{tabular}{cccc}
\hline
$\kappa$ & mass & t range & $\chi^2/d.o.f.$ \\
\hline
0.1600 & 0.6884(8) & 7 - 10 & 0.03 \\ 
0.1613 & 0.6330(8) & 8 - 11 & 0.06 \\ 
0.1625 & 0.5795(8) & 8 - 11 & 0.08 \\ 
0.1638 & 0.5176(10) & 8 - 10 & 0.01 \\ 
0.1650 & 0.4549(11) & 9 - 11 & 0.00 \\ 
\end{tabular}
\caption{
For $\beta$ of 5.70 on a lattice $12^3 \times 10 \times 24$, for each value of
$\kappa$, fitted pseudoscalar meson masses, time range of fit, and fit's
$\chi^2$ per degree of freedom.}
\label{tab:psb57x12}
\end{table}

\begin{table}
\begin{tabular}{cccc}
\hline
$\kappa$ & mass & t range & $\chi^2/d.o.f.$ \\
\hline
0.1625 & 0.5795(4) & 7 - 10 & 0.37 \\ 
0.1650 & 0.4560(5) & 7 - 10 & 0.25 \\ 
\end{tabular}
\caption{
For $\beta$ of 5.70 on a lattice $16^3 \times 24$, for each value of
$\kappa$, fitted pseudoscalar meson masses, time range of fit, and fit's
$\chi^2$ per degree of freedom.}
\label{tab:psb57x16}
\end{table}
 
\begin{table}
\begin{tabular}{cccc}
\hline
$\kappa$ & mass & t range & $\chi^2/d.o.f.$ \\
\hline
0.1600 & 0.6888(5) & 6 - 8 & 0.00 \\ 
0.1650 & 0.4572(3) & 7 - 9 & 0.08 \\ 
\end{tabular}
\caption{
For $\beta$ of 5.70 on a lattice $16^3 \times 24$, for each value of
$\kappa$, fitted pseudoscalar meson masses, time range of fit, and fit's
$\chi^2$ per degree of freedom obtained from propagators not using
random sources.}
\label{tab:psb57x16wdt}
\end{table}

\begin{table}
\begin{tabular}{cccc}
\hline
$\kappa$ & mass & t range & $\chi^2/d.o.f.$ \\
\hline
0.1539 & 0.4835(5) & 6 - 9 & 1.01 \\ 
0.1546 & 0.4456(5) & 6 - 9 & 0.86 \\ 
0.1554 & 0.3996(6) & 6 - 9 & 0.63 \\ 
0.1562 & 0.3496(7) & 6 - 9 & 0.38 \\ 
0.1567 & 0.3154(7) & 6 - 8 & 0.22 \\ 
\end{tabular}
\caption{
For $\beta$ of 5.93 on a lattice $16^2 \times 14 \times 20$, for each value of
$\kappa$, fitted pseudoscalar meson masses, time range of fit, and fit's
$\chi^2$ per degree of freedom.}
\label{tab:psb593x16}
\end{table}

\begin{table}
\begin{tabular}{cccc}
\hline
$\kappa$ & mass & t range & $\chi^2/d.o.f.$ \\
\hline
0.1539 & 0.4820(4) & 8 - 10 & 0.40 \\ 
0.1554 & 0.3982(4) & 8 - 11 & 0.26 \\ 
0.1567 & 0.3147(4) & 8 - 10 & 0.10 \\ 
\end{tabular}
\caption{
For $\beta$ of 5.93 on a lattice $24^4$, for each value of
$\kappa$, fitted pseudoscalar meson masses, time range of fit, and fit's
$\chi^2$ per degree of freedom.}
\label{tab:psb593x24}
\end{table}

\begin{table}
\begin{tabular}{cccc}
\hline
$\kappa$ & mass & t range & $\chi^2/d.o.f.$ \\
\hline
0.1508 & 0.3348(6) & 5 - 14 & 1.05 \\ 
0.1512 & 0.3094(6) & 5 - 14 & 1.22 \\ 
0.1516 & 0.2826(6) & 5 - 14 & 1.39 \\ 
0.1520 & 0.2541(6) & 5 - 14 & 1.56 \\ 
0.1524 & 0.2229(7) & 5 - 14 & 1.69 \\ 
\end{tabular}
\caption{
For $\beta$ of 6.17 on a lattice $24^2 \times 20 \times 32$, for each
value of $\kappa$, fitted pseudoscalar meson masses, time range of fit,
and fit's $\chi^2$ per degree of freedom.}
\label{tab:psb617x24}
\end{table}

\begin{table}
\begin{tabular}{cccc}
\hline
$\kappa$ & mass & t range & $\chi^2/d.o.f.$ \\
\hline
0.1485 & 0.2564(6) & 13 - 16 & 0.96 \\ 
0.1488 & 0.2354(6) & 13 - 16 & 0.88 \\ 
0.1491 & 0.2133(7) & 14 - 16 & 0.00 \\ 
0.1494 & 0.1893(7) & 14 - 17 & 0.04 \\ 
0.1497 & 0.1630(8) & 14 - 17 & 0.12 \\ 
\end{tabular}
\caption{
For $\beta$ of 6.40 on a lattice $32^2 \times 28 \times 40$, for each value of
$\kappa$, fitted pseudoscalar meson masses, time range of fit, and fit's
$\chi^2$ per degree of freedom.}
\label{tab:psb64x32}
\end{table}

\begin{table}
\begin{tabular}{cccc}
\hline
$\kappa$ & mass & t range & $\chi^2/d.o.f.$ \\
\hline
0.1600 & 1.343(14) & 3 - 6 & 0.18 \\ 
0.1613 & 1.316(14) & 3 - 5 & 0.55 \\ 
0.1625 & 1.298(18) & 3 - 5 & 1.16 \\ 
0.1638 & 1.295(13) & 2 - 4 & 0.00 \\ 
0.1650 & 1.293(12) & 2 - 4 & 3.16 \\ 
\end{tabular}
\caption{
For $\beta$ of 5.70 on a lattice $12^3 \times 10 \times 24$, for each value of
$\kappa$, fitted scalar meson masses, time range of fit, and fit's
$\chi^2$ per degree of freedom.}
\label{tab:scb57x12}
\end{table}

\begin{table}
\begin{tabular}{cccc}
\hline
$\kappa$ & mass & t range & $\chi^2/d.o.f.$ \\
\hline
0.1625 & 1.299(11) & 3 - 5 & 0.32 \\ 
0.1650 & 1.287(12) & 2 - 4 & 0.00 \\ 
\end{tabular}
\caption{
For $\beta$ of 5.70 on a lattice $16^3 \times 24$, for each value of
$\kappa$, fitted scalar meson masses, time range of fit, and fit's
$\chi^2$ per degree of freedom.}
\label{tab:scb57x16}
\end{table}

\begin{table}
\begin{tabular}{cccc}
\hline
$\kappa$ & mass & t range & $\chi^2/d.o.f.$ \\
\hline
0.1600 & 1.325(15) & 5 - 9 & 0.19 \\ 
0.1650 & 1.278(3) & 2 - 4 & 0.08 \\ 
\end{tabular}
\caption{
For $\beta$ of 5.70 on a lattice $16^3 \times 24$, for each value of
$\kappa$, fitted scalar meson masses, time range of fit, and fit's
$\chi^2$ per degree of freedom obtained from propagators not using
random sources.}
\label{tab:scb57x16wdt}
\end{table}

\begin{table}
\begin{tabular}{cccc}
\hline
$\kappa$ & mass & t range & $\chi^2/d.o.f.$ \\
\hline
0.1539 & 0.860(4) & 4 - 8 & 2.39 \\ 
0.1546 & 0.837(4) & 4 - 8 & 1.93 \\ 
0.1554 & 0.815(5) & 4 - 8 & 1.48 \\ 
0.1562 & 0.812(6) & 3 - 7 & 1.05 \\ 
0.1567 & 0.818(6) & 3 - 5 & 0.84 \\ 
\end{tabular}
\caption{
For $\beta$ of 5.93 on a lattice $16^2 \times 14 \times 20$, for each value of
$\kappa$, fitted scalar meson masses, time range of fit, and fit's
$\chi^2$ per degree of freedom.}
\label{tab:scb593x16}
\end{table}

\begin{table}
\begin{tabular}{cccc}
\hline
$\kappa$ & mass & t range & $\chi^2/d.o.f.$ \\
\hline
0.1539 & 0.851(9) & 7 - 11 & 0.33 \\ 
0.1554 & 0.806(4) & 4 - 11 & 1.40 \\ 
0.1567 & 0.779(6) & 4 - 7 & 1.48 \\ 
\end{tabular}
\caption{
For $\beta$ of 5.93 on a lattice $24^4$, for each value of
$\kappa$, fitted scalar meson masses, time range of fit, and fit's
$\chi^2$ per degree of freedom.}
\label{tab:scb593x24}
\end{table}

\begin{table}
\begin{tabular}{cccc}
\hline
$\kappa$ & mass & t range & $\chi^2/d.o.f.$ \\
\hline
0.1508 & 0.574(4) & 6 - 9 & 0.25 \\ 
0.1512 & 0.559(5) & 6 - 9 & 0.22 \\ 
0.1516 & 0.546(6) & 6 - 11 & 0.22 \\ 
0.1520 & 0.538(8) & 6 - 10 & 0.08 \\ 
0.1524 & 0.547(7) & 4 - 8 & 0.26 \\ 
\end{tabular}
\caption{
For $\beta$ of 6.17 on a lattice $24^2 \times 20 \times 32$, for each
value of $\kappa$, fitted scalar meson masses, time range of fit,
and fit's $\chi^2$ per degree of freedom.}
\label{tab:scb617x24}
\end{table}

\begin{table}
\begin{tabular}{cccc}
\hline
$\kappa$ & mass & t range & $\chi^2/d.o.f.$ \\
\hline
0.1485 & 0.424(3) & 7 - 13 & 0.91 \\ 
0.1488 & 0.411(3) & 7 - 18 & 0.86 \\ 
0.1491 & 0.404(3) & 7 - 18 & 1.23 \\ 
0.1494 & 0.397(4) & 6 - 13 & 1.03 \\ 
0.1497 & 0.407(5) & 5 - 8 & 1.34 \\ 
\end{tabular}
\caption{
For $\beta$ of 6.40 on a lattice $32^2 \times 28 \times 40$, for each value of
$\kappa$, fitted scalar meson masses, time range of fit, and fit's
$\chi^2$ per degree of freedom.}
\label{tab:scb64x32}
\end{table}

\begin{table}
\begin{tabular}{cccc}
\hline
$\beta$ & lattice & $\kappa_s$ & $\kappa_c$ \\
\hline 
5.70 & $12^2 \times 10 \times 24$ 
 & 0.164382(23) & 0.169538(70) \\
5.70 & $12^2 \times 10 \times 24$ 
 & 0.164392(6) & 0.169652(86) \\
5.93 & $16^2 \times 14 \times 20$ 
 & 0.156391(11) & 0.159062(15) \\
5.93 & $24^4$                     
 & 0.156384(7) & 0.159079(17) \\
6.17 & $24^2 \times 20 \times 32$ 
 & 0.152167(11) & 0.153833(18) \\
6.40 & $32^2 \times 28 \times 40$ 
 & 0.149490(6) & 0.150628(17) \\
\end{tabular}
\caption{
Hopping constant at the strange quark mass and at zero quark mass.}
\label{tab:kskc}
\end{table}

\begin{table}
\begin{tabular}{cccc}
\hline
$\beta$ & lattice & mass \\
\hline
5.70 & $12^2 \times 10 \times 24$ & 1.298(14) \\
5.70 & $16^3 \times 24$           & 1.283(5) \\
5.93 & $16^2 \times 14 \times 20$ & 0.811(6) \\
5.93 & $24^4$                     & 0.784(5) \\
6.17 & $24^2 \times 20 \times 32$ & 0.545(6) \\
6.40 & $32^2 \times 28 \times 40$ & 0.400(3) \\
\end{tabular}
\caption{
Scalar quarkonium mass interpolated to the strange quark mass.}
\label{tab:scs}
\end{table}

\begin{table}
\begin{tabular}{ccccccc}
\hline
$\beta$ & lattice & mass & t range & $\chi^2/d.o.f.$ & lattice & mass \\
\hline 
5.70 & $12^2 \times 10 \times 24$ & 0.945(91) & 2 - 4 & 0.05 &
 $16^3 \times 24$ & 0.955(15) \\
5.93 & $16^2 \times 14 \times 20$ & 0.788(21) & 1 - 4 & 0.53 &
 $16^3 \times 24$ & 0.781(11) \\ 
5.93 & $24^4$ & 0.774(23) & 1 - 4 & 1.06 & 
 $16^3 \times 24$ & 0.781(11) \\
6.17 & $24^2 \times 20 \times 32$ & 0.577(39) & 2 - 4 & 0.00 &
 $24^2 \times 20 \times 32$ & 0.559(17) \\
6.40 & $32^2 \times 28 \times 40$ & 0.397(35) & 4 - 6 & 0.88 & 
 $32^2 \times 30 \times 40$ & 0.432(8) 
\end{tabular}
\caption{For each $\beta$ and lattice, scalar glueball mass, time range of
fit, and fit's $\chi^2$ per degree of freedom, compared with masses
obtained elsewhere from larger ensembles for the same $\beta$ and nearly 
equal lattice sizes.}
\label{tab:glb}
\end{table}

\begin{table}
\begin{tabular}{cccc}
\hline
$\kappa$ & mixing energy & t range & $\chi^2/d.o.f.$ \\
\hline
0.1600 & 0.167(15) & 1 - 4 & 0.43 \\
0.1613 & 0.180(14) & 1 - 4 & 0.39 \\
0.1625 & 0.193(15) & 1 - 4 & 0.37 \\
0.1638 & 0.205(14) & 1 - 4 & 0.38 \\
\end{tabular}
\caption{
For $\beta$ of 5.70 on a lattice $12^3 \times 10 \times 24$, for each value of
$\kappa$, fitted  mixing energy, time range of fit, and fit's
$\chi^2$ per degree of freedom.}
\label{tab:Eb57x12}
\end{table}

\begin{table}
\begin{tabular}{cccc}
\hline
$\kappa$ & mixing energy & t range & $\chi^2/d.o.f.$ \\
\hline
0.1539 & 0.083(10) & 2 - 5 & 0.33 \\
0.1546 & 0.088(10) & 2 - 5 & 0.35 \\
0.1554 & 0.094(10) & 2 - 5 & 0.40 \\
0.1562 & 0.099(10) & 2 - 5 & 0.48 \\
0.1567 & 0.104(11) & 2 - 5 & 0.52 \\
\end{tabular}
\caption{
For $\beta$ of 5.93 on a lattice $16^2 \times 14 \times 20$, for each value of
$\kappa$, fitted  mixing energy, time range of fit, and fit's
$\chi^2$ per degree of freedom.}
\label{tab:Eb593x16}
\end{table}

\begin{table}
\begin{tabular}{cccc}
\hline
$\kappa$ & mixing energy & t range & $\chi^2/d.o.f.$ \\
\hline
0.1539 & 0.105(19) & 2 - 4 & 0.77 \\
0.1554 & 0.115(17) & 2 - 4 & 0.69 \\
0.1567 & 0.126(18) & 2 - 4 & 0.52 \\
\end{tabular}
\caption{
For $\beta$ of 5.93 on a lattice $24^4$, for each value of
$\kappa$, fitted  mixing energy, time range of fit, and fit's
$\chi^2$ per degree of freedom.}
\label{tab:Eb593x24}
\end{table}

\begin{table}
\begin{tabular}{cccc}
\hline
$\kappa$ & mixing energy & t range & $\chi^2/d.o.f.$ \\
\hline
0.1508 & 0.048(9) & 3 - 5 & 0.76 \\
0.1512 & 0.051(9) & 3 - 5 & 0.80 \\
0.1516 & 0.054(8) & 3 - 5 & 0.85 \\
0.1520 & 0.057(8) & 3 - 5 & 0.93 \\
0.1524 & 0.059(9) & 3 - 5 & 1.08 \\
\end{tabular}
\caption{
For $\beta$ of 6.17 on a lattice $24^2 \times 20 \times 32$, for each
value of $\kappa$, fitted  mixing energy, time range of fit,
and fit's $\chi^2$ per degree of freedom.}
\label{tab:Eb617x24}
\end{table}

\begin{table}
\begin{tabular}{cccc}
\hline
$\kappa$ & mixing energy & t range & $\chi^2/d.o.f.$ \\
\hline
0.1491 & 0.033(4) & 2 - 6 & 0.61 \\
0.1494 & 0.036(4) & 2 - 6 & 0.59 \\
0.1497 & 0.039(5) & 2 - 6 & 0.63 \\
\end{tabular}
\caption{
For $\beta$ of 6.40 on a lattice $32^2 \times 28 \times 40$, for each value of
$\kappa$, fitted mixing energy, time range of fit, and fit's
$\chi^2$ per degree of freedom.}
\label{tab:Eb64x32}
\end{table}

\begin{table}
\begin{tabular}{ccccc}
\hline
$\beta$ & lattice & $E(\mu_s)$ & $E(\mu_n)$ & $E(\mu_n)/E(\mu_s)$ \\ 
\hline 
5.70 & $12^2 \times 10 \times 24$ 
 & 0.211(16) & 0.258(19) & 1.22(3) \\
5.93 & $16^2 \times 14 \times 20$ 
 & 0.101(11) & 0.120(11) & 1.18(3) \\
5.93 & $24^4$                     
 & 0.123(18) & 0.142(21) & 1.15(5) \\
6.17 & $24^2 \times 20 \times 32$ 
 & 0.058(9) & 0.069(10) & 1.20(8) \\
6.40 & $32^2 \times 28 \times 40$ 
 & 0.037(4) & 0.046(5) & 1.25(6) \\
\end{tabular}
\caption{Quarkonium-glueball mixing energy in lattice units for each $\beta$ and
lattice.}
\label{tab:Emsmn}
\end{table}

\begin{thebibliography}{9}

\bibitem[*]{LANL}Present address: T-8, LANL, Los Alamos, NM 87545.
\bibitem{Sexton95} J.\ Sexton, A.\ Vaccarino and D.\ Weingarten, Phys.\
Rev.\ Lett.\ 75, 4563 (1995).
\bibitem{latestglue} A.\ Vaccarino and D.\ Weingarten, to appear in
Phys.\ Rev.\ D.
\bibitem{Vaccarino} H.\ Chen, J.\ Sexton, A.\ Vaccarino
and D.\ Weingarten, Nucl.\ Phys.\ B (Proc.\ Suppl.) 34, 357 (1994).
\bibitem{Livertal} G.\ Bali, K.\ Schilling, A.\ Hulsebos,
A.\ Irving, C.\ Michael and P.\ Stephenson,
Phys.\ Lett.\ B 309, 378 (1993).
\bibitem{Weingarten94} D.\ Weingarten, Nucl.\ Phys.\ B (Proc.\ Suppl.)
34, 29 (1994).
\bibitem{Morningstar} C.\ Morningstar and M. Peardon, Phys.\ Rev.\ D56,
4043 (1997); hep-lat/9901004, to appear in Phys.\ Rev.\ D.
\bibitem{Brisudova} M.\ Brisudova, L.\ Burakovsky and T.\ Goldman, LANL
preprint LA-UR-97-3794, hep-ph/9712514.
\bibitem{Weingarten97} D.\ Weingarten, Nucl.\ Phys.\ B
(Proc.\ Suppl.) 53, 232 (1997).
\bibitem{Amsler1} C.\ Amsler {\it et al.}, Phys.\ Lett.\ B355, 425 (1995).
\bibitem{Amsler2} C.\ Amsler and F.\ Close, Phys.\ Lett.\ B353 385 (1995);
Phys.\ Rev.\ D53, 295 (1996).
\bibitem{Weingarten98} W.\ Lee and D.\ Weingarten, Nucl.\ Phys.\ B
(Proc.\ Suppl.) 63 A-C, 198 (1998).
\bibitem{Lee97} W.\ Lee and D.\ Weingarten,
Nucl.\ Phys.\ B (Proc. Suppl.) 53, 236 (1997).
\bibitem{Boglione} M.\ Boglione and M.\ Pennington, Phys.\ Rev.\ Lett.\
79, 1998 (1997).
\bibitem{Lee99} W.\ Lee and D.\ Weingarten, Phys.\ Rev.\ D59, 09508 (1999).
\bibitem{Butler} F.\ Butler, H.\ Chen, J.\ Sexton, A.\ Vaccarino and
D.\ Weingarten, Nucl.\ Phys.\ B 430, 179 (1994).
\bibitem{Abele} A.\ Abele {\it et al.}, Phys.\ Lett.\ B385, 425 (1996).
\bibitem{MarkIII} SLAC-PUB-5669, 1991; SLAC-PUB-7163; W.
Dunwoodie, private communication.
\end{thebibliography}
\end{document}